\documentclass[jog]{igs}

\usepackage[utf8]{inputenc}

\usepackage{graphicx}

\usepackage{siunitx}
\usepackage{amsmath, booktabs, geometry}
\usepackage{changepage}
\usepackage{ragged2e}
\usepackage{longtable}

\jourvolume{421}
\jourissue{0}
\jourpubyear{3000}

\usepackage{xcolor}
\usepackage{hyperref}
\hypersetup{
    colorlinks=true,
    linkcolor=black,
    citecolor=black}

\newcommand{\track}[1]{\textcolor{black}{#1}}
\newcommand{\comment}[1]{\textcolor{violet}{#1}}

\begin{document}

%\title[Glacier Sliding]{Glacier basal motion: a matter of scale and setting}
%\title[What is glacier sliding?]{What is glacier sliding? A question of setting and scale}
\title[What is glacier sliding?]{What is glacier sliding?}

%{What is glacier sliding, and which processes are ice-sheet models inverting for?}
%{What is glacier sliding, and should the parameter(isation)s that define it be effectively invariant?}

\author[Law et al.]{Robert LAW$^{1, 2, 3, 4}$, David CHANDLER$^{4,5}$, Philipp VOIGT$^{3, 4}$, Ivan UTKIN$^{1, 2}$, Andreas BORN$^{3,4}$}%Florent GIMBERT$^{4}$,

\affiliation{%
$^1$Laboratory of Hydraulics, Hydrology and Glaciology (VAW), ETH Zurich, Zurich, Switzerland\\
$^2$Swiss Federal Institute for Forest, Snow and Landscape Research (WSL), Birmensdorf, Switzerland\\
$^3$Department of Earth Science, University of Bergen, Bergen, Norway\\
$^4$Bjerknes Centre for Climate Research, Bergen, Norway\\ 
$^5$NORCE Norwegian Research Centre, Bergen, Norway \\

  \email{roblaw@ethz.ch}}

\begin{frontmatter}

\maketitle

\begin{abstract}

Glacier and ice-sheet motion is fundamental to glaciology. However, we still lack a consensus for the optimal way to relate basal velocity to basal traction for large-scale glacier and ice-sheet models (the 'sliding relationship'). Typically, a single tunable coefficient loosely connected to one or a limited number of physical processes is varied spatially to reconcile model output with observations. Yet, process-agnostic studies indicate that the suitability of a given sliding relationship depends on the setting. Here, we suggest that this arises from myriad overlapping setting- and scale-dependent sliding sub-processes, including complicated near-basal stress states not captured by large-scale models, reviewed here as comprising a basal 'sliding layer'. A corresponding 'bulk layer' then accounts for ice deformation only minimally influenced by bed properties. We provide a framework for incorporating arbitrarily many sub-processes within a given region -- separated into normal ('form drag') and tangential ('slip') resistance at the ice-bed interface, stressing that the maximum scale of cavitation is an important contributor to the division between the two. Under reasonable assumptions, our framework implies that sliding relationships should fall within a sum of regularised-Coulomb and power-law components, with a rough-smooth distinction proving more consequential in dictating sliding behaviour than a traditional hard-soft transition. %We briefly discuss the implications of this behaviour for numerical models.
%, with these distinctions in form significantly and progressively influencing model output over model run time. 

%\law{-We want to explain \textit{why} the sheer diversity of sliding processes makes pinning down one that works everywhere all the time so difficult. }

\end{abstract}

\end{frontmatter}

\section{ Background} \label{s:background}

%Bingham et al. 2017 have a concise description of inversion procedures under "Implications for ice-sheet modelling projections"

\emph{The question then occurs, is the viscosity real or apparent?} \textbf{John Tyndall on glacier motion in 1857.} \nocite{Tyndall1857OnGlaciers}

\vspace{5mm}

Glacier sliding is a phenomenologically distinct problem within earth sciences, separated from simpler plane-on-plane sliding problems between two elastic solids by multiple confounding factors. The relative softness of ice allows for both viscous deformation under a non-Newtonian power-law \emph{and} discrete slip at the ice-bed interface, which may occur in a stick-slip or continuous manner. Slip and deformation within a ‘soft’ underlying sediment phase provides an additional contribution to total ice motion in many instances (and is shaped in turn by glacial processes), while ‘hard’ underlying bedrock can constitute a highly topographically irregular base ranging from <1 cm asperities to dramatic fjords, necessitating ice deformation around rigid obstacles over length scales covering roughly five orders of magnitude. A subglacial hydrological system comprised largely of melted ice and varying significantly in space and time may significantly influence the roughness and material characteristics of the substrate itself. Debris entrained within basal ice interacts with the subglacial environment and influences both basal ice properties and the frictional properties of the glacier sole, which may itself not be a strictly discrete surface. These considerations present a wide parameter space for basal boundary conditions, which may also exhibit considerable spatial and temporal variations. 

The net outcome of this complexity is significant difficulty in the derivation of a ‘universal’ or ‘generalised’ sliding relationship applicable to all subglacial environments. For the purposes of large-scale predictive or palaeo numerical modelling applications covering entire glaciers or ice sheets \track{(hereafter `production models') that are the focus of this paper}, a sliding relationship is usually expressed in the form 
\begin{equation}
\tau_b = f(u_b) \: \: or \: \: \tau_b = f(u_b, N)   
\label{Eq:1}
\end{equation}
where \(\tau_b\) is basal traction, \(u_b\) is ice velocity tangential to the modelled ice-bed interface, \(N=p_i - p_w\) is effective pressure, \(p_w\) is subglacial water pressure, and \(p_i\) is the ice overburden pressure (or with \(u_b\) and \(\tau_b\) vectorised for higher dimensions, e.g., Eq. \ref{eq:constitutive_slip}). Alternative formulations of Eq. \ref{Eq:1} exist but are not in common use in production models: Section \ref{s:stickslip} covers rate-and-state formulations (Eq. \ref{eq:ratestate}) while Appendix \ref{A:differential} briefly covers differential formulations (Eq. \ref{eq:differential}). The main focus of our paper is then:

\vspace{5mm}

\emph{How is sliding represented in production models, and, inversely, what processes are actually incorporated in production models' representation of sliding?}

\vspace{5mm}

Understanding the above, and determining an appropriate way to include sliding in production models is central to producing tractable, physically-based projections of the contribution of ice sheets and glaciers to sea-level rise and fresh-water fluxes over the coming centuries. Extant sliding relationships (Fig. 1, Table. 1) influence ice-sheet model output at catchment to entire ice sheet scales, with the distinction between bounded traction (plastic, regularised-Coulomb) and unbounded traction (all other relationships) being particularly consequential (\citealp{Gillet-Chaulet2012GreenlandModel}; \citealp{Parizek2013DynamicAntarctica}; \citealp{Ritz2015PotentialObservations}; \citealp{Tsai2015MarineConditions}; \citealp{Bons2018GreenlandMotion}; \citealp{Kyrke-Smith2018RelevanceAntarctica}; \citealp{Kazmierczak2022SubglacialForcing}; \citealp{Berends2023CompensatingResponse}; \citealp{Lippert2024Modeling2021}; \citealp{Barnes2022TheAntarctica}; \citealp{Trevers2024ApplicationGreenland}; \citealp{vandenAkker2025CompetingLoss}), resulting in up to a 100\% variation in mass loss over 100 years dependent on initialisation \citep{Brondex2019SensitivityLaw, Akesson2021FutureLaw}. Nonetheless, around five\footnote{Some sliding relationships can be considered as simplified versions of one another. For example power-law and pseudo-plastic can be manipulated into linear or plastic relationships, the \(N\) in Budd sliding is effectively subsumed into the \(C\) of power-law sliding during inversion if it is assumed that subglacial water pressure does not vary in time, making this number subjective.} sliding relationships (Table \ref{tab:eqs}), implicitly or explicitly capturing different sliding processes, are used in the Ice Sheet Model Intercomparison Project 6 (ISMIP6) models \citep{Goelzer2020TheISMIP6, Seroussi2020ISMIP6Century}, which guide worldwide policy on sea level rise mitigation. Determining the appropriate relationship is not immediately straightforward to resolve, as the inversion methods that tune for \(C\), the `traction coefficient', will produce reasonable basal stress states whichever equation in Table \ref{tab:eqs} is chosen as the global stress balance condition must be met \citep{Joughin2009BasalData, Arthern2010InitializationProblem}. There is therefore no simple approach to discounting sliding relationships given solely a snapshot of glacier or ice sheet velocity and geometry (a time series is required, Section \ref{s:heuristic}). Errors associated with an inappropriate choice of basal sliding relationship subsequently increase as a model evolves further from its optimised state \citep{Berends2023CompensatingResponse}.

\begin{figure}
\centering{\includegraphics[width=0.45\textwidth]{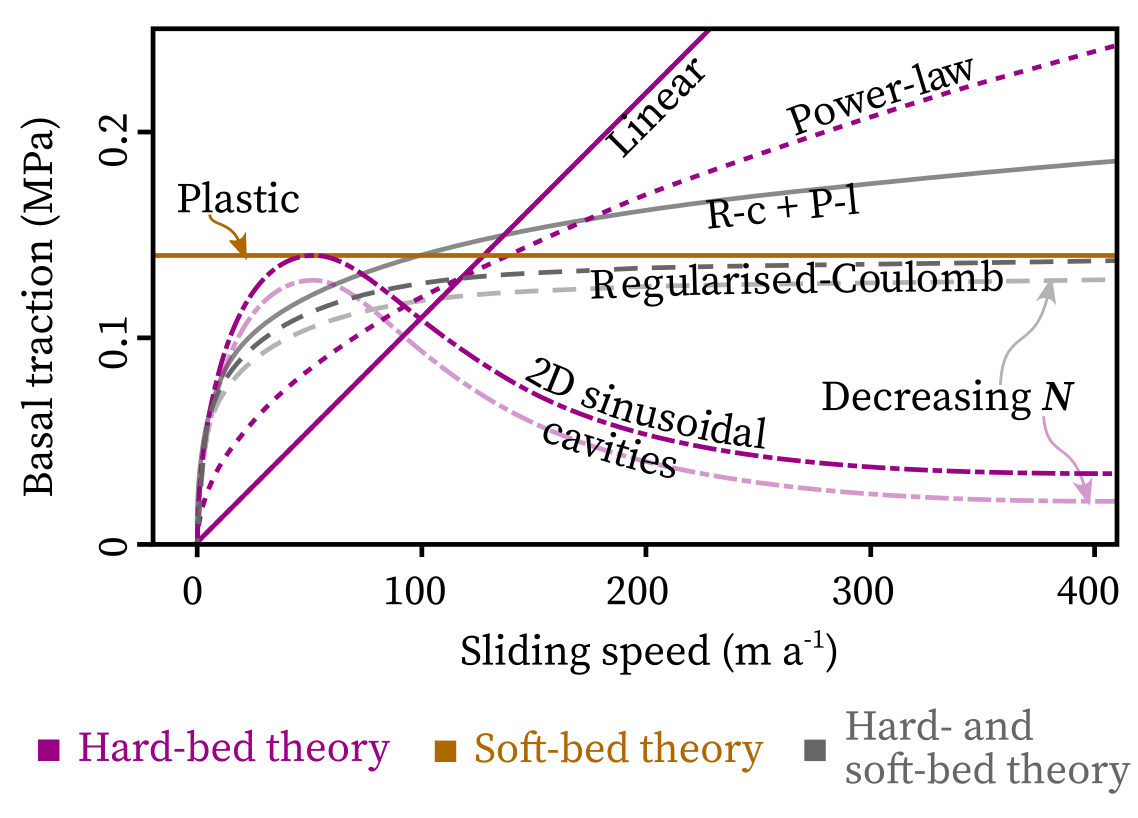}}
\caption{\textbf{Existing sliding parameterisations and how their development relates to soft- and hard-bed theory.} R-c+P-l represents Regularised-Coulomb plus power-law. Traction and velocity values are plausible, but for illustrative purposes only.}
\label{fig:schematic}
\end{figure} 

\begin{table}
    \centering
    \begin{tabular}{|p{2.8cm}|p{4.7cm}|}
    \hline
        Sliding relationship & Equation \\
    \hline
    Linear & \(\tau_b = Cu_{b}\) \ or \ \(\tau_b = NCu_{b}\) \\ 
    Plastic & \( \tau_b = C \) \ or \ \(\tau_b=CN \) \\
    Weertman(-type) \track{\textit{or} power-law} & \(\tau_b = Cu_{b}^{\frac{2}{1+n}}\)  \ or \ \(\tau_b = Cu_{b}^{1/m}\) \\
    Pseudo-plastic & \( \tau_b = \tau_c\frac{u_b^{1/m}}{u_{t}^{1/m}} \) \\
    Budd & \( \tau_b = C\left (  {N}^{q}u_b \right )^{1/m} \) \quad or \quad \(\tau_b=CNu_b^{1/m}\)\\ 
    Regularised-Coulomb & \(\tau_b = C\left(\frac{u_b}{u_b + u_t} \right)^{1/m}\) \\
    \hline
    \end{tabular}
    \caption{\textbf{Selected sliding relationships applied in production models, expressed in one-dimensional form}. All listed sliding relationship are used in \citet{Goelzer2020TheISMIP6} or \citet{Seroussi2020ISMIP6Century}, or \citet{Seroussi2024EvolutionEnsemble} ISMIP6 experiments. In all equations \(\tau_b\) is basal traction and \(u_b\) is the basal ice velocity tangential to the bed. Where present, \(n\) is the exponent used in Glen's flow law (usually 3) and is only used where an explicit link to Glen's flow law is made in the paper proposing the sliding relationship, \(m\) is an exponent often related to Glen's flow law, but not always explicitly, \(N\) is the effective pressure \track{(with more information on implementations in Section \ref{s:hydrology})}, \(C\) is the traction coefficient which may be adapted from its use in the original paper for simplified intercomparison (for example we replace \(1/C\) with \(C\) in the Budd relationship), \(\tau_c\) is the yield stress in the pseudo-plastic relationship, \(q\) is a fitting parameter in the Budd relationship, and \(u_t\) is the threshold velocity in the pseudo-plastic and regularised-Coulomb relationships. The rate-weakening two-dimensional cavities relationship in Fig. \ref{fig:schematic} is not included, as no \track{production} models include rate-weakening behaviour above a given velocity threshold. Depending on convention, the RHS may be negative in the original paper to indicate a traction force opposing velocity. In dimensions higher than 1 the sliding relationship is vectorised (e.g., Eq. \ref{eq:constitutive_slip}). \textbf{Notes: Linear:} Adapted from \citet{Nye1969AApproximation} and \citet{Kamb1970SlidingObservation}. As applied in e.g., \citet{Morlighem2013InversionModel}. The version with \(N\) is used in some \citet{Goelzer2020TheISMIP6} experiments. \textbf{Weertman(-type) or power law:} Equation featuring \(n\) is adapted from \citet{Weertman1957OnGlaciers} where \(n\) is set as 4.2. \(m\) is often used in models instead of \(n\) where \(m\) is usually between 3 and 4 (referred to as power-law here). \citet{Greve2013ResolutionSheet} use an exponential multiplier related to basal temperature (covered further in Section \ref{s:frozenthawed}). \citet{Furst2015Ice-dynamicWarming} include a multiplier connected to surface runoff rates. \textbf{Pseudo-plastic:} Effectively the same as Weertman if \(\tau_c/u_t^{1/m}=C\) but included for frequent use in PISM applications (e.g., \citealp{Aschwanden2016ComplexCaptured}), the pseudo-plastic relationship can be varied between plastic (\(m = \infty\)), linear (\(m = 1\)), or power-law (\(2 \lesssim m \lesssim 4\)) behaviour. \textbf{Regularised-Coulomb:} The simplest form, excluding \(N\), adapted from \citep{Joughin2019RegularizedAntarctica}. Other formulations exist which produce non-negligible differences (Fig. \ref{fig:helanow-zoet}). \textbf{Plastic:} As used in e.g., \citet{Bougamont2014SensitiveBed} and the PISM default \citep{Winkelmann2011TheDescription}. \textbf{Budd:} As developed in \citet{Budd1979EmpiricalSliding} where \(q=1\) and \(m=3\) and used in \citet{Budd1984ASheet} with \(q=2\) and \(m=1\). The simplified Budd formulation is as used in \citet{Choi2022UncoveringDecade}. These formulations are sometimes referred to as Weertman sliding \citep{Greve2013ResolutionSheet}. \citet{Tsai2021ASliding} also produces a relationship that has similarities to Budd sliding.}
    \label{tab:eqs}
\end{table}

%\lawcomment{can a bit more be made about 'scalability' of relationships here as it's used extensively later in the paper.}

In numerical ice-sheet models, these sliding relationships form a critical thread between the processes occurring at the sub-grid-scale and the numerical solution defined at the grid scale (where `grid' refers to discretised model grid cells or mesh elements). In discretised continuum mechanics models, the distinction between sub-grid-scale and grid-scale is usually made using a ‘representative volume element’ that is large enough to be representative of the bulk behaviour of micro-scale processes within the constitutive material it represents. For ice deformation, this is straightforward -- an ice volume of 1 m\textsuperscript{3}, much below the resolution of even the most detailed numerical models, will contain sufficient variability in crystal orientation, size, and impurities as to have effectively the same response to an applied stress field as a nearby 1 m\textsuperscript{3} of the same material. Lab-based experiments using small volumes of ice (e.g., \citealp{Glen1952ExperimentsIce}) can therefore be reasonably applied to much larger-scale models, even if separate issues such as anisotropy and the importance of tertiary creep remain \citep{Adams2021SofteningWater}. %An upper limit also applies as the rheological properties should be consistent across the representative volume. This condition is breached by, for example, large changes in ice temperature over an ice sheet profile in a shallow ice approximation, but is not covered further in this paper.  

There is not, however, an obvious `representative surface element' size for glacier sliding processes, with investigations into glacier sliding covering a range of scales extending over seven orders of magnitude from millimetres to 10s of kilometres (Fig. 2). For example, lab experiments for ice-sediment shearing behaviour may be conducted at a scale of tens of centimetres (e.g., \citealp{Iverson1998Ring-shearBeds}; \citealp{Zoet2020ABeds}), up-scaled to field settings covering an entire glacier where subglacially entrained clasts, variable subglacial hydrology, or bedrock obstacles further influence the relationship between traction and velocity (e.g., \citealp{Iverson1995FlowBeds}; \citealp{Hedfors2003InvestigatingAnalysis}; \citealp{Gimbert2021DoSpeed}), and ultimately applied to ice-sheet models where individual grid cells can exceed 10 km (or have a highly variable size across a given domain) and cover an uncertain and/or heterogeneous subglacial landscape (e.g., \citealp{Kyrke-Smith2018RelevanceAntarctica, Holschuh2020LinkingAntarctica, Paxman2021Neogene-QuaternaryMorphology, Williams2025CalculationsResolution}, Fig. 3). Nonetheless, routinely-used sliding parameterisations in production models \track{are exclusively derived from} one or a limited set of sliding processes at small scales \citep{Goelzer2020TheISMIP6, Seroussi2020ISMIP6Century}, without representing the full range of scale-dependent processes.

%A tau as function of ub is not actually that common in other fields. FEM models will also naturally invalidate scale assumptions \emph{within the same model} if mesh resolution varies

\begin{figure*}
\centering{\includegraphics[width=1\textwidth]{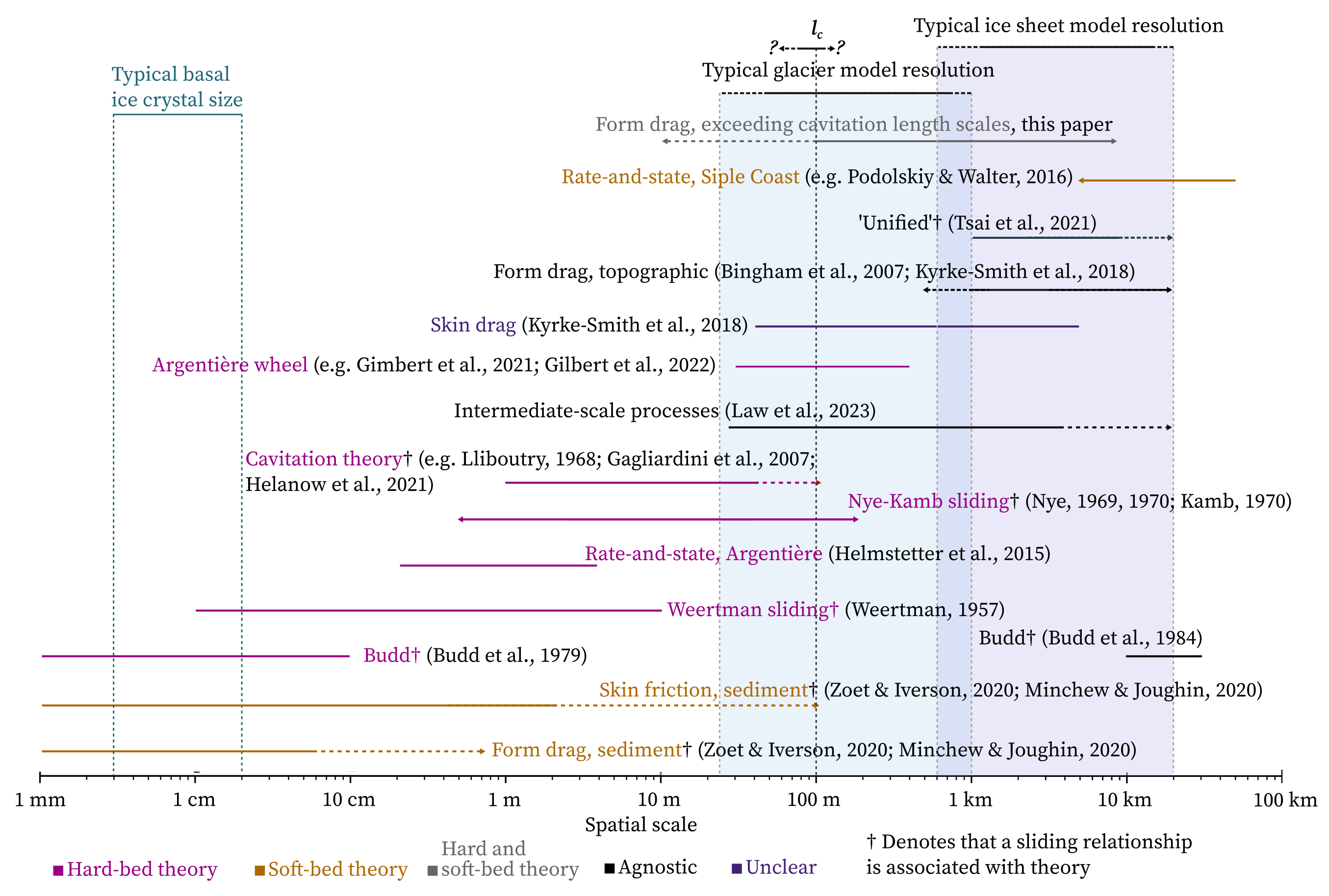}}
\caption{\textbf{Spatial scales for theories and parameterisations of basal sliding.} A solid line indicates coverage as explicitly defined in the associated paper paper while a dashed line indicates probable situation dependent coverage and an arrow indicates extension beyond the scale bar or an uncertain coverage beyond the given limit. \(l_{c}\) is the (likely setting-dependent) upper-limit length-scale for cavitation (Sections \ref{s:innerouterinner}, \ref{s:ikens}). Reasoning behind the positioning of other spatial ranges is provided in Appendix \ref{A:scale}.} 
\label{fig:scale}
\end{figure*}

\begin{figure*}
\centering{\includegraphics[width=0.8\textwidth]{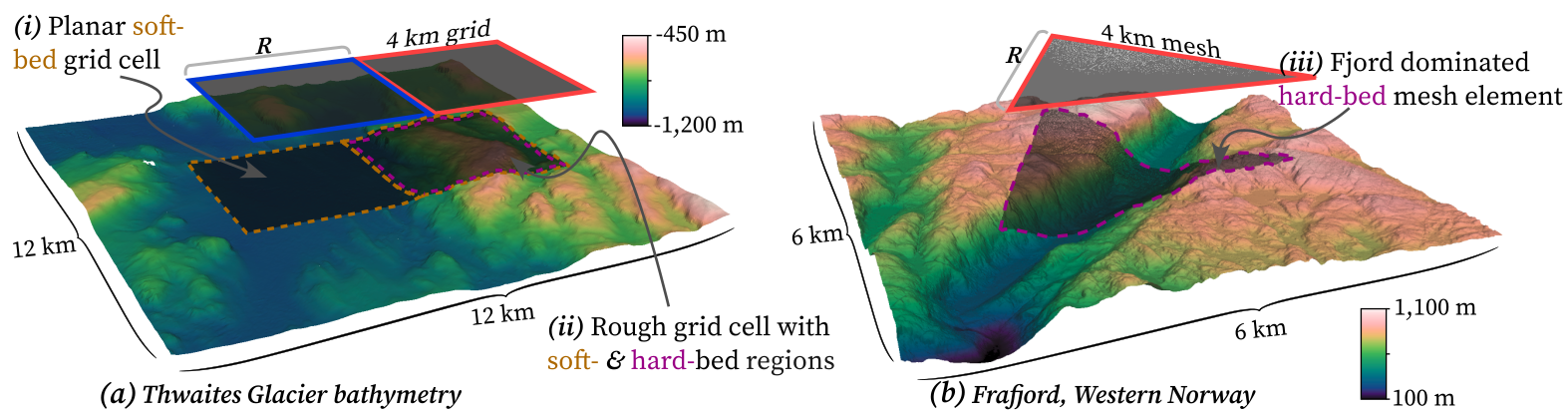}}
\caption{\textbf{Typically sized grid cells and mesh element overlain on previously glaciated regions}. Thwaites Glacier bathymetry is from \citet{Hogan2020RevealingButtressing}. Frafjord topography is from Kartverket.no.} %[\comment{Should label \(R\) as \(\tilde{\Gamma}_m\)}]
\label{fig:3D}
\end{figure*}

Representation of sliding in glacier and ice-sheet models is further hindered by a tendency to classify the bed either as `soft sediment' (less rigid than basal ice and subject to sediment deformation and transport under glacier slip and subglacial hydrology even at short time scales) or `hard bedrock' (more rigid than basal ice and only appreciably modified by sliding or hydrology at glacial-cycle time scales). \track{Existing sliding parameterisations are generally not formulated} to represent settings that may feature both soft- and hard-bed characteristics within a single grid cell or across a model domain (Fig. \ref{fig:3D}). This `one or the other' classification is challenged by geological and geomorphological evidence of mixed soft- and hard-bed regions in both recently deglaciated landscapes (e.g., \citealp{Kleman2008PatternsExplanation, Hogan2020RevealingButtressing, Garcia-Oteyza2022LateGreenland}) and in active subglacial settings \citep{Jordan2023GeologicalObservations} but has received very little attention in glaciological studies \citep{Koellner2019TheFlow}. A hard-soft distinction may also be condition dependent if, i.e. the substrate is only deformable with high subglacial water pressure. Sub-grid-scale topography is furthermore known to significantly influence ice flow (\citealp{Kyrke-Smith2018RelevanceAntarctica, Hoffman2022TheSliding, Law2023ComplexIceb, Barndon2025IceLandscapes}), and may even dominate grid-scale basal traction in certain settings (Fig. \ref{fig:3D}b, \citealp{Barndon2025IceLandscapes}), but is also lacking from present sliding parameterisations. Last, while a sliding relationship should simply be a mathematical relationship strictly between sliding and basal traction -- their inclusion as one of the only tuning tools in production models means that in practice they may additionally be representing a hard-to-gauge spread of model errors arising from sources such as surface mass balance, bed topography, or bulk ice physics.

\begin{figure*}
\centering{\includegraphics[width=0.7\textwidth]{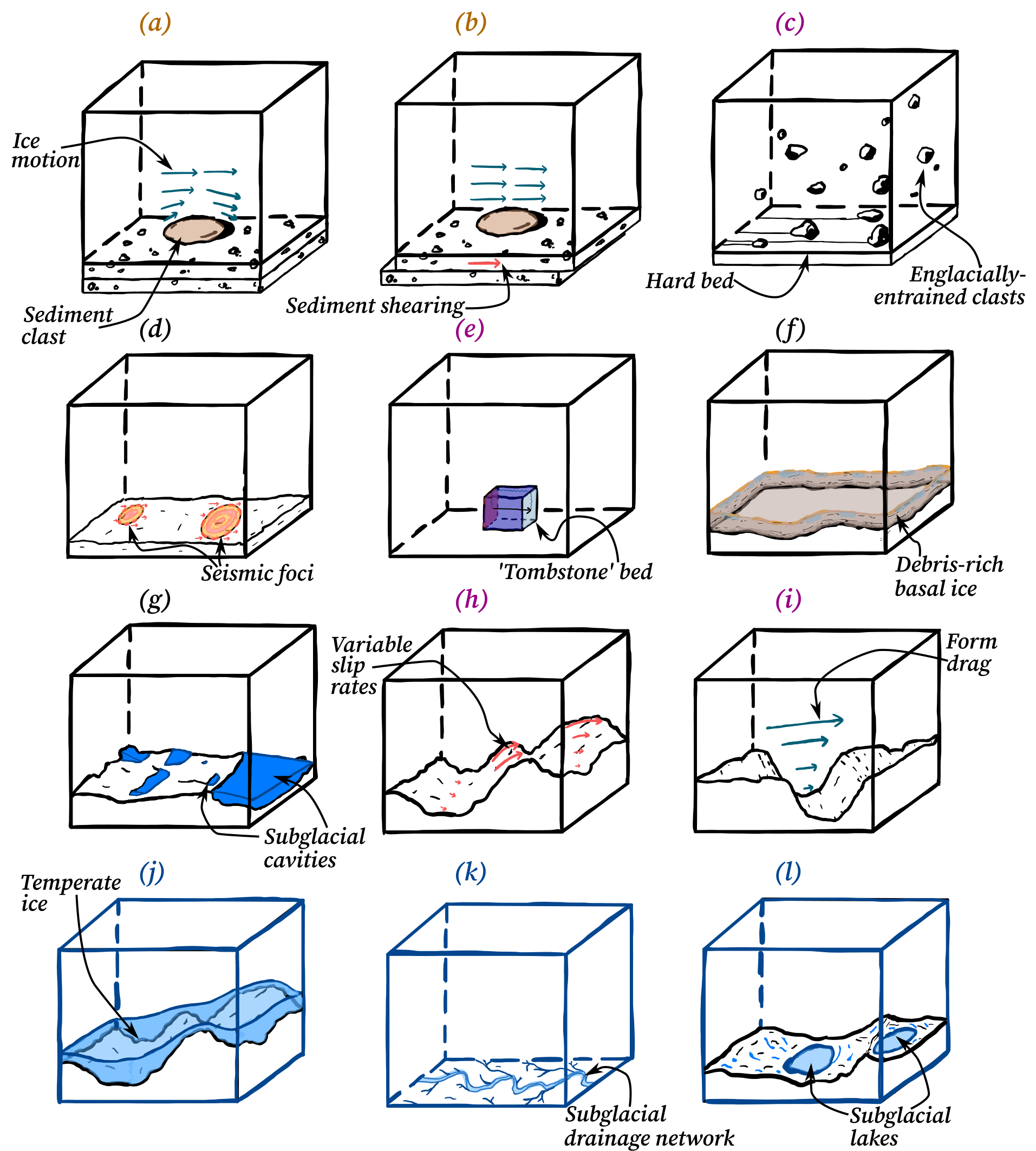}}
\caption{\textbf{Schematics of glacier sliding sub-processes and controls discussed in the text.} Scale is intentionally omitted but scale generally increases left to right and top to bottom through sub-processes. Hydrological processes and enthalpy field/temperate ice in \textbf{j}, \textbf{k}, and \textbf{l} are considered as controls, since they do not contribute directly to ice transport \track{(even if hydrology is still a mass transport process in the strictest sense)}; these are distinguished with blue outline boxes. \textbf{a} Form drag over sedimentary clasts. \textbf{b} Shear of sediment. \textbf{c} Ice with clasts sliding over a flat hard bed. \textbf{d} Stick-slip events. \textbf{e} Regelation. \textbf{f} Deformation within the basal ice layer. \textbf{g} Cavitation. \textbf{h} Spatially variable slip rates resulting from sliding over rough topography.  \textbf{i} Spatially variable ice deformation over rough topography. \textbf{j} Temperate layer processes. \textbf{k} Hydrology and channelisation. \textbf{l} Subglacial lakes.} 
\label{fig:modes}
\end{figure*}

In this paper, we suggest that these complicating factors may explain the diversity of sliding relationships -- from plastic to power-law -- found to be more/less effective in process-agnostic heuristic studies using decadal or longer data series (covered in Section \ref{s:heuristic}). We review the myriad processes that may fall within a `sliding layer' (Fig. \ref{fig:modes}, Section \ref{s:slidingprocesses}), including briefly covering those that fall within the more standard categories of `soft' (Section \ref{s:soft}) and `hard' (Section \ref{s:hard}). 
In doing so we make the inclusion of some ice deformation processes -- such as `topographic' form drag (Fig. \ref{fig:scale}, Section \ref{s:form}) and the basal ice layer (Section \ref{s:BIL}) -- explicitly possible within the sliding layer in contrast to some previous discussions where they may be excluded. We also briefly cover the main ways in which effective pressure, \(N\), is incorporated within production models, and how a complex subglacial hydrology system may influence sliding at the scale of model grid cells (Section \ref{s:hydrology}). 

We focus on annual timescales as these are the most relevant for \track{production} models, and largely avoid the transient influence of e.g., supraglacial lake drainages \citep{Das2008FractureDrainage.}, rainfall events \citep{Doyle2015AmplifiedRainfall}, or seasonal modulation \citep{Bartholomew2011SeasonalElevations, Sole2013WinterSummers}. Throughout this paper we use \emph{sliding} to refer to the net effect of processes that fall within the sliding layer, and \emph{slip} to refer to slip at the ice-bed interface where a clear ice-bed interface can be discerned. 

\track{In Section \ref{s:inout}, we build a framework to describe the net effect of an arbitrary number of sliding sub-processes acting over a 3-dimensional region,} % building from the idea of an `inner-outer' flow division in \citet{Fowler1977GlacierDynamics},
separating sub-processes into (1) slip at the ice-bed interface itself and (2) the normal resistance provided by obstacles to flow. Separating sliding into these components allows us to infer the range of forms that a `unified' sliding relationship may take under reasonable simplifying assumptions (Section \ref{s:unified}) and to discuss whether this can then be related to a simple implementation within production models. %Finally, we discuss the settings and scales under which power-law or regularised-Coulomb sliding behaviour may be anticipated, \track{and directions for future work} (Section \ref{s:discussion}).} 
We hope our coverage of sliding processes is useful to those seeking an introduction to glacier sliding and an overview of advancements over the last decade, but we do not attempt a fully exhaustive review. Earlier reviews \citep{Clarke1987AGlaciers, Clarke2005SubglacialProcesses, Fowler2010WeertmanTheory} and book chapters \citep{Benn2010GlaciersGlaciation, Cuffey2010TheGlaciers} offer more comprehensive coverage. Our discussion is limited to continuum-scales in ice and we do not cover crystal-scale deformation or molecular-level interaction -- \citet{Schulson2009CreepIce} and \citet{Krim1996FrictionScale} provide more information on these topics.

%\lawcomment{Comparison of sliding relationships in ISMIP experiments, and general lack of hydrology in predictive models (I think Aschwanden et al. 2016 \emph{does} include it should also go somewhere in the intro}

\section{ Heuristic studies} \label{s:heuristic}

There are two approaches to exploring glacier sliding. From the ground up (i.e., process-based through analytical, numerical, laboratory, or field studies -- Section \ref{s:slidingprocesses}) or from the top down (hereafter `heuristic') by using remotely-sensed or in-situ observations of externally or internally forced variations in the glacier system. The heuristic approach typically involves at least some time variation to circumvent the problems inherent to steady-state inversions, which link surface velocity or geometry to basal traction. In such inversions, any relationship in Table \ref{tab:eqs} will produce a reasonable basal stress state in order to meet the global stress balance condition. A time series with significant variation in surface geometry is best suited to the decadal time scales over which mass loss is most often considered \citep{Gillet-Chaulet2016AssimilationGlacier, Gilbert2023InferringGlacier}, but several other variations have been exploited (Table \ref{tab:heuristic}). Generally, these studies test for the most appropriate time-invariant \(m\) in a power-law relationship (Table \ref{tab:eqs}) and/or whether a regularised-Coulomb or power-law relationship is more appropriate overall. A heuristic approach is also process agnostic, that is, it will not directly provide information about the processes contributing to the form of the sliding relationship that produces the best match with observations. However, this approach is beneficial when demonstrating if/where a sliding relationship works in practice.

Broadly considered, these studies indicate that variation in the form of the sliding relationship across variable glacier settings is not captured solely through just one sliding relationship featuring a single term with a single spatially-variable single tunable coefficient (which we give the moniker of a `single-term universal sliding law' here and in Section \ref{s:unified}), \(C\), obtained through an inversion procedure  (where \(C\) may be used in any of the Eqs. in Table \ref{tab:eqs}). For example, \citet{Gimbert2021DoSpeed} and \citet{Gilbert2023InferringGlacier} find \(m = 3.1\pm0.3\) in power-law sliding for Argenti\`ere Glacier in the French Alps using a combination of surface velocity changes, length changes, and direct sliding measurements while \citet{Gillet-Chaulet2016AssimilationGlacier} find \(m\geq5\) and up to 50 well reproduces surface velocities at Pine Island Glacier, Antarctica. Meanwhile, \citet{Maier2021BasalGreenland} indicate that a single \(m\) value, also for power-law sliding, cannot be applied to the entire Greenland Ice Sheet with, for example, an \(m\) of 8 in northeast Greenland contrasting an \(m\) of 4 in central southwest Greenland. Disagreement between even two of these studies can dispel the notion that there is a `single-term universal sliding law'. %as outlined at the start of this paragraph.
 %\track{(}Note that in standard applications using power-law sliding, \(C\) is the subject of the inversion and varies in space, while \(m\) is kept constant\track{)}.

\renewcommand{\arraystretch}{1.15}
\begin{table*}
    \centering
    \begin{tabular}{|p{4.2cm}|p{4.2cm}|p{4.2cm}|p{2cm}|}
    \hline
        Setting & Variation & Study & \(m\) \\
    \hline
    Hofsj\"okull, Iceland -- ice cap\textsuperscript{\S}  & Summer-winter velocity comparison & \citet{Minchew2016PlasticFlux} & \(\infty\)/plastic \\
    High Mountain Asia -- valley glaciers & Velocity fields 2000-2017 & \citet{Dehecq2018Twenty-firstAsia} & 4 \\
    High Mountain Asia -- >5 km length valley glacier & Velocity fields during surge cycle & \citet{Beaud2022SurgeLaw} & Rate-weakening* \\
    Argenti\`ere, French Alps -- valley glacier\textsuperscript{\ddag} & Length (velocity) over 100 (15) years to \(\sim\) 2020 & \citet{Gilbert2023InferringGlacier} & 3 \\ 
    Greenland Ice Sheet -- ice sheet & Single velocity field averaged 2005-2015 & \citet{Maier2021BasalGreenland} & 3-10 \\
    Northwest Greenland -- outlet glaciers & Forward modelling of mass loss 2007-2018 & \citet{Choi2022UncoveringDecade} & 6 in modified Budd, or r-C \\
    Northeast Greenland Ice Stream -- ice stream & Forward modelling of mass change 2006-2021 & \citet{Khan2022ExtensiveStream} & r-C \\
    West Greenland -- land-terminating outlet glacier & Annual velocity and hydrology variations 2006-2013 & \citet{Tsai2021ASliding} & 4 in modified Budd \\
    West Greenland -- outlet glacier & Front position from 1985-2019 & \citet{Jager2024ValidatingElevation} & r-C \\
    Pine Island Glacier, Antarctica -- outlet glacier & Geometry 2003-2008 & \citet{Joughin2010SensitivityAntarctica} & mix of 3 and \(\infty\)/plastic \\
    Pine Island Glacier, Antarctica -- outlet glacier\textsuperscript{\S} & Velocity in 1996 and 2007-2010 & \citet{Gillet-Chaulet2016AssimilationGlacier} & \(\geq\)5 \\
    Pine Island Glacier, Antarctica -- outlet glacier & Present day velocity inversion, statistical inference, seismic measurements & \citet{Hank2025TheObservations} & r-C \\
    Pine Island Glacier, Antarctica -- ice stream & Displacement over fortnightly tidal cycles & \citet{Gudmundsson2011Ice-streamLaw} & 3 \\
    Rutford Ice Stream, Antarctica -- ice stream & Displacement over fortnightly tidal cycles & \citet{Gudmundsson2007TidesAntarctica} & 3 \\

    \hline
    \end{tabular}
    \caption{\textbf{Heuristic studies on sliding parameters}. r-C refers to regularised-Coulomb. \textsuperscript{\ddag} indicates a largely hard bed is inferred, \textsuperscript{\S}} indicates a largely soft bed is inferred. Absence of a superscript indicates ambiguous or mixed conditions. \textbf{Notes:} \citet{Gudmundsson2011Ice-streamLaw} did not test \(m\) values above 3. \citet{Dehecq2018Twenty-firstAsia} combine ice deformation and sliding in a 1D model. \citet{Choi2022UncoveringDecade} also include the influence of \(N\), defined as ice pressure above hydrostatic equilibrium with a sheet perfectly connected to the ocean, finding best fitting with \(N\) included. *\citet{Beaud2022SurgeLaw} do not include numerical modelling in their approach, with spatio-temporal variation in \(N\) liable to perform an important but unquantified role in their generalized sliding relationship. 
    \label{tab:heuristic}
\end{table*}

Heuristic studies provide a test for the efficacy of a given sliding relationship, but a match to a given theory does not on its own demonstrate that the physical processes underlying it are actually occurring in the study location (a case of correlation \(\not=\) causation). This is straightforward in the case of \(m\geq5\) from \citet{Gillet-Chaulet2016AssimilationGlacier} at Pine Island Glacier. No existing theory (except perhaps \citealp{Koellner2019TheFlow}) explicitly formulates this relationship (Section \ref{s:slidingprocesses}) so the utility of \(m\geq5\) in power-law sliding comes either from the fact that it emulates regularised-Coulomb sliding which does have a large body of theory behind it, or it represents a hitherto unappreciated set of processes not presently incorporated into a process-based sliding relationship. Conversely, the interpretation of plastic deformation in \citet{Minchew2016PlasticFlux} based on observations from a setting known to have significant till coverage and relatively smooth topography \citep{Bjornsson2003SurgesIceland} more straightforwardly connects heuristic findings with physical processes. However, this heuristic-process link is perhaps more complicated where \citet{Gimbert2021DoSpeed} and \citet{Gilbert2023InferringGlacier} suggest that their findings of \(m\sim3\) reflects only the ice deformation component of the original \citet{Weertman1957OnGlaciers} paper without the regelation component where \(m=1\), while the original Weertman theory is dependent upon a combination of both regelation and ice deformation, and making only limited allowance for the slip component actively observed beneath Argenti\`ere \citep{Vincent2016SlidingArea, Gimbert2021DoSpeed}. \citet{Maier2021BasalGreenland} also invoke \textit{Weertman-type} hard-bed physics as justification for lower values of \(m\), despite well-documented issues underlying the original theory (Section \ref{s:hard}). % and the fact that no theory building upon \citet{Weertman1957OnGlaciers}, or the original study, presently provide a physical justification for \(n=4\) or above. 

Therefore, while heuristic studies presently offer one of the more powerful practical methods for constraining the sliding relationship for a given modelling setting, particularly when compared with steady-state inversions, they also emphasise a heterogeneity between settings that is not easily reconciled through a single `single-term universal sliding law' tuning for \(C\) as defined above.

\section{ Sliding processes} \label{s:slidingprocesses}

\subsection{ Sliding over `soft' (sediment) beds} \label{s:soft}

Early work on soft-bed sliding suggested that subglacial till behaves as a mildly non-linear Bingham viscous fluid (such that, above a yield stress, resistance to deformation increases with deformation rate), with distributed deformation extending around 50 cm below the ice-bed interface \citep{Boulton1987SedimentConsequences}. This idea received support from some contemporaneous studies of subglacial conditions \citep{Blake1992TheProperties, Humphrey1993CharacteristicsAlaska} but most subsequent studies -- including laboratory, direct measurement, and analogous laboratory experiments from different material science disciplines -- find strain-localisation close to the ice-bed interface, or manifested in multiple discrete shear zones within the subglacial till, and a Coulomb-plastic rheology (e.g., \citealp{Biegel1989TheDistribution}; \citealp{Kamb1991RheologicalMotion}; \citealp{Hooke1997RheologySweden}; \citealp{Iverson1997AContents}; \citealp{Iverson1998Ring-shearBeds}; \citealp{Iverson2001DistributedSlip}; \citealp{Kavanaugh2006DiscriminationModel}; \citealp{Damsgaard2013DiscreteDeformation}; \citealp{Cuffey2010TheGlaciers}), and we therefore do not discuss viscous sediment deformation further. %Deformation at greater depths within the till has furthermore been shown to be possible under high effective pressure with an elastoplastic rheology \citep{Iverson2001DistributedSlip, Damsgaard2013DiscreteDeformation}.

In laboratory experiments of glacial-till deformation beneath a rigid shear ring, the till reaches its yield strength and then fails at a uniform rate under steady applied shear stress, followed by either a near-constant or modest decrease in basal traction over a large range of shear velocities (0-400 m a\textsuperscript{-1}; \citealp{Iverson1998Ring-shearBeds, Fischer2001HydraulicMotion, Iverson2010ShearRules}). Field observations using strain-gauges emplaced in actively deforming subglacial sediment repeatedly support Coulomb-plastic deformation \citep{Iverson1994In-situTill, Iverson1995FlowBeds, Fischer2001HydraulicMotion}, with an inverse relationship between pore-water pressure and sediment strength \citep{Fischer2001HydraulicMotion} -- note however that the absence of varying slip velocity makes it challenging to link slip rates to traction. Numerical models indicate that the pore-water pressure-strength relationship is further modulated by till permeability properties \citep{Damsgaard2017TheSheets, Damsgaard2020WaterStreams}, with till properties influenced in turn by crushing and compression from overlying ice \citep{Iverson1999CouplingResults}. Field observations additionally emphasise the importance of ploughing of clasts lodged at the ice-till interface, particularly where sediment pore pressure is high, as the pressure exerted by the clast on the downflow sediment can locally increase the water pressure and weaken the sediment \citep{Iverson1994In-situTill, Rousselot2005EvidencePloughing}. In the case where sediment is not weakened by an increase in pore pressure, a clast may be pushed downwards into the sediment and no longer be ploughed through the matrix \citep{Clark1989ClastBed}.

More recently, \citet{Zoet2020ABeds} included a ring of 20 cm-thick temperate ice  between the rigid shear ring and underlying sediment (meaning some ice deformation is included in the sliding velocity). In this setting, rate-strengthening behaviour occurs at lower velocities (below around 50 m a\textsuperscript{-1} in sediment with clasts, Fig. \ref{fig:modes}a) as ice viscously deforms and regelates around static clasts before the clasts begin to plough through the finer-grained matrix at a stress limited by the till’s Coulomb strength (Fig. \ref{fig:modes}b). This can be represented as \(\tau_b=\textrm{min}(C_p N, C_{pl}u_b^{1/m_{pl}})\) where \(C_{p,pl}\) are coefficients for plastic and power law components, respectively, and \(m_{pl}\) is the power-law sliding exponent. This form has been previously suggested in \citet{Tsai2015MarineConditions}, or as a regularised-Coulomb relationship (Fig. 1, Table 1).  

Sedimentary bedforms, including drumlins, mega-scale glacial lineations, and ribbed moraine \citep{Stokes2018GeomorphologyProcess} are ubiquitous in areas of continuous sediment and likely represent a continuum of features formed by glacier dynamics and subglacial hydrology \citep{Ely2023NumericalTransitions}. These bedforms are generally streamlined and have a low aspect ratio (height/length) in the flow direction, but their contribution to up-flow resistance has not been quantified to our knowledge. 

\subsection{ Sliding over `hard' (bedrock) beds} \label{s:hard}

%\emph{We still have rough, incomplete and wrong theories developed in the 60’s by Weertman, myself, Nye, Röthlisberger and Budd, that have but a historical interest}. \textbf{Louis Lliboutry, 2005}

Hard-bed sliding studies can be divided into those considering slip across planar surfaces of limited extent and those concerned with sliding over rough or undulating beds at longer wavelengths. In studies considering planar surfaces water pressure is typically kept spatially constant, while in studies considering a rough or undulating bed subglacial water may be theorised as either a microscopically thin and continuous layer, or as a pressurised system of subglacial cavities occupying local bedrock depressions. Undulating beds with a frictionless ice-bed interface were the first to attract attention in sliding theory, but we begin with macroscopically flat (i.e. flat to the naked eye) surfaces due to their simplicity. Roughness at length scales beyond 25 m is covered in Section \ref{s:form}.

\subsubsection{ Planar hard beds}

The slip of ice over a planar surface (Fig. \ref{fig:modes}c) where friction is generated via embedded clasts is closer to the assumptions of standard Coulomb behaviour \citep{Coulomb1785TheorieCordages, Desplanques2014Amontons-CoulombManuscript}, though plane-normal velocity components, englacial clasts, and possible clast-scale cavitation present complications \citep{Hoffman2022TheSliding}. Rather than a sliding relationship of the form of Eq. 1, basal traction in these studies is often reported as a static \(\mu_s\) or transient \(\mu_t\) coefficient of friction, \(\mu_{s} \: \textup{or} \: \mu_{t}=\frac{\tau_b}{N}\) where \(\mu_s\) and \(\mu_t\) describe shear-to-normal stress ratios required to set or keep the plane in motion, respectively. Shear-ring studies using temperate ice and englacial sediment content of 0-20\% over macroscopically flat granite find a very small \(\mu_s\) of 0.02-0.05 \citep{Barnes1971TheIce, Zoet2013TheGlacier, McCarthy2017TemperatureConditions, Thompson2020AnSliding}. %, lower than teflon-on-teflon with no lubrication \citep{Fetfatsidis2013DesignProcess}. 
In the above studies, sliding may be stick-slip or steady-state (outlined further in Section \ref{s:stickslip}) depending on water pressure and particular material properties. Transient coefficients of friction between 0.05-0.08 were obtained from subglacial access to the ice-rock interface of the Engabreen outlet glacier in Norway \citep{Iverson2003EffectsFlow, Cohen2005Debris-bedGlaciers}. Unexpectedly high shear traction values of up to 0.5 MPa accompany the low \(\mu_t\) values at Engabreen and these are still not fully explained, suggesting a possible gap in understanding -- see \citet{Thompson2020ControlsIce} for a more in depth discussion (and note that, in calculations of transient friction, in contrast to static friction, the line defined by \(\mu_t\) will not necessarily pass through the graph origin of \(N\) and \(\tau_b\)). The transient coefficient of friction is much higher for cold ice (0.5 at -20\textsuperscript{o}C) making sub-temperate slip very low, but potentially non-negligible (\citealp{Barnes1971TheIce, McCarthy2017TemperatureConditions, Atkins2013GeomorphologicalAntarctica, Mantelli2019IceAdvection}; Section \ref{s:frozenthawed}).

Where an ice block instead rests unconstrained on a plane, and water pressure beneath the ice is close to atmospheric pressure, \(\mu_s\) can reach 0.6 for a rough (up to 0.25 cm roughness) pebbly surface or 0.2 for smooth concrete \citep{Budd1979EmpiricalSliding}. These are not typical stress or hydrological conditions for glacier beds (e.g., \citealp{Hubbard1997AlpineHydrology, Woodard2021VariationsGlaciers}) but were used by \citet{Budd1979EmpiricalSliding} to derive an empirical sliding relationship (Table 1), with later modifications to exponents (including \(q\) for \(N\), Table \ref{tab:eqs}) to produce a better fit with field data at an ice-sheet scale \citep{Budd1984ASheet} that are arguably large enough to mark a departure from the empirical underpinnings. To summarise planar hard-bed sliding, it is reasonable to treat the friction of ice on rock in the presence of pressurised water as low, but it is not negligible, and may in certain configurations be an important local control on basal traction. There is significant scope to improve understanding in the processes responsible for ice-rock friction across a wide parameter space \citep{Thompson2020ControlsIce}.

\subsubsection{ Geometric hard beds}

Sliding over rough hard beds has also received much attention. This began with the classical work of \citet{Weertman1957OnGlaciers} (and which we henceforth intend when we specify `Weertman sliding' unless otherwise stated), who considered ice motion over isolated cuboids (akin to the geometry in Fig. \ref{fig:modes}e) with free-slip and no ice-bed separation through two processes (which were first emphasised for glaciological settings in \citealp{Deeley1913TheIce}): regelation and enhanced creep (Fig. \ref{fig:modes}e). In regelation, ice moves without deformation, and motion is accommodated by melting and refreezing. Temperate ice approaching an obstacle is subject to greater pressure, lowering its melting point. Consequently, some of the ice melts at the upstream face, and the meltwater is driven to the downstream face by the pressure gradient. Here, where the pressure is lower and the melting point is higher, the meltwater refreezes. The higher temperature at the downflow side of the obstacle results in heat flow back through the obstacle, completing the cycle. Regelation is an ongoing process in many subglacial settings \citep{Kamb1964DirectBedrock, Kamb1970SlidingObservation, Hallet1978TheLimestones, Hubbard1993WeertmanLayer, Iverson2000SedimentBed, Cook2011OriginIceland, Rempel2019PremeltingMagnitude}, including in soft bed settings outside the scope of the original \citet{Weertman1957OnGlaciers} paper. However, the requirement for return heat flow causes regelation rate to vary inversely with obstacle wavelength, limiting the importance of regelation as a sliding process to obstacles smaller than a few cm \citep{Weertman1957OnGlaciers}. Regelation is therefore generally not considered a significant contributor to overall sliding rates \citep{MacAyeal2019RevisitingBed} and has received comparatively little attention since the mid-1990s, though renewed study could help clarify its role when interacting with other processes.

Enhanced creep is prompted by stress concentrations at bedrock obstacles, leading to locally increased strain rates facilitated by the rate-weakening rheology of ice. Functionally, this means enhanced creep can be considered as a synonym for form drag (Section \ref{s:form}). Conversely to regelation, the contribution of enhanced creep to glacier sliding in Weertman's \citeyear{Weertman1957OnGlaciers} study increases with obstacle size. Focusing on the obstacle sizes that produce the greatest rate of sliding when regelation and enhanced creep act in unison (millimetre- to metre-scale) \citet{Weertman1957OnGlaciers} reaches a value of \(m=(n+1)/2=2.6\), with \(n\)=4.2 and \(C\) representing a combination of bed geometry and thermal characteristics. Technically, values of \(m\) deviating from 2.6 when \(n=4.2\) (or 2 when \(n=3\)) are therefore not physically based on the original Weertman theory making `power-law', which we use henceforth, or `Weertman-type' sliding a more suitable descriptor. Sliding over a sinusoidal two-dimensional bed with no ice-bed separation and a linear ice rheology produces a linear sliding relationship ($m=1$ in power-law sliding; \citealp{Nye1969AApproximation}), as does Weertman sliding with $n=1$. 

Power-law sliding, usually with \(m=3\), is the most commonly used sliding relationship in the latest round of ISMIP6 experiments (with linear sliding also frequently used) \citep{Goelzer2020TheISMIP6, Seroussi2020ISMIP6Century}, yet the theoretical underpinnings supporting these relationships in the original papers are not well-matched with subsequent progress in glaciology \citep{Lliboutry1968GeneralGlaciers, Fowler2010WeertmanTheory} -- including by Weertman's own reappraisal \citep{Weertman1979TheProblem}. Although Weertman's original theory features a non-linear rheology following \citet{Glen1955TheIce}, it is apparent that no glacier beds are actually characterised by a controlling obstacle size or the cuboid morphology used by \citet{Weertman1957OnGlaciers}. Changes to the assumptions and geometry used to derive a relationship for non-Newtonian flow with regelation and no sliding or cavitation also result in a materially different relationship \citep{Lliboutry1979LocalOpenings}. Weertman sliding is further challenged by ubiquitous evidence and theoretical support for subglacial cavitation (introduced in an English-language journal in \citealp{Lliboutry1968GeneralGlaciers}), or regions of bounded ponded water at high pressure between the glacier sole and underlying hard or soft bed, which form when the pressure on the downstream face of an obstacle falls below a critical level (Fig. \ref{fig:modes}g and references including \citealp{Walder1979GeometryCavities, Iken1981TheModel, Kamb1987GlacierSystem, Hooke1989EnglacialReview, Helanow2021ATopography}). Weertman sliding also does not provide an explanation for increased sliding rates when subglacial water pressure increases. The persistent utility of power-law sliding in heuristic studies, though generally with \(m\geq3\), despite these major problems is covered in Sections \ref{s:heuristic} and \ref{s:unified}.

%work in https://agupubs.onlinelibrary.wiley.com/doi/10.1029/2021GL097507 and https://agupubs.onlinelibrary.wiley.com/doi/full/10.1029/2023GL104503 if not already

Subglacial cavities (Fig. \ref{fig:modes}g) reduce the area of direct ice–bed contact and expand in response to increasing basal velocity or water pressure. Unlike soft beds, where higher \(p_w\) weakens the ice-sediment interface and underlying sediment itself, in cavitation theory increasing \(p_w\) instead reduces the effective roughness at the de facto ice base by opening cavities and diminishing ice contact. In the two-dimensional formulations of \citet{Lliboutry1968GeneralGlaciers}, \citet{Iken1981TheModel}, \citet{Schoof2005TheSliding}, and \citet{Gagliardini2007Finite-elementLaw}, amongst others, where the bed is comprised of frictionless sinusoids, this produces an upper bound for basal traction as region-averaged basal slip increases, followed by rate-weakening behaviour (Fig. \ref{fig:schematic}). Put mathematically, `Iken's bound' as termed by \citet{Schoof2005TheSliding} is given as 
\begin{equation}
    \tau_b \leq N\textup{tan}(\theta)
    \label{Eq:iken}
\end{equation}
where \(\theta\) is the maximum up-slope angle between the bed and the mean flow direction of the ice \citep{Iken1981TheModel, Schoof2005TheSliding}. Numerical and analytical modelling by \citet{RoldanBlasco2025ImpactSliding} confirms the presence of a traction bound in the case of plastic slip over a sinusoidally constructed bed in two-dimensions, who modify Eq. \ref{Eq:iken} to \(\tau_b \leq N\textrm{tan}(\theta') + \tau_p\) where \(\tau_p\) comes from an area-averaged plastic (Coulomb) relationship and \(\theta'\) refers to the maximum ice-water slope (rather than ice-bed). %\citet{Roldan-Blasco2022TheLaw} suggest that in the simplified two-dimensional framework of compound sinusoids, non-negligible friction at the ice-bed interface will not significantly alter the form of the sliding relationship or the existence of Iken's bound.
However, the geometry used in these studies can also be considered unrealistic, with the absence of a third dimension preventing lateral escape of subglacial water which would (1) act to reduce the cavity size \citep{Gimbert2021DoSpeed} and (2) may place a limit on the maximum obtainable subglacial water pressure averaged over a large region at long (i.e. annual) time scales after the system has had time to equilibrate (e.g., \citealp{Dow2015ModelingDrainage, DeFleurian2018SHMIPProject,  Hart2022TheGlacier}). \citet{Helanow2019SlidingBeds} and \citet{Helanow2021ATopography} address the three-dimensional problem for smaller scales, and show that realistic three-dimensional bed topographies up to 25 m still result in bounded basal traction, but that significant rate-weakening behaviour is not expected (the regularised-Coulomb relationship in Fig. \ref{fig:schematic}). We cover some further issues concerning Eq. \ref{Eq:iken} within the sliding layer in Sections \ref{s:innerouterinner}, \ref{s:ikens}. 

More recent work on cavitation includes \citet{Gilbert2022AGlaciers} who use the long-duration Argentière record as an observational basis for treating seasonal variations in cavitation and basal traction as a function of water discharge, rather than water pressure -- expanded upon in \citet{Togaibekov2024ObservingGlacier} to connect rainfall events to GPS station movement. Separately, \citet{Schoof2023TheLaws} and \citet{Schoof2023TheModel} consider, for two-dimensional beds with bounded slopes, the hysteretic behaviour of hydraulically isolated cavities and the influence upon this of viscoelastic ice rheology.

%[\comment{Need to neaten this up a bit, and emphasise that \(N\) in these theories is changing the geometry and normal forces at the bed, rather than the bed properties themselves. }]

\subsection{ Roughness, form drag, and temperate ice} \label{s:form}

%[\comment{Mauro's comment that "two things come into play, model resolution and observation resolution, both typically >>25 m, should be layed out again"}]

Glaciated and deglaciated landscapes are characterised by roughness not only at scales of centimetres to metres (as featured in the previous section), but also by roughness at scales of tens to thousands of metres. This includes cnoc-and-lochan landscapes, drumlins, incised fjords, and other erosional or depositional features that bridge a very subjective boundary between `roughness' and `topography' (Figs. \ref{fig:modes}h-j). Radar flight lines reveal extensive and variably rough beds beneath the Greenland and Antarctic ice sheets (e.g., \citealp{Bingham2009QuantifyingHistory, Rippin2013BedSheet, MunevarGarcia2023CharacterizingMargin}), often quantified through a Fast Fourier Transform or variogram (e.g., \citealp{MacKie2021StochasticGlacier}) which pares back the geomorphological uniqueness of a landscape but enables regional intercomparison. At present, only roughness up to a length scale of 25 m is explicitly incorporated in a process-based sliding relationship \citep{Helanow2021ATopography}, but it is clear from borehole observations  \citep{Ryser2014SustainedDeformation, Doyle2018PhysicalGreenland, Maier2019SlidingSheet, Law2021ThermodynamicsSensing} and simulations of ice motion over rough terrain at multi-kilometre scale \citep{Hoffman2022TheSliding, Law2023ComplexIceb, Liu2024SpontaneousBedrock, Barndon2025IceLandscapes} that roughness at larger scales significantly influences patterns of ice motion. Deformation rates are highly variable in proximity to the bed (Fig. \ref{fig:modes}i) with basal slip ranging between 5\%-95\% of total surface velocity (Fig. \ref{fig:modes}h), or even reversed in some cases through Moffat eddies \citep{Meyer2017FormationValleys, Barndon2025IceLandscapes}, with much higher slip rates over topographic highs \citep{Law2023ComplexIceb,Barndon2025IceLandscapes}. Existing soft- and hard-bed sliding theories and parameterisations do not incorporate roughness at larger (>25 m) scales -- yet the limited studies that have investigated the influence of roughness on bed properties suggest a notable control \citep{Wilkens2015ThermalStudy, Falcini2018QuantifyingStreams}.

In polythermal glacier and ice-sheet settings, a layer of lower-viscosity basal temperate ice (Fig. \ref{fig:modes}j) may furthermore modulate the complex motion patterns caused by topographic roughness \citep{Luthi2002MechanismsBedrock, Luthi2009CalvingGreenland, Krabbendam2016SlidingStreaming, Law2021ThermodynamicsSensing, Law2023ComplexIceb}, with greater thicknesses of temperate ice found in topographic troughs than peaks, and the formation of a shear band at the top of the temperate zone \citep{Law2021ThermodynamicsSensing, Law2023ComplexIceb, Liu2024SpontaneousBedrock, Barndon2025IceLandscapes}. The rheology of pure (or with a small concentration of sodium chloride) temperate ice has recently been constrained as likely linear viscous \citep{Lliboutry1971PermeabilityIce, Adams2021SofteningWater, Schohn2025Linear-viscousIce}, though the rheology of temperate ice samples with impurity contents similar to those of glacier ice remains unconstrained through laboratory experiments to our knowledge (phenomenologically, it has been described in the field under low confining pressure as `like cheese'; \citealp{Carol1947TheMoutonnees}).

The influence of roughness on sliding can be theorised through form drag. \citet{Kyrke-Smith2018RelevanceAntarctica} consider the influence of increasing the spatial resolution of a numerical model from 5 km to 500 m (and therefore also increasing the fidelity of the underlying high-resolution radar-derived topography) on the basal traction obtained from an inversion procedure. Intuitively, \citet{Kyrke-Smith2018RelevanceAntarctica} find that as spatial resolution is decreased, basal traction increases. Form drag is then described as the area-averaged basal traction with high topographic fidelity subtracted from the area-averaged basal traction with low topographic fidelity, i.e. the basal traction arising from topographic obstacles not explicitly represented by the basal boundary position at coarser resolutions. 
%This aligns with the idea that form drag can be theorised as part of a sliding layer which collapses to the modelled bed surface (covered in Section \ref{s:inout}). 
At a much smaller scale, \citet{Minchew2020TowardLaw} comment on \citet{Zoet2020ABeds} and invoke form drag as the viscous resistance that arises as ice deforms over entrained clast `micro-topography' in laboratory soft-bed sliding experiments before the sediment reaches its yield strength described previously in Section \ref{s:soft}. There are some challenges here in describing the scale at which form drag should descend to -- for example at a resolution of 500 m \citet{Kyrke-Smith2018RelevanceAntarctica} will still miss some topographic features that are then subsumed within their skin drag, while the form drag defined in \citet{Minchew2020TowardLaw} is easily encapsulated within \citet{Kyrke-Smith2018RelevanceAntarctica}'s skin drag. We discuss these issues further in Sections \ref{s:scalecavities}, \ref{s:appeq8}, \ref{s:scale}.

Form drag is therefore a useful concept in descriptions of glacier sliding, but it is actually challenging to arrive at a `standard' definition. In aerodynamic applications for example, form drag refers to the force acting on a solid body moving through a viscous and locally-turbulent fluid that is opposite and parallel to its velocity and which is not accounted for by induced drag, where the induced drag is a force resulting from turbulent vorticies connected to the lift on the body (\citealp{Batchelor1967AnDynamics}, sections 5.11, 7.8). Form drag has also been used in atmospheric and ocean dynamics, where it typically refers to the entire drag force resulting from turbulence around an obstacle or obstacles fixed to the surface, rather than a division of it (e.g., \citealp{Arya1973ContributionIce, Renfrew2019AtmosphericResponses, Jagannathan2023EvolutionInstabilities}). To address this, we offer a stipulative re-definition of form drag specific to glaciological production models in Section \ref{s:deformation}. This allows us to emphasise the importance of form drag in glacier modelling as the response to topographic roughness that exceeds the typical length-scale of cavitation. % Difficulties are still encountered when applying the concept of form drag to settings where there is a fixed continuous surface however, even if turbulence is present \citep{MacCready2003FormFlows}.

%[\comment{Will need to define form drag as resistance over \(\Gamma_b\) from obstacles above \(l_c\). So sediment-clast powered form drag is still technically form drag, but falls within tangential resistance in our framework.}]

%To circumvent the problem of defining form drag only at a specific scale -- i.e. for `topographic' roughness \citep{Kyrke-Smith2018RelevanceAntarctica} or for `sedimentary-clast' roughness \citep{Minchew2020TowardLaw} -- we refer to form drag here as \(\boldsymbol{\mathcal{D}}+\boldsymbol{\mathcal{G}}\): the sum of all distributed viscous deformation and geometric contributions to the resistive stress operating over a given region within the inner layer. This requires separating ice deformation and slip components from existing theories (Sections \ref{s:soft}, \ref{s:hard}) before recategorising them as two distinct terms, but allows for a definition of form drag comprising myriad ice rheologies (this Section, Section \ref{s:BIL}) across a very wide spatial scale (Fig. \ref{fig:scale}).% We make this grouping lacking a consistent rule to meaningfully make distinctions between these processes.   

\subsection{ The basal ice layer and ice rheology} \label{s:BIL}

\begin{figure}
\centering{\includegraphics[width=0.35\textwidth]{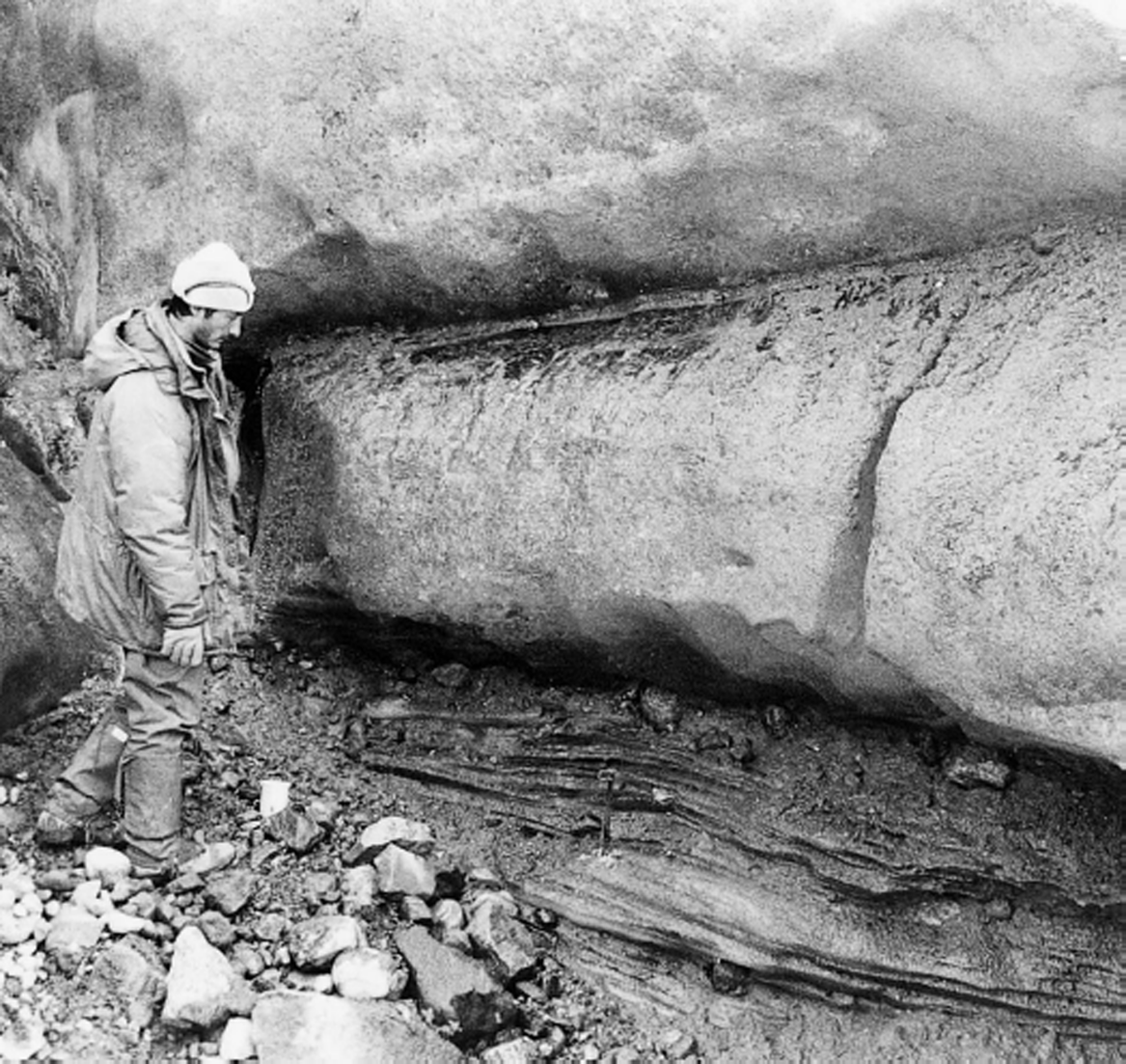}}
\caption{\textbf{The basal ice layer at Russel Glacier, Greenland.} From \citet{Knight2002DischargeSheet} with permission showing characteristic stratigraphy.}  %with written permission from Peter Knight.}
\label{fig:BIL}
\end{figure}

A basal ice band of a distinctly different nature to overlying debris-poor meteoric ice, referred to here as the basal ice layer (Figs. \ref{fig:modes}f, \ref{fig:BIL}), but which may also be called the frozen fringe \citep{Meyer2018Freeze-onGlaciers, Hansen2024PresenceRelationship}, is frequently a feature of the lowermost layer of glaciers and ice sheets. The basal ice layer is characterised by entrained debris and the diagenetic modification (or metamorphism) of meteoric ice by hydrologic processes, melting, refreezing, and intense strain \track{and may result in decreased \textit{or} increased ice viscosity} (see \citealp{Hubbard1989BasalReview}; \citealp{ Knight1997TheSheets}; \citealp{Souchez2000BasalData}; \citealp{Hubbard2009BasalApproach}; and \citealp{Hansen2024PresenceRelationship}). A basal ice layer is observed in deep ice-divide Greenlandic and Antarctic ice sheet ice cores  (e.g., \citealp{Gow1979OnSheet, Gow1996NatureProcesses, Souchez1998AGreenland, Tison1998IsCore, Souchez2002HighlyAntarctica}), Greenlandic and Antarctic ice sheet margins (e.g., \citealp{Swinzow1962InvestigationGreenland, Tison1993DebrisAntarctica}), and across many alpine and ice-cap settings (e.g., \citealp{Hubbard1995BasalAlps, Lawson1978AnAlaska}). Proposed formation mechanisms for the basal ice layer are basal freeze on entraining sediment through regelation (to a vertical thickness of around one meter; e.g., \citealp{Weertman1961MechanismSheets, Hubbard1993WeertmanLayer}), a pressure-driven heat pump of temperate ice (up to a scale of several meters; \citealp{Robin1976IsPoint}), or movement of subglacial meltwater along hydro-potential pathways to a region of the glacier bed below the pressure melting point (e.g., \citealp{Bell2011WidespreadBase}), with the basal ice layer potentially of first-order importance for the transport of previously subglacial sediment \citep{Pierce2024ModelingLayers}. Basal melting associated with frictional heat generated through slip is expected to decrease the thickness of the basal ice layer and impede its growth \citep{Hubbard1989BasalReview, Knight1997TheSheets}. However, basal ice layer thicknesses up to 30 m are recorded at the western terminus of the GrIS (Fig. 6, \citealp{Knight2002DischargeSheet}) and it is not conclusive if these sequences formed from freeze on in the immediate vicinity of the margin, or survived passage through up to 100 km of thawed bed conditions and slip-driven basal melt \citep{MacGregor2016ASheet}. Limited observations suggest ongoing strain within the basal ice layer (including stick-slip behaviour in shear zones) even at the western Greenland Ice Sheet margin \citep{Chandler2005BasalGreenland}. Recent modelling and radar mapping of Antarctica suggests thick (>100 m) sequences of sediment-laden basal ice may persist for tens of kilometres at least beyond a transition from frozen to thawed bed settings \citep{Franke2024SedimentAntarctica} but further work is required to fully explore the parameter space of these processes. Further, spatially and temporally varying basal ice characteristics ranging from debris-rich ice to frozen sediment, will blur the distinction between soft- and hard-bedded regions. 

Shear-ring laboratory experiments for the deformation of a 1-2 cm thick ice-sediment melange between thawed sediments and sediment \citep{Hansen2024PresenceRelationship} suggest similar regularised-Coulomb behaviour to shear-ring experiments of clean ice over sediment \citep{Zoet2020ABeds} and limited borehole observations indicate lowered viscosity in temperate ice \citep{Chandler2008OptimisingModelling}. \citet{Meyer2018Freeze-onGlaciers} also construct an analytical model arguing that ice infiltration into previously thawed sediments limits bed strength to a narrow range in pervasively soft-bedded regions. Beyond these studies however, we are not aware of experiments considering the influence upon basal traction relationships of basal ice sequences exceeding a few cm in thickness or that of the large diversity of facies found in the basal ice layer \citep{Hubbard2009BasalApproach}.  %Further work is needed to quantify the rheology of debris-rich basal ice and its contribution to form drag across a range of typical settings. %It is reasonable to suggest however, that impurity contents that can exceed 60\% \citep{Christoffersen2006ASediment} will result in significant weakness compared with clean meteoric ice, and that, where present, the basal ice layer will interact with processes of soft- and hard-bed sliding (Sections \ref{s:soft}, \ref{s:hard}), roughness, and stick-slip (Sections \ref{s:form}, \ref{s:stickslip}). 

Three further rheological factors present significant complications for ice deformation in the close vicinity of the bed (and which may therefore fall within the sliding layer we describe in Section \ref{s:inout}). % and are hence captured at least partially within the inner flow. 
\textbf{First}, pre-Holocene ice with a higher dust concentration and typically smaller grain size deposited during the last glacial period deforms at a rate around 2.5 times that of Holocene ice under the same stress and temperature conditions \citep{Paterson1991Whysoft}. Such ice is widely present in the strata of the Greenland and Antarctic Ice Sheets \citep{Macgregor2015RadiostratigraphySheet, Winter2019AgeHorizons, Ashmore2020EnglacialSheet} but uncommon in glaciers and ice caps where it is limited to very slow moving areas \citep{Thompson1997TropicalCore}. Valley glaciers generally feature much higher concentrations of englacial debris impurities in any case (e.g., \citealp{Goodsell2005DebrisSwitzerland}). To our knowledge, no production models incorporate age/depth-dependent rheology. \textbf{Second}, ice at depth has a highly anisotropic rheology due to the development of a strong crystallographic preferred orientation under consistent uni-directional deformation \citep{Lile1978TheIce, Baker1981TexturalMasses, Duval1981CreepCompression, Wilson2002TheIce}. Such mechanical anisotropy presents major challenges in implementation, bench-marking, and interpretation within numerical modelling \citep{Martin2004Three-dimensionalAntarctica, Gillet-Chaulet2005AModeling} and may substantially alter the basal stress state of glaciers and ice sheets when compared to non-basal ice, though to our knowledge these effects are not covered by any existing glaciological literature. %within viscous deformation of ice in the inner flow where strain rates are both high and spatially complex \citep{Law2023ComplexIceb}. \textbf{Third}, folds close to the bed may arise from mechanical differences in ice layers \citep{Whillans1987FoldingSheet, Bons2016ConvergingSheet, Zhang2023IceSheet} with the possibility to significantly influence the bulk rheological behaviour of the inner flow. 
\textbf{Third}, while \(n=3\) is the default in glacier and ice-sheet models (in fact already rounded down from the 3.2 reported in \citealp{Glen1955TheIce}), there are numerous field-observation based studies suggesting that for cold (i.e. not temperate) ice \(n=4\), representing a dislocation creep regime, is more appropriate in higher stress settings (away from ice-sheet ice divides, for example; \citealp{Bons2018GreenlandMotion, Millstein2022IceAssumed, Ranganathan2024ASheets}) and perhaps also at areas dominated by pure shear \citep{Gillet-Chaulet2011In-situRadar}. Considering a smooth BedMachine lower boundary in northern Greenland, using \(n=4\) instead of \(n=3\) dramatically decreases the area over which basal sliding is expected to contribute significantly to total surface displacement \citep{Bons2018GreenlandMotion}. The influence of \(n=4\) on form drag and sliding parameterisations, and its interaction with rough topography is to our knowledge entirely unexplored. These factors are not presently accounted for in the bulk ice rheology of production models, nor included in any existing sliding relationships, yet they will play an outsized role in overall glacier motion due to the adjacency of the basal ice layer to the actual bed. 

%We note that as the above factors are not included in the bulk ice rheology of production models (and play an outsized role in overall glacier motion due to their proximity to the bed) they will, where present, already be implicitly and unavoidably accounted for via inversion procedures within basal-sliding relationshiops where they are not explicitly accounted for. As we explore in the Discussion, the potential heterogeneity of their interactions creates significant complexity that future work should seek to disentangle.  

\subsection{ Stick-slip and continuous sliding} \label{s:stickslip}

In most cases, glacier and ice-sheet models neglect acceleration in the Navier-Stokes equations and treat ice as inelastic \track{(exceptions exist; \citealp{Bassis2023BeyondModels})}. \track{In production models,} these reasonable assumptions at typical modelling time-scales are also accompanied by the use of a continuous relationship in the form of Eq. 1 connecting basal traction to basal velocity, with rapid (e.g., order of seconds) changes in velocity not accounted for.

However, this implementation is sometimes inconsistent with field observations of sliding, where near-instantaneous stick-slip behaviour (Fig. \ref{fig:modes}d) is frequently recorded by seismometers and geophones as basal icequakes (e.g., \citealp{Neave1970IcequakesGlacier, Graff2021ChangingGlacier, Hudson2023HighlyBed}) and also by direct observation at the glacier bed or margin \citep{Theakstone1967BasalNorway, Hubbard2002DirectSwitzerland, Chandler2005BasalGreenland}. Icequakes are also generated englacially by processes such as extensional faulting near the surface, but we do not cover these phenomena further here.

%add https://doi.org/10.1016/j.epsl.2011.02.052 to the above

Basal icequakes occur due to a rapid release of elastically stored energy, resulting in ice motion that is transiently far higher than the long-term average \citep{Winberry2009BasalAntarctica, Winberry2011DynamicsAntarctica, Walter2013DeepGlaciers, Podolskiy2016Cryoseismology}. Moment magnitudes have been recorded from negative in alpine settings \citep{Helmstetter2015BasalMotion} to magnitude seven at Whillans Ice Stream in West Antarctica \citep{Wiens2008SimultaneousStream}. For a basal icequake to occur, the local frictional shear strength of the bed must be lower when the sliding interface is in motion than when it is static, i.e. \(\mu_t\), the dynamic friction coefficient, must be lower than \(\mu_s\), the static friction coefficient \citep{Bahr1996Stick-slipGlacier, Rice2001RateSolids}, describing slip-weakening behaviour. Stick-slip is usually associated with diurnal or tidal variations in subglacial water pressure allowing a clear build-up period for elastic energy \citep{Bahr1996Stick-slipGlacier, Bindschadler2003TidallyStream, Walter2008BasalSwitzerland, Stevens2024IcequakeDrainage} but basal icequakes are also recorded outside of such cycles \citep{Hubbard2002DirectSwitzerland}. A rate-and-state framework is commonly invoked to describe stick-slip behaviour \citep{Rice2001RateSolids} -- as often applied in geological-fault settings (\citealp{Gomberg2000OnModels}; Appendix \ref{A:geodynamics}) -- where a state variable, \(\psi\), is introduced to account for the strength of the fault \citep{Rice2001RateSolids, vandenEnde2018ASlip}, usually based on the slip displacement of the fault \citep{Ruina1983SlipLaws} or its time-dependent evolution \citep{Dieterich1979ModelingEquations} giving
\begin{equation}
    \tau_b = f(u_b, N, \psi) \: .
    \label{eq:ratestate}
\end{equation}
In contrast to Eq. \ref{Eq:1}, the expectation is that \(u_b\) in Eq. \ref{eq:ratestate} can vary rapidly in both time and space. A rate-and-state friction relationship can still account for the aseismic rate-strengthening steady slip of a fault \citep{vandenEnde2018ASlip} but is not well-suited for models entirely excluding rapid velocity changes and elastic behaviour by design. 

Quantifying the contribution of stick-slip basal sliding to glacier motion is challenging due to large uncertainties in fault dimensions, slip displacement, and stress calculated from seismic data \citep{Abercrombie2015InvestigatingParameters}. Some estimates suggest stick-slip displacement accounts for all basal sliding over an annual period \citep{Helmstetter2015BasalMotion}, while in other cases, coseismic and aseismic regions coexist \citep{Barcheck2020IcequakeStream, Kufner2021NotAntarctica} -- analogous to coexisting seismogenic and creeping faults in actively deforming geological faults (e.g., \citealp{Azzaro2020Stick-slipItaly}). Stick-slip patterns may also lead or follow large-scale velocity changes -- for example deceleration of the Whillans Ice Stream in the 2000s was about three times greater at a central sticky spot than in surrounding non stick-slip locations \citep{Winberry2014TidalAntarctica}. The central sticky spot also skipped the low-tide rupture event much more frequently following the slow down, with displacement only 150\% greater for the subsequent slip event as catch-up, indicating some degree of compensation through a viscoelastic ice rheology. 

Despite the growing interest in stick-slip motion in glaciers %(\citealp{Podolskiy2016Cryoseismology, Zoet2020ApplicationSeismicity}) 
there are limited connections between rate-and-state theory (of form Eq. \ref{eq:ratestate}) and continuous basal traction theory relationships (of form Eq. \ref{Eq:1}). We are only aware of the statistical mechanical approach of \citet{Bahr1995TheoryFlow}, where the glacier sole is represented as numerous elastically interconnected blocks with stochastically varying roughness parameters and the similar but simplified 1D approach of \citep{Kopfli2022HydraulicGlacier}. (Elastically interconnected blocks have also been used to investigate glacier stability; \citealp{Faillettaz2010GravitydrivenCracking, Faillettaz2012InstabilitiesSwitzerland}.) Bahr and Rundle's approach, intended for thin, steep, valley glaciers, finds that the average total force on an individual block (including elastic forces from adjacent blocks) is proportional to the area-average basal traction. If this holds for greater ice thicknesses and larger scales then stick-slip behaviour alone will not invalidate continuous basal traction relationships over a sufficiently large area, but further work is required to integrate this theory with recent developments in understanding the stick-slip behaviour of glaciers, and how this can relate to the spatial scales of production model grid cell resolution.

\subsection{ Overlap between soft- and hard-bed sliding} \label{s:overlap}

Most glacier and ice sheet modelling studies use basal traction relationships that implicitly assume the bed is either soft or hard across the entire domain with the ostensibly hard-bed power-law sliding relationshiop (tracing its lineage from Weertman sliding, Section \ref{s:hard}) with \(2\geq m \geq4\) being the modal choice for models contributing to ISMIP6 \citep{Goelzer2020TheISMIP6, Seroussi2020ISMIP6Century}. In some situations a hard-soft division is appropriate, for example most studies are in agreement that extensive, relatively planar sediment lies beneath the ice streams draining into the Ronne and Ross ice shelves in Antarctica \citep{Tulaczyk2000BasalMechanics, Vaughan2003AcousticStreams, Peters2007ExtensiveStream}, while geographically limited areas of Canadian continental shield may qualify as `true', relatively planar, sediment-free hard beds \citep{Slaymaker2017PleistoceneCanada}. Many individual valley glaciers can also be reasonably categorised as predominantly soft- (e.g., \citealp{Murray2001BasalSvalbard}) or hard-bedded (e.g., \citealp{Hubbard2002DirectSwitzerland}). However, taking Isunnguata Sermia in west Greenland as one well-studied example, borehole data mostly from topographic highs are used to suggest hard-bed conditions are dominant \citep{Harper2017BoreholeSheet, Maier2019SlidingSheet} while nearby seismic surveys are used to suggest the opposite \citep{Booth2012Thin-layerGreenland, Dow2013SeismicGreenland, Kulessa2017SeismicFlow}. Meanwhile, ice-marginal studies in west Greenland \citep{Klint2013LineamentGreenland} find a complex mix of bedrock, sediment-filled depressions, and extensive sandurs filling valley bottoms in front of the major land-terminating outlets \citep{Grocott2017BasinGreenland}. In the case of the seismic survey of \citet{Kulessa2017SeismicFlow} the recorded seismograms also allow for hard-bed conditions at topographic highs which would be in agreement with \citet{Harper2017BoreholeSheet}. Similarly, recent work at Thwaites Glacier and the Wilkes and Aurora subglacial basin in Antarctica further suggests the coexistence of soft- and hard-bed regions over distances less than 10 km \citep{Muto2019RelatingAntarctica, Muto2019Bed-typeAntarctica, Holschuh2020LinkingAntarctica, Jordan2023GeologicalObservations,Liebsch2025ApplicationAntarctica}. While soft- or hard-bed categories can be a reasonable distinction for individual glaciers, a more appropriate default for present and palaeo ice-sheets is therefore a mixed bed condition lying between the two end-members.   

Some recent attention has been directed towards the interaction between soft- and hard-beds. \citet{Koellner2019TheFlow} consider two-dimensional ice flow over sinusoidal beds with wavelengths from 8-60 km and setting the value of \(m\) in the power-law sliding relationship  (Table \ref{tab:eqs}) to 8 to approximate regularised-Coulomb sliding in sinusoid troughs, while keeping it at 3 over the sinusoid peaks. This demonstrates that a hybrid \(m = 3\leftrightarrow8\) Weertman rheology results in a modelled glacier response between \(m=3\) and \(m=8\) end-members. Given sliding rates at the modelled ice-bed interface can also vary significantly over much shorter distances \citep{Law2023ComplexIceb,Barndon2025IceLandscapes}, further work is still needed to more comprehensively isolate the influence of realistic variations in soft- and hard-bed sliding, and to assess whether hard-bed highs and soft-bed depressions are a reasonable model for the subglacial landscape of the Greenland and Antarctic ice sheets. \citet{Joughin2010SensitivityAntarctica} also apply a plastic sliding relationship to inferred soft-bedded regions and power-law sliding with \(m=3\) to inferred hard-bedded regions \citet{Joughin2009BasalData} at Pine Island Glacier, Antarctica, finding good agreement in this mixed model over an 8-year period (Section \ref{s:heuristic}). %(More accurately this work considers the interaction between Weertman and regularised-Coulomb sliding relationships given the uncertainty surrounding the place of Weertman sliding in hard-bed sliding covered in Section \ref{s:hard}) 

From empirical studies at small (<25 m) scales, both the derived sliding relationship for hard-bed cavitation \citep{Schoof2005TheSliding, Gagliardini2007Finite-elementLaw, Helanow2021ATopography} and slip over sediments \citep{Zoet2020ABeds} can yield a bounded traction relationship, leading to suggestions that regularised-Coulomb represents a `single-term universal sliding law' for both hard and soft beds \citep{Minchew2020TowardLaw}. As explored in Section \ref{s:heuristic}, there is evidence that there is not a universally applicable single-term single-spatially-tunable-coefficient sliding relationship applicable across all settings, but there are also some important differences between \citet{Zoet2020ABeds} and \citet{Helanow2021ATopography}. In order to match the relationships of \citet{Zoet2020ABeds} and \citet{Helanow2021ATopography}, an unrealistically low friction angle is required in the equation of \citet{Zoet2020ABeds} (Fig. \ref{fig:helanow-zoet}), indicating that the set-up of \citet{Zoet2020ABeds} produces a stronger bed than \citet{Helanow2021ATopography} under equivalent effective pressure and velocities, despite the general conception that soft beds are weaker than their hard-bed equivalents (e.g., \citealp{Koellner2019TheFlow, Joughin2009BasalData, Gowan2023TheScales}). This could be explained by (i) a natural requirement for lower subglacial water pressure in hard beds, %(by around 4\%), which is supported by available evidence (e.g., \citealp{Engelhardt1990PhysicalStream, Doyle2018PhysicalGreenland}), 
(ii) non-negligible ice-rock interface friction, which is not included in \citet{Helanow2021ATopography}, or (iii) unaccounted for form drag (covered in Sections \ref{s:inout}, \ref{s:discussion}) occurring in many hard-bed settings resulting in a view of a stronger bed. The \(u_t\) (threshold velocities) values obtained through these studies also differ enough to yield meaningfully different production-model behaviours (Section \ref{s:threshold}). Thus, while both planar (at the <25 m scale) soft- and hard-bed settings conform to bounded traction, there is still some distance to go before we reach a genuinely complete understanding of the differences between planar soft- and hard-bed sliding.

Last, \citet{Tsai2021ASliding} take an agnostic approach to the soft- and hard- bed division, focusing instead on the region of the bed where the hydrological system is expected to be active (where \(\tau_b\) is locally approximated as 0) and inactive (where \(\tau_b\) is locally determined using power-law sliding). Under this formulation \citet{Tsai2021ASliding} find a relationship similar to \citet{Budd1984ASheet} sliding (their Eq. 6) which matches well with annual variations in velocity and moulin discharge at a land-terminating sector of the Greenland Ice Sheet. %This relationship may be well-applicable in large-scale models (Section \ref{s:unified}).

\subsection{ Frozen-thawed transitions and sub-temperate sliding} \label{s:frozenthawed}

Slip of ice over a frozen bed is expected to be small, but non-negligible in some settings, with field studies finding slow (<0.2 m a\textsuperscript{-1}) slip rates even with temperatures some degrees below the pressure-melting point for both soft- and hard-bed settings \citep{Fowler1986Sub-TemperateSliding, Echelmeyer1987DirectTemperatures, Cuffey1999Interfacial-17C, Fitzsimons1999StructureTemperatures}. The mechanical process or processes responsible for sub-temperate sliding in these instances remains uncertain (\citealp{Cuffey2010TheGlaciers}, 7.2.8) but may be related to a very thin (nm) pre-melt layer at the ice bed \citep{Shreve1984GlacierTemperatures}, the inclusion of solid friction (\citealp{Fowler1986Sub-TemperateSliding}; Section \ref{s:hard}), or sediment deformation despite freezing conditions \citet{Waller2001TheGlaciers}. Regelation is also possible even when bulk ice temperature is below zero, provided local stress concentrations (e.g. on the stoss faces of bedrock roughness elements) locally raise the pressure melting point above the ice temperature.

While some simpler theories treat sliding at the model bed surface as strictly zero in frozen settings (\citealp{Lliboutry1966BottomSheet}; \citealp{Cuffey2010TheGlaciers}, 7.1.1), most process-based and palaeo models take a heuristic approach to scaling basal traction with temperature to allow a small degree of sub-temperate sliding (e.g., \citealp{Fowler1978OnAnalysis}; \citealp{Blatter1991PolythermalGlaciers}; \citealp{Tarasov2007CoevolutionAmerica}; \citealp{Mantelli2019IceAdvection}) giving
\begin{equation}
    \tau_b=f(u_b, N, T*)
\end{equation}
as an adaptation of Eq. \ref{Eq:1} where \(T*=T_{\textrm{pmp}}-T_b\) is calculated as the temperature at the base, \(T_b\) relative to the pressure melting point, \(T_{\textrm{pmp}}\). \track{For example}
\begin{equation}
    \label{eq:frozencuzzone}
    \tau_b = CNe^{-T*/\alpha}u_b
\end{equation}
\track{modified from \citet{Cuzzone2019TheISSM} where \(\alpha\) is a constant scaling factor or following alternative formulations in \citet{Hindmarsh2001DynamicalSheets}, \citet{Pattyn2017Sea-levelV1.0}, and \citet{Calov2018SimulationModel} amongst others.}

\begin{figure}
\centering{\includegraphics[width=0.35\textwidth]{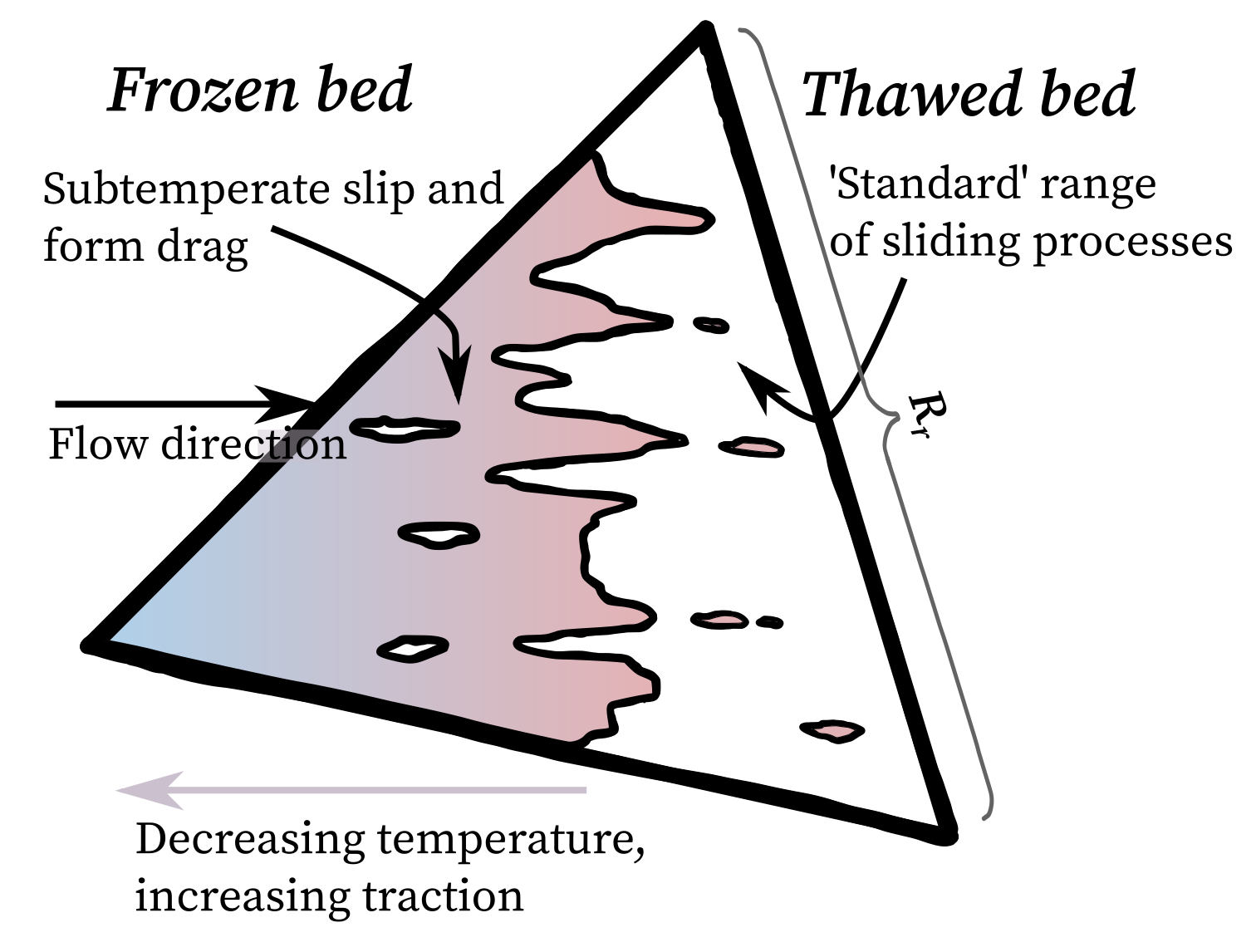}}
\caption{\textbf{Schematic representation of a mesh element featuring temperate and subtemperate sliding}. Triangle represents an area \(R_r\) under consideration. Solid black line represents frozen-thawed transition at the bed and cooler colours represent decreasing temperature of frozen portion of bed. Sub-temperate sliding is not sufficiently well-studied to provide a definite spatial range, but speculatively this may range from 100 m - 10 km. Adapted from \citet{Wilch2000CalculatingSheet}.}
\label{fig:subtemp}
\end{figure}

However, all models inverting for basal traction with present-day velocity fields in ISMIP6 experiments neglect any temperature control on sliding, resulting in model-prescribed sliding over likely frozen sections of the Greenland and Antarctic ice sheet beds \citep{Scofield1991EvidenceAntarctica, Atkins2013GeomorphologicalAntarctica, Dawson2022IceState}, including areas they model as frozen themselves (\citealp{Macgregor2022GBaTSv2:Sheet, Raspoet2025EstimatesSheet}). The recently-developed Instructed Glacier Model used for valley glacier \citep{Jouvet2022DeepMagnitude, Cook2023CommittedAssimilation} and paleo ice sheet modelling \citep{Leger2025AAIb} also presently neglects a temperature dependence on \(\tau_b\). This means that there is no distinction between the physics and sub-processes for cold- and warm-bedded regions in these models (or for grid cells featuring combinations thereof, Fig. \ref{fig:subtemp}). Generally, a pattern of greater basal traction in interior ice-sheet regions where a frozen bed is considered more likely is obtained in ice-sheet wide inversions \citep{Larour2012ContinentalISSM, Morlighem2013InversionModel}, but this does raise the question as to what the production-model sliding is representing here. Of course, form drag is one possibility. We explore some possible answers to this question, such as that this `sliding' can be treated as enhanced deformation or form drag near the bed \citep{Weertman1967SlidingGlaciers} coupled with a maximum slip yield strength, in Section \ref{s:localfreeze}.

\section{ Subglacial Hydrology} \label{s:hydrology}

%Outsanding Q is why Weertman can expl. hard bed with cavities in Gilbert and Gimbert GRL papers. Are cavities just much less important in real 3D settings? This is outlined fairly well in 4.2 of the Gilbert GRL paper. So for them it's more of a mathematical simplification, i.e. the form just happens to work -- rather than an actual absence of cavities. but none of these papers will include form drag. Honestly I should probably just omit the ikens stuff in the gimbert grl paper, probably not worth including that complication here. 

%[\comment{I think this is getting there now, but wondering if it could be made slightly shorter}]

Subglacial hydrology and basal sliding are closely linked through feedbacks in many settings. Water pressure influences sliding velocity through modification of subglacial till properties (soft beds, Section \ref{s:soft}), bed separation (perhaps mainly hard beds, Section \ref{s:hard}), combinations thereof (hard-soft overlap), and drainage network evolution (both soft- and hard-beds). Reviews including  \citet{Clarke2005SubglacialProcesses, Flowers2010GlacierColleagues,Irvine-Fynn2011PolythermalReview, Ashmore2014AntarcticChallenges, Chu2014GreenlandHydrology, Flowers2015ModellingSheets, Nienow2017RecentSystem, Davison2019TheSheet} cover these processes in detail and with a greater emphasis on diurnal and seasonal variations, so we focus here on the main methods for incorporating hydrology into production models and the question of what value effective pressure, \(N\), actually represents within a grid cell.

%the main controls %at greater-than annual timescales
%and how variable subglacial hydrology is incorporated into production models in the context of sliding.

Importantly, even when \(N\) is not explicitly included in the sliding relationship of choice in a given production model (for example, the power law in Table \ref{tab:eqs}), the influence of hydrology is still indirectly incorporated through the value of the sliding coefficient \(C\). This becomes particularly consequential if \(C\) is obtained in an inversion procedure using a snapshot of velocity and then remains constant through time while the hydrological conditions change. For example, the effective pressure will be zero at the grounding line/zone by definition, but will increase inland in most circumstances. Consequently, a shift in grounding line position will change the local value of \(N\) even if this is not captured by a temporally constant \(C\). A shift from marine to land terminating glaciation might also significantly influence the subglacial hydrological system \citep{Maier2022ThresholdGreenland}. A common way to address this problem is to include \(N\) through a height above buoyancy approach with
\begin{equation}
    N = \rho_i gH_i - \rho_wg\textrm{max}(0, -z_r)
    \label{eq:HIAB}
\end{equation}
where \(z_r\) is the ice-bed position relative to sea level and \(H_i\) is the ice thickness -- as frequently implemented within ISMIP6 experiments \citep{Goelzer2020TheISMIP6, Seroussi2020ISMIP6Century} and elsewhere (e.g. \citealp{Winkelmann2011TheDescription, Tsai2015MarineConditions, Pattyn2017Sea-levelV1.0, Ruckamp2020SensitivityISSM}). The problem here is that this sets \(N=\rho_igh_i\) when \(z_r\) reaches 0, but subglacial water pressure often remains close to overburden (\(\gtrsim0.95\rho_i gh_i\)) far inland (for instance, \citealp{Hubbard1995BoreholeSwitzerland, Hermann1998BasalAntarctica, Fudge2005DiurnalUSA, Wright2016MeasuredPotential, Meierbachtol2018ShortSystem}). \citet{Kazmierczak2022SubglacialForcing, Kazmierczak2024AGlaciers} show that the steep gradient in \(N\) inherent to a transition from \(N=0\) to \(N=\rho_i gh_i\) artificially increases grounding-line sensitivity compared to more realistic alternative parameterisations (Figs. 5b, 7 of \citealp{Kazmierczak2024AGlaciers}). %An inverted or adjusted \(C\) field will compensate for this discrepancy at initialisation, but compensating errors may be anticipated as the ice sheet evolves away from its optimised state. 
\citet{Kazmierczak2024AGlaciers} instead advocate for relatively straightforward calculation of \(N\) based on soft- or hard-bed conditions that results in a lower gradient in effective pressure sustained over a much shorter distance. In alternative approaches, hydrology models based on hydraulic potential (\citealp{Bueler2015Mass-conserving0.6}) or immediate transfer of annually averaged surface melt to the bed (\citealp{Furst2015Ice-dynamicWarming}) also feature in ISMIP6 experiments. Models such as GlaDS \citep{Werder2013ModelingDimensions}, which is implemented within Elmer/Ice and the Ice Sheet System Model (ISSM), calculate subglacial water pressure fields at scale through a distributed sheet and channels approach. However, while there have been hydrology model intercomparison model efforts \citep{DeFleurian2018SHMIPProject} and numerous setting-specific studies utilising these models, we are not aware of these more complicated models being used within predictive or paleo ice sheet or glacier models. This is with some justification, as large-scale inclusion of a hydrology model becomes a compromise between improved realism and the additional introduced complexity and parameter uncertainty.

%[\comment{Advocate for not dropping \(p_w\) below about 90 percent of overburden here or in discussion? I think in discussion. Ideally, the mismatch should be as small as possible, while not introducing unecessary complications. N at frozen section will be rhogh, but less dynamically active, so a mismatch here is less of a problem.  }]

Observations evidence widely varying \(p_w\) time series in adjacent boreholes \citep{Doyle2018PhysicalGreenland, Rada2018ChannelizedYukon, Doyle2021WaterGreenland}, consistent with theoretical studies finding well-connected low-pressure conduits adjacent to poorly-connected higher-pressure regions  \citep{Engelhardt1997BasalObservations, Hewitt2018TheFormation}. Given such variability, which \(N\) should be applied to a given grid cell? Or, what is the value of \(N\) in a production model actually representing -- the mean \(N\) across a grid cell, the local maximum \(N\) (assuming that the correspondingly higher local traction controls regional traction across), or some other weighting? How \(N\) is incorporated within a sliding relationship is also not a settled manner. Both \citet{Zoet2020ABeds} and \citet{Helanow2021ATopography} advocate for \(N\) as a linear prefactor alongside \(C\) but also include it within the \(u_t\) term (see also Section \ref{s:threshold}), while in Budd sliding \(N\) is raised to a power \(q\) \citep{Bindschadler1983TheBed., Budd1984ASheet, Schweizer1992TheBed}, even though most modern applications employing Budd sliding set \(q\) to 1 \citep{Choi2022UncoveringDecade}. Aside from the derivation of \citet{Fowler1987SlidingFormation}, the physical validity of which is contested by \citet{Schoof2005TheSliding}, it is worth noting that we are aware of no process-based sliding model that yields an effective-pressure exponent where \(q\not=1\). 

Therefore, while the influence of \(N\) is fairly well understood for specific sliding sub-processes in isolation, %-- for example, cavitation over a relatively flat bedrock area \citep{Helanow2021ATopography}, and sediment shearing in a laboratory setting (e.g., \citealp{Iverson1998Ring-shearBeds}) or through a numerical continuum model (e.g., \citealp{Damsgaard2020WaterStreams}) -- 
further work is needed to fully quantify the influence of a spatially variable \(N\) within a given model grid cell. Specifically, we need to understand better how to relate \(N\) to other parameters and variables within a standard sliding relationship. What constitutes an optimum method of incorporating \(N\) within a production model at the grid-scale and up, whether through a simple approach related to bed elevation or a more complex secondary model (such as \citealp{Kazmierczak2024AGlaciers} or those featured in \citealp{DeFleurian2018SHMIPProject}) also remains an important open question. 

\section{ Sliding-bulk flow separation} \label{s:inout}

In this section we propose basal sliding at the production model bed as the sum of three terms: (i) a regularised-Coulomb component dependent on effective pressure and slip velocity at an ice-bed interface that has been smoothed to preclude cavitation; (ii) a power-law component only indirectly influenced by effective pressure representing normal forces resulting from roughness exceeding the scale of cavitation but not represented within the production model bed topography; and (iii) an error term including both the error inherent to the approximations of (i) and (ii) and to compensatory errors arising from the use of a sliding relationship as a tuning mechanism for other uncertain processes and fields within the production model. In 1D form this is written as
\begin{equation}
    \label{eq:slidingcompound1D}
    \tau_b = NC_{rC}\left(\frac{u_b}{u_b + u_t}\right)^{1/m_{rC}} + C_{pl}{u_b}^{1/{m_{pl}}} + \epsilon_t \,
\end{equation}
where \(C_{rC}\) and \(m_{rC}\) are the sliding coefficient and exponent for the regularised-Coulomb term, respectively, and \(\epsilon_t\) is the total error term (with its potential components covered in Section \ref{s:errors}). In effect, this defines sliding for production models as the negative of modelled bulk ice deformation. The regularised-Coulomb term accounts for soft- and hard-beds, stick slip, and soft-hard overlap (Sections \ref{s:soft}, \ref{s:hard}, \ref{s:stickslip}, \ref{s:overlap}) and the power-law term accounts for roughness (Section \ref{s:form}) while temperate ice and the basal ice layer (Sections \ref{s:form}, \ref{s:BIL}) influence both terms and may introduce errors if misrepresented within modelled bulk ice deformation.

\subsection{ Background for sliding-bulk flows}

To obtain Eq. \ref{eq:slidingcompound1D} we start from ideas in the early work of \citet{Fowler1977GlacierDynamics}, \citet{Fowler1978OnAnalysis}, and \citet{Fowler1981AJSTOR} (hereafter collectively Fowler1977, outlined in Appendix \ref{A:fowler}) who introduce an inner-outer flow framework for analytical models in two-dimensions with a reduced set of processes. In our application, we introduce a `sliding' layer, which accounts for the relevant subset of processes outlined in Fig. \ref{fig:modes}, and in the preceding paragraph, and a `bulk' layer which comprises englacial ice deformation with minimal influence from basal processes. 

\begin{figure*}
\centering{\includegraphics[width=0.85\textwidth]{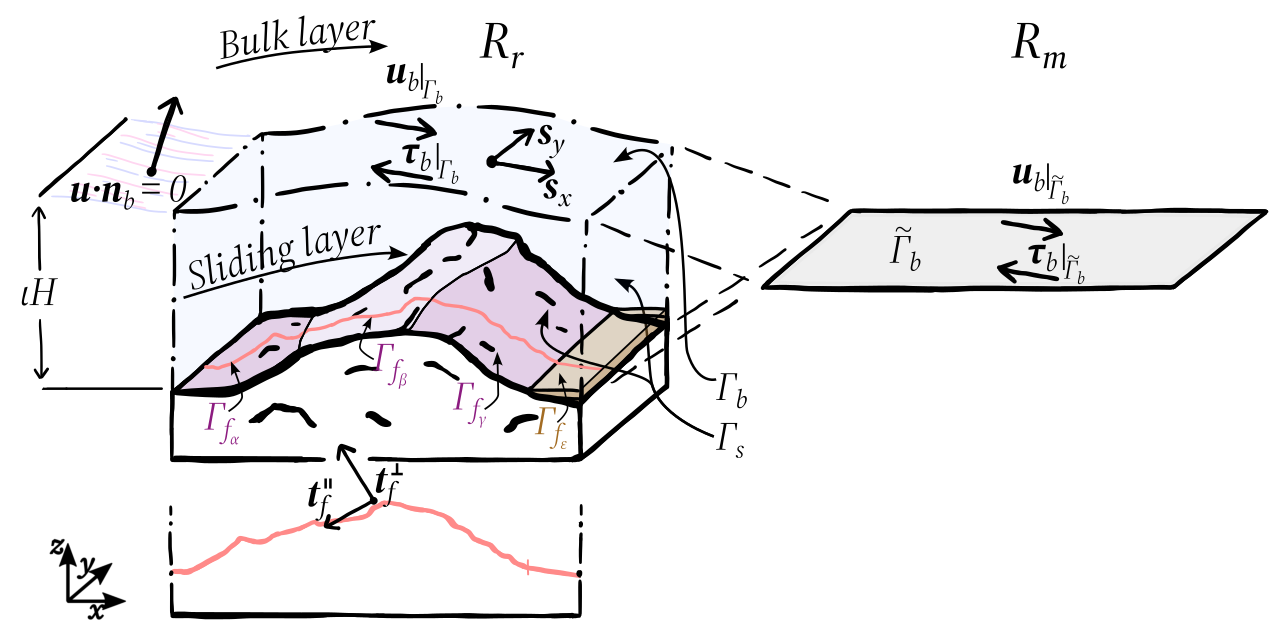}}
\caption{\textbf{Bulk and sliding layers.} Bulk and sliding layers for a `real' (and in this example, rough) glacier region, \(R_r\), and its production model mirror, \(R_m\), showing the sliding-bulk interface (\(b_{b}\)), regions of varying basal slip parameters (\(\Gamma_{f_\alpha}, \Gamma_{f_\beta}, ...\)), and other quantities described throughout Section \ref{s:inout}. Depending on setting the sliding layer may feature any of the processes in Section \ref{s:slidingprocesses}, Fig. \ref{fig:modes}.}
\label{fig:inout}
\end{figure*}

As visualised in Fig. \ref{fig:inout} we refer to two regions: a \emph{real} three-dimensional volume, \(R_r\), that represents the theoretical ideal of what is actually happening in reality and a zero-thickness production \emph{model} surface, \(R_m\), at the modelled ice-bed interface, where \(R_r\) and \(R_m\) have the same lateral extent. The challenge then, is how the \emph{real} can best be wrought into a production \emph{model} grid cell. In practice, obtaining a fully detailed \emph{real} stress/velocity field and bed topography is not actually possible for an existing glacier, so here \emph{real} is primarily used as a theoretical ideal that we attempt to approximate through the production \emph{model} representation. Alternatively, this can be viewed as a hierarchy of models, or progression of assumptions. Returning to the bulk rheology of ice as our glaciological analogy, observations or modelling of grain-scale processes (such as \citealp{Steinbach2017TheAnalysis,Behn2021TheLaw}) function as the \emph{real}, the continuum implementation of this behaviour as the production \emph{model}. As previously discussed, the simplifying factor there is that a representative volume element scale distinguishing ice rheology modes is fairly neatly defined, whereas it is not defined at all for sliding. Italicised \emph{real} and \emph{model} terms are retained throughout when referencing these regions to emphasise the distinction.

Using the the ice-bed interface itself to determine a sliding relationship is quite common (i.e. \citealp{Helanow2021ATopography, deDiego2022NumericalSliding}), but there are two reasons for giving our \(R_r\) finite thickness. \textbf{First}, we want to allow for rheological differences in the lower ice column between the \emph{real} and the \emph{model} (for example, the basal ice layer, temperate ice). \textbf{Second} while traction summed over the region's bed will reasonably match the traction at the roof of the region %(minus the integral over non-hydrostatic pressure, see Section \ref{s:innervolume}),
the same cannot be said of the summed slip velocity if slip itself is low but shear bands that would not be captured by the production \emph{model} are present \citep{Law2023ComplexIceb, Liu2024SpontaneousBedrock, Barndon2025IceLandscapes}. Shear bands are probably only ubiquitous over beds exhibiting roughness at scales larger than the cavitation length scale (described in Section \ref{s:scalecavities}), but if we want a system that can capture all sliding behaviours then a finite thickness \(R_r\) is necessary. In an ideal case the shear stress at our sliding-bulk interface in the \emph{model} then effectively emulates the shear stress present in the \emph{real} glacier (Fig. \ref{fig:inout}) but we are forced by the discretised geometry to break from Fowler1977 by relaxing the requirement for exact equivalency between the \emph{real} and \emph{model} regions. 

%Recognising that the smooth basal boundary used in larger-scale analytical models is not a realistic representation of a rough glacier bed, 
Using a sliding-bulk layer distinction means that we can incorporate an arbitrary number of sliding sub-processes within the sliding layer and more closely align \emph{model} and \emph{real} regions, but a universally applicable definition for the position of the sliding-bulk interface is not immediately clear. In Fowler1977, the boundary is set as an impenetrable surface (that is, no normal flow) at \textasciitilde{}1--10\% of the overall ice thickness based on an assumed roughness wavelength of the \emph{real} bed and is then taken as the \textit{de facto} model bed. As perturbations in strain rates, and the strain rates themselves, decrease in amplitude with increasing height above the bed in most settings (e.g., \citealp{Sharp1953DeformationGlacier, 
Balise1985TransferGlacier, Arenson2002BoreholeSwitzerland, Harper2001SpatialBoreholes, Ryser2014SustainedDeformation, Doyle2018PhysicalGreenland, Lee2019AAlaska,Law2021ThermodynamicsSensing}) %either exponentially \citep{Balise1985TransferGlacier} or as a power-law \citep{Chandler2006ACalculations}
a sliding-bulk layer boundary at around the \textasciitilde{}1--10\% mark is a reasonable starting point and we continue with this usage here. Nonetheless, realistic conditions such as vertically dominated flow near ice divides, basal melting (which would gradually decrease the thickness of the sliding layer), %mismatches between the \emph{real} and \emph{model} bed, 
and varying appropriate positions for the sliding-bulk interface complicate the global validity of this treatment. %Any approach that takes a non-zero sliding layer thickness will also encounter problems regarding treating a 3D volume as a 0-thickness surface. 
We can at least partially mitigate the issue of spatial variability of the sliding-bulk interface by defining it on a piece-wise grid-cell by grid-cell basis.

These concerns are covered further in Section \ref{s:notoneiota} [\comment{todo}], but we emphasise that even if this problem is not widely discussed, it is already present in every production model implementation of glacier sliding. The \emph{model} bed will always be a misrepresentation of \emph{reality}, and, as a result of this and further rheological/bed condition misrepresentations, the \emph{real} basal stress state of a glacier will never be ideally matched to its \emph{model} implementation. Our approach therefore does not generate these misalignments (they already exist) but provides a framework to more clearly describe them.

\subsection{ Region and surface definitions}

We define the sliding-bulk interface, \(b_b(x,y)\) for the \emph{real} region through the use of flow-advecting streamlines originating at a uniform normalised height, \(\iota\), at the region's upstream boundary (shown in Fig. \ref{fig:inout}). \(x\) and \(y\) are the horizontal axes with \(z\) vertical. We distinguish the sliding-bulk interface from the ice-bed interface, \(b_f(x,y)\), (\(_f\) for 'floor') and use \(\tilde{b_{b}}(x,y)\) for the zero-thickness \emph{model} bed, where a \(\tilde{\,}\) henceforth denotes a variable attached to the \emph{model} region. Eddies, folds, and overturns (e.g. \citealp{Bons2016ConvergingSheet, Zhang2024FormationSheet}) are neglected. The upstream seed position for the streamlines is 
\begin{equation}
    %z_{\textrm{seed}}(x, y)=b_f(x,y) + \iota H(x,y) \: 
    b_{b,0}(x, y)=b_f(x,y) + \iota H_i(x,y) \,.
    \label{eq:seed}
\end{equation}

We then use \(\Gamma_f\), \(\Gamma_b\), and \(\Gamma_s\) for the ice-bed interface, bulk-sliding interface, and sides of the \emph{real} region, \(R_r\), and \(\tilde{\Gamma}_b\) for the \emph{model} bed. \(V_r\) represents the volume of \(R_r\) while \(a_b\) and \(\tilde{a}_b\) represent the areas of \(\Gamma_b\) and \(\tilde{\Gamma}_b\). \(\Gamma_f\), \(\Gamma_b\), and \(\tilde{\Gamma}_b\) are subsets of \(b_f\), \(b_b\), and \(\tilde{b}_b\) respectively. To describe basal sliding at a given point, we seek the basal shear stress \(\boldsymbol{\tau}_b=(\tau_{b,x}, \tau_{b,y})\), and basal velocity, \(\boldsymbol{u}_b=(u_{b,x}, u_{b,y})\), defined at \(b_{b}(x,y)\) and transferred to \(\tilde{b_{b}}(x,y)\) (with \(x,y\) dependence dropped hereafter).

\subsection{ Extracting \(\boldsymbol{u}_b\) and \(\boldsymbol{\tau}_b\) from known fields} \label{s:ioposition}

While a full stress and velocity field of a \emph{real} region is beyond the scope of available observations a detailed numerical model (e.g. \citealp{Helanow2021ATopography, Law2023ComplexIceb, Barndon2025IceLandscapes}; and separate to the italicised production \emph{model} region introduced above) can provide a closer approximation of these fields for a given setting. If a stress field is obtained from the output of a detailed numerical model and taken as the \emph{real} field then \(\boldsymbol{\tau}_b\) at the sliding-bulk interface, \(b_{b}\), can be simply calculated directly from the overall deviatoric stress field. First, \(\boldsymbol{n}_{b}\), the upwards-pointing normalised vector normal to \(b_{b}\) is 
\begin{equation}
    \hat{\boldsymbol{n}}_{b} = \begin{bmatrix}
        -\frac{\partial b_{b}}{\partial x} \\
        -\frac{\partial b_{b}}{\partial y} \\
        1 \\
    \end{bmatrix}, \quad
    \boldsymbol{n}_{b} = \frac{\hat{\boldsymbol{n}}_{b}}{||\hat{\boldsymbol{n}}_{b}||} \: .
\label{eq:normalnormal}
\end{equation}

To obtain components aligned tangential to the interface, i.e., aligned with the local tangent plane at \(b_b\), we define two unit vectors perpendicular to one another, \(\boldsymbol{s}_x\) and \(\boldsymbol{s}_y\) satisfying \(\boldsymbol{s}_x\cdot\boldsymbol{s}_y=0\) and \(\boldsymbol{s}_i\cdot\boldsymbol{n}_{b}=0\) where \(\boldsymbol{s}_i \in \{ \boldsymbol{s}_x, \boldsymbol{s}_y \}\) (which could be constructed via a Gram-Schmidt orthogonalization procedure). 
The tangential velocity and traction vectors can then be projected onto \(\boldsymbol{s}_i\) to yield scalar components:
\begin{equation}
    \boldsymbol{u}_b =
    \begin{bmatrix}
        \boldsymbol{u} \cdot \boldsymbol{s}_x \\
        \boldsymbol{u} \cdot \boldsymbol{s}_y
    \end{bmatrix}
\end{equation}
and 
\begin{equation}
    \boldsymbol{\tau}_b =
    \begin{bmatrix}
        (\boldsymbol{\sigma} \cdot \boldsymbol{n}_{b}) \cdot \boldsymbol{s}_x \\
        (\boldsymbol{\sigma} \cdot \boldsymbol{n}_{b}) \cdot \boldsymbol{s}_y
    \end{bmatrix} \, 
\end{equation}
where \(\boldsymbol{u}\) is the full three-dimensional velocity vector and \(\boldsymbol{\sigma}\) is the total Cauchy stress tensor.

%The most straightforward way to probe relationships between \(\boldsymbol{u}_b\) and \(\boldsymbol{\tau}_b\) may then be through an ensemble of detailed numerical models, but we continue to explore the general behaviour here in order to consider given settings where different sub-processes may come to dominate. 

\subsection{ Sliding layer force balance}
\label{s:innervolume} 

Our aim is to include a flexible number of processes within the sliding layer, but we first define the forces operating over the \emph{real} and \emph{model} regions. In practical applications for production models the computational grid is topographically fitted and projected vertically from the bed in three-dimensional applications, or remains planar in two-dimensional applications. \(R_r\) and \(R_m\), will therefore have the same horizontal extent, even if \(R_r\) is the only region to represent a volume and not a surface. We assume that the forces acting on opposing \(\Gamma_s\) of the thin \(R_r\) volume are small enough compared to that on \(\Gamma_b\) and \(\Gamma_f\) that these can be neglected, similar to the assumptions of the shallow ice approximation (but restricted to near-basal flow). %and to the assumptions in cavitation theories (e.g. \citealp{Schoof2005TheSliding}) that bed properties are macroscopically continuous outside the region of interest. 

The main aim of this approach is to match the shear forces defined by \(F_{\boldsymbol{s}_i}^{\Gamma_{b}}\) and \(\tilde{F}_{\boldsymbol{s}_i}^{\tilde{\Gamma}_{b}}\) which act tangentially across \(\Gamma_{b}\) and \(\tilde{\Gamma}_b\) (and \(R_r\) and \(R_m\)) in the directions \(\boldsymbol{s}_x\) and \(\boldsymbol{s}_y\) as closely as possible. These steps are provided in Appendix \ref{a:forcebalance} and present \(\boldsymbol{\tau}_b\) averaged over \(\Gamma_b\) as
\begin{equation}
\label{eq:taubAio}
    \boldsymbol{\tau}_b|_{\Gamma_b} = \frac{1}{A_{b}} \begin{bmatrix}
        F_{\boldsymbol{s}_x}^{\Gamma_{b}} \\
        F_{\boldsymbol{s}_y}^{\Gamma_{b}} 
    \end{bmatrix} = \frac{1}{A_{b}} \begin{bmatrix}
        F_{\boldsymbol{s}_x}^{\parallel} + F_{\boldsymbol{s}_x}^{\perp} \\
        F_{\boldsymbol{s}_y}^{\parallel} + F_{\boldsymbol{s}_y}^{\perp}  
    \end{bmatrix} \, 
\end{equation}
for the region \(R_r\) where \(F_{\boldsymbol{s}_i}^{\parallel}\) and \(F_{\boldsymbol{s}_i}^{\perp}\) represent the forces tangential and perpendicular to the ice-bed interface respectively in the directions \(\boldsymbol{s}_i\). These come from the traction components projected at the ice base
\begin{equation}
\label{eq:normaltangential}
    \boldsymbol{t}_f\cdot\boldsymbol{s}_i = \underbrace{\boldsymbol{t}_f^{\perp} \cdot\boldsymbol{s}_i}_{\text{normal component}} + \underbrace{\boldsymbol{t}_f^{\parallel} \cdot \boldsymbol{s}_i}_{\text{tangential component}} \, 
\end{equation} 
where \(\boldsymbol{t}_f=\boldsymbol{\tau}\cdot(-\boldsymbol{n}_f) - p'(-\boldsymbol{n}_f)\) is the overall traction -- with \(\boldsymbol{n}_f\) the upwards pointing normal at \(b_f\) obtained as for \(\boldsymbol{n}_b\), \(\boldsymbol{\tau}\) the deviatoric stress tensor, and \(p'\) the non-hydrostatic pressure -- and \(\boldsymbol{t}_f^\perp\) and \(\boldsymbol{t}_f^\parallel\) are its normal and tangential components, respectively (Appendix \ref{a:forcebalance} for further details). The requirement of a strictly planar surface for \(\tilde{\Gamma}_b\) for production model purposes means that it will not be directly equivalent to \(\Gamma_b\) and that \(\tilde{a}_b\) may not match \({a}_b\). Ideally, we can then set 
\begin{equation}
\label{eq:taubRmRr}
    \boldsymbol{\tau}_b|_{\tilde{\Gamma}_b} = \boldsymbol{\tau}_b|_{\Gamma_b} + \boldsymbol{\epsilon}_{\boldsymbol{\tau}_b} \:,\:\:\:
    \boldsymbol{u}_b|_{\tilde{\Gamma}_b} = \boldsymbol{u}_b|_{\Gamma_b} + \boldsymbol{\epsilon}_{\boldsymbol{u}_b}
\end{equation}
where \(\boldsymbol{\epsilon}_{\boldsymbol{\tau}_b}\) or \(\boldsymbol{\epsilon}_{\boldsymbol{u}_b}\) are the error terms that characterise the mismatch between the \emph{model} and the \emph{real}.

\subsection{ Including sliding layer processes}
\label{s:innerouterinner}

\subsubsection{ Scale and cavities}
\label{s:scalecavities}

In this sub-section, we present one way to bridge smaller-scale laboratory and numerical studies of sliding processes with the typical resolutions of production glacier and ice sheet models. However, to do this we need to be careful in stating which processes can contribute to \(\boldsymbol{t}_f^\parallel\) and which surface we take \(b_f\) to represent. For sediment processes this is simple -- there is a separation of orders of magnitude between sediment slip processes and typical glacier model resolution (Fig. \ref{fig:scale}) making it clearly appropriate to sub-parameterise slip at any aerially obtained \(b_f\) (resolutions \(\sim\)0.5 m and greater) with relationships derived for ice-sediment friction. This is in keeping with the long-established principle of applying scale-separating filters to represent sub-grid processes in continuum models \citep{Schumann1975SubgridAnnuli}. %It is furthermore clearly appropriate to take \(b_f\) as smooth at the metre-scale, rather than have it feature the milimetre- to centimetre-scale undulations that may be inherent to the sediment surface.

However, hard-bed cavitation theory (and observations) slightly overlap with typical glacier model resolution (though not with ice-sheet model resolution, Fig. \ref{fig:scale}). If the velocity field and subglacial water pressure within \(R_r\), and therefore also cavity geometry, are temporally constant, then this presents no problem: Section \ref{s:innervolume} still provides a useful way to think about the forces contributing to \(\boldsymbol{\tau}_b|_{\tilde{\Gamma}_b}\) at \(\tilde{\Gamma}_b\) -- we just treat the ice-water interface as a result of cavitation as the \emph{de facto} ice-bed interface and the local normal force \(\boldsymbol{t}_f^\perp\) becomes a projection of the local subglacial water pressure. However, this clearly does not work if the geometry of \(b_f\) is a function of \(\boldsymbol{u}_b\) and \(N\). 

For our purposes in this sub-section, we then place cavitation processes within \(\boldsymbol{t}_f^\parallel\) in our hierarchy of models and consider a smoothed \(b_f\) and \(\Gamma_f\), from here denoted \(\bar{b}_f\) and \(\bar{\Gamma}_f\) that does not feature cavitation (Fig. \ref{fig:iken}). To note, a reduction from \(b_f\) to a smoother bed not featuring cavitation is already inherent to all transformations from process-based cavitation studies to larger-scale models (and not limited to production models). To formalise this simplification, the \emph{real} basal topography may be smoothed to filter roughness components shorter than a prescribed transition wavelength. One practical way to achieve this is to apply low-pass filter as widely deployed to remove digital elevation model processing artefacts \citep{Arrell2008SpectralData} or more recently to accentuate and identify crater impacts \citep{Gonzalez-Diez2021TheFeatures,Gonzalez-Diez2023ImprovingArea}. The transition between the scale where this approximation is and is not appropriate is an outstanding question covered in Section \ref{s:ikens}, but we for now take it as reasonable at least for ice-sheet models whose spatial resolution is typically an order of magnitude greater than cavitation theory (Fig. \ref{fig:scale}) and offer \(l_{c}\)=100 m. As we shall see, this simplification still allows us to make interesting assertions about the role of cavitation at scale. 

 % wide a spectral (Fourier) filter, in which \(b_f\) is decomposed into its spatial wavelengths and a smooth low-pass function is used to attenuate or remove components shorter than a chosen cutoff \(l_c\). [\comment{Omit the following sentence as unecessary? Appendicise it?}] Specifically, this would involve computing the Fourier transform \(\hat{b}_f = \mathcal{F}\{b_f\}\), multiplying by a filter \(H(|\boldsymbol{k}|)\) that transitions from one at long wavelengths (small \(|\boldsymbol{k}|\)) to near zero around \(|\boldsymbol{k}| = 2\pi/l_c\), and then inverse-transforming to obtain a smoothed field \(\bar{b}_f = \mathcal{F}^{-1}\{H(|\boldsymbol{k}|)\,\hat{b}_f\}\). This technique is widely deployed to remove digital elevation model processing artefacts \citep{Arrell2008SpectralData} or more recently to accentuate and identify crater impacts \citep{Gonzalez-Diez2021TheFeatures,Gonzalez-Diez2023ImprovingArea}. %If a more gradual transition is desired, two length scales \(l_{c_1}\) and \(l_{c_2}\) can be specified to form a tapering window (such as a Butterworth function, Butterworth, 1930) that progressively reduces power between those wavelengths. A real-space Gaussian filter could also be used, where the kernel width sets an equivalent smoothing length \citep{Hoffman2022TheSliding}, although this offers less explicit control over the retained wavelengths. 

\subsubsection{ Slip}
\label{s:inclslip}

Looking first at slip processes, we follow most work on glacier sliding in assuming that slip velocity is continuous in time and an algebraic function of \(\boldsymbol{t}_f^\parallel\) and represented through a Robin boundary condition. Slip will then have a co-interaction with the stress field within \(R_r\) --- even a perfectly flat bed could result in a complicated stress field through `stickier' or `slipperier' bed patches \citep{Ryser2014Caterpillar-likeSheet} that will in turn influence both the velocity and traction at the ice-bed interface. Therefore while slip, \(F_{\boldsymbol{s}_i}^{\parallel}\), and drag, \(F_{\boldsymbol{s}_i}^{\perp}\), forces influence one another (but remain separate in the force balance), treating them as independent is a convenient simplification if we wish to divide the processes contributing to \(\boldsymbol{\tau}_b\) into slip and drag components.

In Appendix \ref{a:slip} we define multiple, spatially independent, slip sub-processes within \(R_r\) as being a function of space. This can can function using a variety of sliding relationships but as evidence in Section \ref{s:slidingprocesses} suggests that a regularised-Coulomb relationship (Eq. \ref{eq:constitutive_slip}) may be the most appropriate relationship at length scales below 25 m we progress only with regularised-Coulomb. Over \(\Gamma_f\) this gives
\begin{equation}
    \boldsymbol{t}_f^\parallel|_{\Gamma_f} \approx -P_{R_a}N_{R_a} \left( \frac{||\boldsymbol{u}_f||}{||\boldsymbol{u}_f||+{u_t}_{R_a}} \right)^{\frac{1}{m_{R_a}}} \frac{\boldsymbol{u}_f}{||\boldsymbol{u}_f||}
    \label{eq:constitutive_slipsumaggregate}
\end{equation}
where the subscript \(R_a\) refers to the aggregate or representative value over \(\Gamma_f\)  for parameters \(P\), \(N\), \(u_{t}\), and \(m\). These can be reduced or approximated to single regionally-representative values as discussed in Appendix \ref{a:slip}.

\subsubsection{ (Locally) frozen beds}
\label{s:localfreeze}

There are two ways to treat slip at the ice-bed interface for frozen beds. In the first, no slip displacement is permitted, following limited observations suggesting true slip is close to zero. We treat the first case where \(\boldsymbol{u}_f\) is set as 0 for a 1D simple shear column in Appendix \ref{A:frozencolumn} and focus in this Section on what we propose here as a second case. Specifically, we suggest \(\boldsymbol{u}_f\) is exponentially reduced as temperature relative to the melting point decreases. Then a yield stress is imposed, which can be written as:
\begin{equation}
    \label{eq:frozenn}
\boldsymbol{t}_f^\parallel = \textrm{min}(CNe^{-T*/\alpha}\boldsymbol{u}_b^{1/m}, \tau_{\textrm{yield}}) \, 
\end{equation}
where \(\tau_\textrm{yield}\) is the yield strength of the frozen boundary which is similar in form to Eq. 8 of \citet{Tsai2015MarineConditions}. \(N\) is still included here as in  Eq. \ref{eq:frozencuzzone}. Although a frozen bed will not have water pressure the position of the cold-thawed transition is highly uncertain \citep{Macgregor2022GBaTSv2:Sheet, Raspoet2025EstimatesSheet}, denoting a boundary where hydrology is turned 'off' presents its own complications, and the mismatch from compensating for \(C\) through an inversion in frozen areas where \(N\) equals overburden is less dynamically consequential than not including \(N\) close to grounding lines. (\(CNe^{-T*/\alpha}\boldsymbol{u}_b^{1/m}\) could also be replaced in favour of a regularised-Coulomb term though this makes less sense if the maximum \(\tau_b\) in this substitution falls below \(\tau_\textrm{yield}\).) In contrast to a hard no-slip boundary (Appendix \ref{A:frozencolumn}), Eq. \ref{eq:frozenn} can be approximated as a regularised-Coulomb relationship, with a comparatively very low \(u_t\) and very high \(C\). Treating \(\boldsymbol{t}_f^\parallel\) in frozen subregions as a regularised-Coulomb approximation then allows us to retain the same approach to \(\boldsymbol{t}_f^\parallel\) as in Eq. \ref{eq:constitutive_slipsumaggregate} (and also Eq. \ref{eq:constitutive_slip} in Appendix \ref{a:slip}) regardless of whether part, or the entirety of \(R_r\) is frozen to its base. 

We continue under the assumptions of the second approach (Eq. \ref{eq:frozenn}), it being perhaps the more physically motivated of the two, and also the more compatible with the rest of our framework. Tests under varied settings will elucidate if the two approaches produce meaningfully different outcomes. The question of whether frozen beds can `slide' is addressed in Section \ref{s:frozenthawed}. In either case the tangential component, \(\boldsymbol{t}_f^\perp\), will function the same as in Section \ref{s:deformation}.

%Additionally, expanding sub-temperate sliding out to a grid cell under consideration also allows for the situation of both freezing and thawed areas within the same grid cell (Fig. \ref{fig:subtemp}), which may be a common situation in frozen-thawed transition zones (e.g., \citealp{Oswald2012MappingSheet}; \citealp{Dawson2024HeterogeneousAntarctica}). 

\subsubsection{ Subglacial lakes and ice caves}
\label{s:frictionlake}

The opposite situation to Section \ref{s:localfreeze} arises in the case that a sub-region of \(R_r\) is occupied by water or air (excluding the cavities covered in Section \ref{s:scalecavities}) where we can put \(\boldsymbol{t}_f^\parallel=0\). If an ice shelf of lake (such as Lake Vostok in Antarctica) largely or wholly occupies \(R_r\), and if \(\boldsymbol{s}_i\) defines a plane parallel to the ice-water interface and/or perpendicular to the direction of gravity, respectively, then \(F_{\boldsymbol{s}_i}^\perp\) will go to 0. This is a roundabout way of saying that our treatment of a sliding layer returns the expected results under these conditions, but also allows the case of regions partly occupied by water or air to be formalised. 

\subsubsection{ Deformation and form drag}
\label{s:deformation}

The processes contributing to the composite normal traction field, \(\boldsymbol{t}_f^\perp(x, y, \theta_{p_{d}})\), cannot be separated in the same manner as \(\boldsymbol{t}_f^\parallel\) as the parameters, \(\theta_{p_{d}}\), that may influence \(\boldsymbol{t}_f^\perp\) locally (including topography, rheological variations, and slip properties) span the entire three-dimensional region of \(R_r\). 

However, \(F_{\boldsymbol{s}_i}^\perp\) should be predominantly controlled by ice flow over the bed. At very low Reynolds numbers low-angled obstacles provide minimal resistance, but steeper or more abrupt obstacles can present important resistance contributions which cannot be calculated analytically except in special cases (\citealp{Fox2006IntroductionMechanics}, 9.7). \citet{Helanow2021ATopography} indicate a power-law for the case where cavity formation is not permitted (their Fig. S6) over small-scale (<25 m across) domains, in agreement with earlier analytical and modelling studies considering simplified geometry and no cavity evolution \citep{Weertman1957OnGlaciers, Lliboutry1968GeneralGlaciers, Lliboutry1979LocalOpenings, Fowler1981AJSTOR, Lliboutry1987VeryGlaciology., Gudmundsson1997Basal-flowBedrock}. Through dimensional analysis (Appendix \ref{A:dimensional}) we can write
\begin{equation}
    \label{eq:dimensional}
    F_D\sim A^{\frac{-1}{n}}U_{\textrm{far}}^{\frac{1}{n}}e^{\frac{2n-1}{n}}\Phi\left(h/e, n\right)
\end{equation}
where \(F_D\) is the drag force per unit width generated by flow around roughness elements of characteristic height \(e\), \(U_\textrm{far}\) is the characteristic sliding velocity, \(h\) is the height from the roughness element baseline to the point where \(U_\textrm{far}\) is taken, and \(A\) comes from the Nye-Glen isotropic flow, which in invariant form is
\begin{equation}
    \label{eq:glenn}
    \dot{\epsilon}_e=A\tau_e^n
\end{equation}
where \(\dot{\epsilon}_e=1/2\textrm{tr}(\dot{\boldsymbol{\epsilon}}^2)\) and \(\tau_e=1/2\textrm{tr}(\boldsymbol{\tau}^2)\) are the effective strain rate and effective stress, respectively. Last, \(\Phi(h/e, n)\) is a dimensionless function of the aspect ratio \(h/e\) encapsulating the geometric dependence arising from the near-bed flow structure, and a power-law exponent.

%This is functionally similar to the classical example of a sphere stationary against a background flow of a Newtonian fluid with very low (<<1) Reynolds number with basic scaling arguments producing a power-law relationship such as %where one-third of total resistance arises from pressure drag and the remainder comes from viscous drag (\citealp{Stokes1851OnPendulums}; \citealp{Happel1983LowHydrodynamics}, 4-17.28). 
%\begin{equation}
%    F^\perp \propto U_\textrm{far}^{1/n}
%\end{equation}
%where \(U_\textrm{far}\) is the far-field velocity, analogous in some respects to \(\boldsymbol{u}_b\) (further details in Appendix \ref{A:pressuredrag}).

Therefore, while obtaining \(\boldsymbol{t}_{f}^\perp\) for complicated bed geometries would require detailed fluid mechanical modelling, it is sufficient to say for now that \(\boldsymbol{t}_{f}^\perp\) is of little importance for very smooth beds, but an increasing function of the pressure-drag provided by the bed geometry following a power law for most bed geometries. Approximating \(F_D\) in Eq. \ref{eq:dimensional} as \(F_{\boldsymbol{s}_i}^\perp\) and \(U_\textrm{far}\) as \(\boldsymbol{u}_b\) gives
\begin{equation}
    \label{eq:powerlawD}
    %F_{\boldsymbol{s}_i}^\perp\approx C_{R_a} A_{R_a}^{-1/n_{R_a}} \boldsymbol{u}_b^{\frac{1}{n_{R_a}}}
    \boldsymbol{t}_f^\perp|_{\Gamma_b}\approx \frac{1}{a_b}C_{R_a} A_{R_a}^{-1/n_{R_a}} \boldsymbol{u}_b^{\frac{1}{n_{R_a}}}
\end{equation}
where \(C_{R_a} = e_{R_a}^{(2n_{R_a}-1)/n}\,\Phi(h_{R_a}/e_{R_a}, n_{R_a})\) represents aggregate geometric properties and \(_{R_a}\) refers to representative aggregate as for the slip component. Note that many studies preceding modern production models and inversion methods were interested in deriving \(C_{R_a}\) analytically. Understanding geometric controls on \(C_{R_a}\) is valuable, but secondary to obtaining the scaling relationship as, excluding compensating errors, \(C_{R_a}\) can be revealed through inversion procedures. Nonetheless, detailed future modelling will reveal the expected magnitude of variation in \(C_{d_a}\) based on realistic landscape configurations and parameter sets, \(\theta_{p_d}\). %Also note that Eq. \ref{eq:powerlawD} is not dependent on effective pressure.

As explored in Sections \ref{s:hard}, \ref{s:form} glaciological definitions for form drag vary. For the purposes of this paper we define form drag as \(\boldsymbol{t}_{f}^\perp\) (or as \(F_{\boldsymbol{s}_i}^\perp\) when defined as a force over a grid cell), meaning we treat resistance arising from ice motion around sedimentary clasts as part of the tangential resistance. 

%This treatment of course raises more philosophical questions over what glacier sliding is if \(\boldsymbol{u}_b > 0\) when \(\boldsymbol{u}_f = 0\) across \(R_r\). Production-model sliding is non-zero in the interiors of the Greenland and Antarctic ice sheets when an inversion procedure based on present-day surface velocity fields is used \citep{Goelzer2020TheISMIP6, Seroussi2020ISMIP6Century}. We suggest then that the situation described here is a reasonable answer to the question of what this production-model sliding represents. The situtation described here is also consistent with our treatment of sliding in the rest of this Section and with the original Weertman theory where slip is neglected. 

%If a subregion, or potentially all, of \(R_r\) is frozen (Section \ref{s:frozenthawed}) and we neglect sub-temperate slip (or the basal temperature is low enough that it is negligible) then \(\boldsymbol{u}_f=0\) as a local Dirichlet condition and \(\boldsymbol{t}_f^\parallel\) becomes potentially unbounded (meaning even a small frozen patch within \(R_r\) could substantially alter the resultant \(F_{\boldsymbol{s}_i}^\perp\)). This results in a breakdown of the treatment of \(\mathcal{S}_{R_i}\) as a regularised-Coulomb relationship (Eq. \ref{eq:constitutive_slip}). 

\subsection{ Sliding processes into a sliding relationship at \(\tilde{\Gamma}_b\)}

\subsubsection{ Traction at the sliding-bulk layer interface as a simplified function of ice velocity}

We can obtain approximations for \(\boldsymbol{t}_f^\parallel\) and \(\boldsymbol{t}_f^\perp\) through Section \ref{s:innerouterinner} as Eqs. \ref{eq:constitutive_slipsumaggregate} and \ref{eq:powerlawD}, respectively. In the case of the slip component and Eq. \ref{eq:constitutive_slipsumaggregate} the major approximation is that we must replace \(\boldsymbol{u}_f\) with \(\boldsymbol{u}_b\) as \(\boldsymbol{u}_f\) is not simply obtainable. This approximation is then motivated by the projection of \(\boldsymbol{t}_f\) to \(\boldsymbol{t}_b\) (Section \ref{s:innervolume}). In the case of Eq. \ref{eq:powerlawD} the approximation comes from setting \(U_\textrm{far}\) as \(\boldsymbol{u}_b\). Then, setting \(P_{R_a}\approx C_{rC}\) and \(m_{R_a}\approx m_{rC}\) as the first sliding coefficient and exponent respectively and \(a_b^{-1}C_{R_a}A_{R_a}^{1/n_{R_a}}\approx C_{pl}\) and \(n_{R_a}\approx n_{pl}\) as the second sliding coefficient and exponent respectively we write
\begin{equation}
\label{eq:slidingcompound}
\begin{aligned}
    \boldsymbol{\tau_b}\big|_{\tilde{\Gamma}} =\;&
    NC_{rC}
    \left(
        \frac{\lVert \boldsymbol{u_b}\rvert_{\tilde{\Gamma}} \rVert}
             {\lVert \boldsymbol{u_b}\rvert_{\tilde{\Gamma}} \rVert + u_t}
    \right)^{1/m_{rC}}
    \frac{\boldsymbol{u_b}\rvert_{\tilde{\Gamma}}}
         {\lVert \boldsymbol{u_b}\rvert_{\tilde{\Gamma}} \rVert}
    \\
    &\quad
    + C_{pl}\,{\boldsymbol{u_b}\rvert_{\tilde{\Gamma}}}^{\,1/n_{pl}}
    + \boldsymbol{\epsilon}_{\boldsymbol{\tau}_b}
    + \boldsymbol{\epsilon}_a .
\end{aligned}
\end{equation}
where \(\boldsymbol{\epsilon}_a\) represents the additional errors inherent to the approximations for Eqs. \ref{eq:constitutive_slipsumaggregate} and \ref{eq:powerlawD} and where the \(|_{\tilde{\Gamma}}\) makes it explicit that this is the average outcome over the given grid-cell. Finally, dropping the \(|_{\tilde{\Gamma}}\), simplifying further to 1D, and setting total error \(\boldsymbol{\epsilon}_t=\boldsymbol{\epsilon}_{\boldsymbol{\tau}_b} + \boldsymbol{\epsilon}_a\) gives Eq. \ref{eq:slidingcompound1D} introduced at the beginning of this section (with further consideration to the components of the total error term given in Section \ref{s:errors}). Building upon the assumptions and simplifications introduced throughout Section \ref{s:inout} this expresses basal traction as the sum of an effective-pressure dependent regularised-Coulomb term for \(\boldsymbol{t}_f^\parallel\) and a power-law term that is not effective-pressure dependent for \(\boldsymbol{t}_f^\perp\). The adjustable parameters in Eq. \ref{eq:slidingcompound} %(\(C_{rC}\), \(u_t\), \(m_{rC}\), \(C_{pl}\), \(n_{pl}\)) 
are now only loosely connected to underlying processes and bulk sliding-relevant properties. Future detailed numerical modelling for velocity and stress fields of the \emph{real} region can determine more closely (i) the validity of the regularised-Coulomb and power-law approximations and (ii) reasonable and realistic values for parameters that are presently poorly constrained my targeted studies (\(C_{pl}\), \(n_{pl}\)).

\subsubsection{ Velocity at the sliding-bulk layer interface as a simplified function of traction}

Generally, \(\boldsymbol{u}_b\) and \(\boldsymbol{\tau}_b\) cannot be prescribed independently but form part of the solution of an ice sheet model. The default in most models is to set \(\boldsymbol{\tau}_b\) as the quantity derived through a sliding relationship. We cannot invert Eq. \ref{eq:slidingcompound} analytically as it combines distinct terms in \(\boldsymbol{u}_b\). However, Eq. \ref{eq:slidingcompound} is monotonic for positive constants \(C\), \(u_t\), and \(\boldsymbol{\epsilon}\) values so a unique \(\boldsymbol{u}_b\) exists for any \(\boldsymbol{\tau}_b>0\). A solution can therefore be obtained numerically if required, at negligible computational cost relative to a global finite-difference or finite-element solution. It may also be possible to build a full framework based on a summation of velocity components (\citealp{Cuffey2010TheGlaciers} Section 7.2.1) but we have not attempted that here. 

\subsection{ General error terms incorporated within sliding parameterisations} \label{s:errors}

%[\comment{Be careful with the error terms following ivan chat. But should make kind of talk about how the xi will be different if taubtildegamma represents an ideal model or a production model (where compensating errors become more important). But it should be enough to just talk about that and not do lots of formal maths. }.]

In most production models initialised using a present-day velocity field, the sliding coefficient is either the only spatially-variable free parameter inverted for, or it is inverted for alongside or in sequence with \(A\) in Glen's flow law (Eq., \ref{eq:glenn}; \citealp{Morlighem2013InversionModel, Gagliardini2013CapabilitiesModel, Ranganathan2020AStreams, Barnes2021TheModels}). This means that the sliding coefficient does more than just represent sliding at the \emph{model} bed, it also incorporates the \emph{misrepresentations} of fields such as ice temperature, surface mass balance, and basal topography, as well as misrepresentations of ice physics such as ice rheology or approximations of the Stokes equations (see \citealp{Barnes2021TheModels} and \citealp{Berends2023CompensatingResponse} for more detailed introductions to inversion procedures and associated problems). Even relatively minor adjustments to these physics/fields can then lead to compensating errors that  exceed spatial variation in the traction coefficient before the adjustments are introduced \citep{Berends2023CompensatingResponse}. To be comprehensive in describing what a tunable sliding coefficient, \(C\), is actually representing within a given production model, it is then appropriate to include a term related to compensating errors. This is a really poorly understood problem. Initialisation procedures, which include the inversion for the sliding coefficient, can modify model outputs over time \citep{Aschwanden2021BriefLevel} but the role played by the sliding relationship has not been investigated to our knowledge.

For example, making bulk ice rheology too stiff will result in incorrectly low basal traction obtained through an inversion by way of compensation, but it is unclear whether or how each of these errors now compensated within \(C\) will affect the relationship between \(\boldsymbol{\tau}_b|_{\tilde{\Gamma}_b}\) and \(\boldsymbol{u}_b|_{\tilde{\Gamma}_b}\). 
We denote this term \(\boldsymbol{\epsilon}_c(\theta_{p_c})\) where \(\theta_{p_c}\) is the set of parameters influencing the relationship and avoid speculating too far on its nature but note that if it is sufficiently large, \(\boldsymbol{\epsilon}_c(\theta_{p_c})\) may become a dominant term in a sliding relationship in a production model. \(\boldsymbol{\epsilon}_c\) is not included in Eq. \ref{eq:slidingcompound1D} or \ref{eq:slidingcompound} -- to do so we need to treat \(\boldsymbol{\tau_b}\big|_{\tilde{\Gamma}}\) as the traction across an actually operational production model grid cell (rather than a \emph{model} mirror of the \emph{real}), which we have not done previously to avoid complication and additional notation. However, it is simple to follow this alternative definition for \(\boldsymbol{\tau}_b|_{\tilde{\Gamma}_b}\) and thereby include \(\boldsymbol{\epsilon}_c\) within \(\boldsymbol{\epsilon}_t\).
%\law{Considering adding an appendix where we just wildly speculate on this, but probably not that wise}

\subsubsection{ Varying \(\iota \) and issues related to a non-zero \(R_r\) thickness}
\label{s:notoneiota}

For a given velocity and deviatoric stress field (and excluding unusual behaviour that would lead to a decrease in velocity, or increase in deviatoric stress with height above the bed), increasing \(\iota\) and therefore the height of the bulk-sliding interface will lead to an increasing \(\boldsymbol{u}_b\) and decreasing \(\boldsymbol{\tau}_b\). With low \(\iota\) this is a reasonable trade off that allows us to relate a \emph{model} grid cell to complex \emph{real} topography and rheology and is in keeping with previous studies (Fowler77). However, taken to extremes (say, an \(\iota\) of 0.5) this clearly renders a sliding layer unworkable. If varying \(\iota\) does not influence the form of the constituent components Eq. \ref{eq:slidingcompound} then its variation becomes less important as it will be corrected by an inversion procedure anyway. %Or, a compensation term similar to \(F_{\boldsymbol{s}_i}^{\nabla p' }\) can be included. 
%Increasing the model bed position, \(\tilde{b}_b\), by \(\eta H\) would also remove double-counting of bulk-layer processes. 
Including forces at the \emph{real} sides, \(\Gamma_s\), would address this problem, but then the form of Eq. \ref{eq:slidingcompound}, becomes more complicated, with more parameters to account for (and, being related to ice dynamics these terms may be power-laws in any case). The height at which sliding-layer processes become negligible will also vary as a function of the spatial scale under consideration, heterogeneity in bed properties, the roughness characteristics of the bed topography, and the resolution used in the numerical model. For example, sliding-layer processes will be important at a greater height above the bed if considering an area of rough topography around a planar soft-bed region, than if considering the planar soft-bed region in isolation. We outline these issues here, but leave full quantification for a future study. 

\section{ Discussion} \label{s:discussion}

%[\comment{Some discussion requried about how the complications in the classic papers often come from working out geometry parameters, but we're not even trying that.}]

\subsection{ Applying a sliding-bulk layer framework to multiple sub-processes: bridging hard and soft beds and highlighting the importance of topographic roughness} \label{s:appeq8}

%We have separated resistance arising from sliding at the scale of grid-cells into tangential and normal components operating at a (slightly) smoothed \emph{real} bed, \(\bar{b}_f\). 
In keeping with past studies \citep[for instance][]{Minchew2020TowardLaw, Trevers2024ApplicationGreenland, Zhao2025SubglacialRise} there is strong evidence to treat the tangential resistance, \(\boldsymbol{t}_f^\parallel\), over the smoothed \emph{real} bed, \(\bar{b}_f\), using a regularised-Coulomb relationship related to \(N\). As explored in previous work and again in Section \ref{s:overlap}, the general form holds for both sediment shearing and cavitation, though with complications relating to \(u_t\) and the physical meaning of a single spatially-tunable sliding coefficient (Sections \ref{s:threshold}, \ref{s:physicalmeaning}). 

%https://www.nature.com/articles/s41467-025-58375-4

New (or re-introduced) here however, is the setting-dependent importance of a normal resistance, \(\boldsymbol{t}_f^\perp\), contribution to traction across a model grid cell surface, \(\tilde{\Gamma}_b\). This is hinted at or implied in multiple papers investigating the importance of roughness on basal traction \citep{Bingham2017DiverseFlow, Kyrke-Smith2018RelevanceAntarctica,Hoffman2022TheSliding} where -- as explored earlier -- decreasing fidelity (and therefore decreasing the roughness of the model bed) leads to increasing traction. It is clear that the source of the increased traction is not an increase in \(\boldsymbol{t}_f^\parallel\), but an increase in \(\boldsymbol{t}_f^\perp\) even as \(\tilde{\Gamma}_b\) itself remains planar as it grows in area. These studies deal with resolutions \(\geq\)100 m, and at this scale a possible role for cavitation is reasonably implicitly omitted. The question then, is what scale this form drag descends to, before cavitation partially or wholly submerges its influence.

\begin{figure}
\centering{\includegraphics[width=0.42\textwidth]{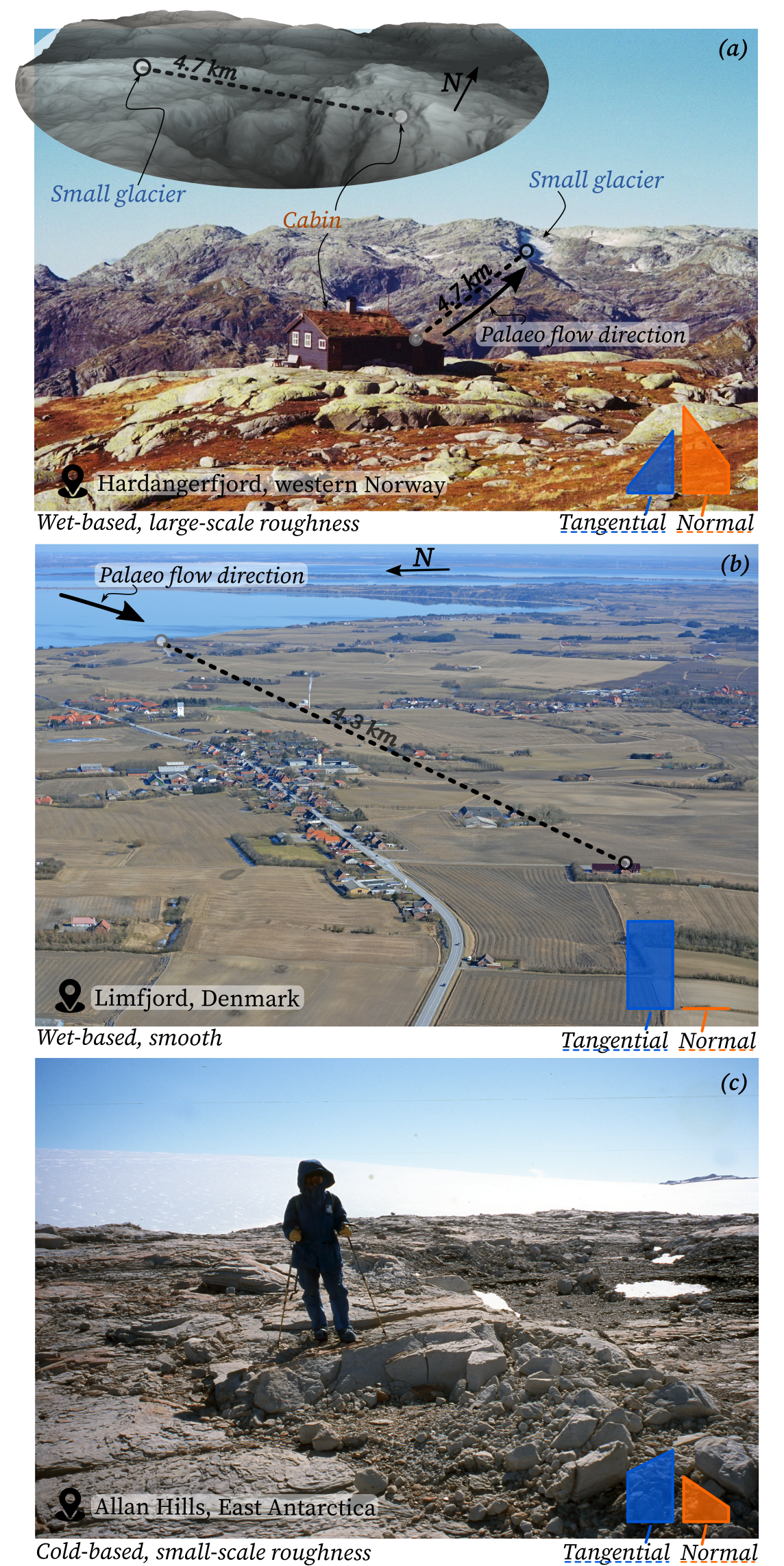}}
\caption{\textbf{Varied ice sheet and glacier beds with different dominant sliding modes.} (\textbf{a}) Mountains on the northern flank of Hardangerfjord in western Norway. Hut location is 60.4526 N, 6.1406 E, inset shows ArcticDEM topography for the valley between the hut and the small glacier in the distance with no vertical exaggeration \citep{Porter2018ArcticDEM}. Palaeo flow direction at around 20 kyr ago available from \citet{Mangerud2019Ice-flowNorway, Jungdal-Olesen2024TheDimensions}. Numerous very steep surfaces opposing the direction of palaeo ice sheet motion can be seen at both a micro and macro scale. Photo taken by RL in September 2024. (\textbf{b}) Limfjord in northern Denmark with a view over Nørre Nissum from lemvig.eu. The Main Stationary Line of the Fennoscandinavian Ice Sheet was around 10 km south along the flow direction \citep{Nielsen2007ExtentDenmark}. (\textbf{c}) Exposed bed adjacent to a cold-based section of the East Antarctic Ice Sheet at Allan Hills. Photo from \cite{Atkins2013GeomorphologicalAntarctica} provided by Cliff Atkins. Tangential (blue) and normal (orange) resistance components are illustrative proportional contributions. In (\textbf{a}), slip rate varies from low in troughs to high at crests in counter-sync to strain rate \citep{Law2023ComplexIceb}. In (\textbf{b}), parallel resistance is present while perpendicular resistance is effectively absent. In (\textbf{c}) both parallel and perpendicular components could be important.}  
\label{fig:hardanger}
\end{figure}

We discuss the cavitation scale in the next Section. For now, we emphasise that `non-cavitating' roughness can have a major influence upon the relationship between \(\boldsymbol{u}_b|_{\tilde{\Gamma}_f}\) and \(\boldsymbol{\tau}_b|_{\tilde{\Gamma}_f}\) and may assist in explaining the variety of sliding relationship forms found in Table \ref{tab:heuristic}. Palaeo ice sheet beds help to make this point. In some situations (such as many locations across western Norway which functions as a reasonable proxy for Greenland Ice Sheet bed conditions; Fig. \ref{fig:hardanger}a; \citealp{Barndon2025IceLandscapes}) there are clearly landforms with reverse bed-slopes that would not be captured by a discretised production model bed even if one had full knowledge of the bed conditions. Contributions to \(\boldsymbol{\tau}_b|_{\tilde{\Gamma}_f}\) for a given grid-cell region from both \(\boldsymbol{t}_f^\parallel\) and \(\boldsymbol{t}_f^\perp\) are then expected if the roughness exceeds the cavitation scale. Alternatively, an entirely planar region -- or more strictly a region where \(\Gamma_f\) and \(\Gamma_b\) align -- such as northern Denmark in Fig. \ref{fig:hardanger}b will have negligible \(\boldsymbol{t}_f^\perp\). If \(\boldsymbol{t}_f^\perp\) is well-represented through a power-law (Eq. \ref{eq:powerlawD}) then the net effect of increasing non-cavitating roughness in a given grid cell is that \(\boldsymbol{\tau}_b|_{\tilde{\Gamma}_f}\) will transition from bounded to unbounded. A side-by-side comparison of planar and rough grid-cells is Fig. \ref{fig:3D}. Note that this framework will still work frozen regions with a rough bed (Fig. \ref{fig:hardanger}c; Section \ref{s:localfreeze}). We suggest that this places non-cavitating roughness as a much more important determiner of the form of the sliding relationship than a traditional hard-soft divide. The power-law component related to non-cavitating roughness is furthermore only indirectly connected to effective pressure, \(N\), through its influence on tangential resistance. 

%From Hoffman et al., 2022 "Theoretical descriptions of the basal boundary condition often separate the contributions from the material friction at the ice-bed interface (skin drag) and the enhanced stresses within the ice required to overcome geometric obstacles to flow (form drag; Kyrke-Smith et al., 2018). Models that lack sufficient resolution to capture small-scale features of the ice-sheet bed have no way to disentangle these two controls on ice flow. The result is a basal friction proxy that misrepresents sources of stress at the ice-bed interface."
%Quick paragraph or two to say that most sliding studies focus on \(\boldsymbol{t}_f^\parallel\). Studies varying resolution show that there is a clear component from \(\boldsymbol{t}_f^\perp\) and this demonstrates that topographic roughness is important. It perhaps just happens that hard beds happen to contain more topographic roughness. 

\subsection{ Maximum cavitation sizes, and revisiting Iken's bound}
\label{s:ikens}

Here we address two issues: the maximum cavitation length-scale and possible caveats to Iken's bound regardless of scale. First, the question of maximum cavitation length-scale has not been explicitly explored previously to our knowledge and we therefore have limited understanding of where \(l_c\) ought to lie, with the upper limit in non-dimensionalised cavitation studies only limited by the two-dimensional geometry used (and therefore abstracted away from real glacier beds). Dimensional, or field-based cavitation studies generally do not insinuate cavities larger than around 15 m (Fig. \ref{fig:hydro}; \citealp{Vivian1973SubglacialFrance}; \citealp{Walder1979GeometryCavities}; \citealp{Helanow2021ATopography}) but this is not the main motivation behind these studies, so further intentional quantification would be welcome. We highlight here however, that maximum cavitation length in a given setting will be controlled by geometry -- a single protuberance in an otherwise flat region will be able to support a longer cavity than a pervasively rough region -- and may further be co-related to hydrology.
This also invites the question of when a cavity become a lake. We suggest lakes can be distinguished as features where subglacial water pressure balances overburden, while cavities require ongoing basal slip to sustain their geometry and typically have a water pressure slightly below overburden. For now, and largely as a conservatively high placeholder value in lieu of future studies, we set \(l_c\) at 100 m (shown in Fig. \ref{fig:scale}) and leave the important task of improved quantification to a future study.  

Beyond the scale for \(l_c\) there are additional issues that may influence the formulation of Iken's law as outlined in \citet{Schoof2005TheSliding}. First, slip resistance is not included. In \citet{RoldanBlasco2025ImpactSliding} plastic slip resistance increases Iken's bound as defined in Eq. \ref{Eq:iken} by a slip-rate-independent value (Section \ref{s:hard}). If \(\tau_p\) was instead taken from a regularised-Coulomb relationship then this %update on Iken's bound 
would introduce an additional basal slip velocity dependence. Second, the term \(\mathrm{tan}(\theta)\) in Iken's bound (Eq. \ref{Eq:iken}) will tend towards infinity as slopes opposing the prevailing approach perpendicular such that it is no longer an effective bound in the situation of steeply oriented faces, with steep faces not uncommon in deglaciated landscapes characterized by extensive bedrock exposure (e.g. Fig. \ref{fig:hardanger}, \citealp{Barndon2025IceLandscapes}). Third, cavitation models assume that \(N\) remains constant and is not influenced by the change in cavity geometry as basal velocity increases. This is not an unreasonable assumption, but it remains untested whether this holds as cavities grow and coalesce, or whether the increased hydraulic conductivity enables lower-resistance escape pathways and hence a reduction in \(N\). Even if these three mechanisms adjust the bound upwards rather than removing it, the question then is if the bound has the possibility of being reached for realistic glacier driving stresses, which are below 0.1 MPa in \(\sim\)80\% of cases \citep{Meyer2018Freeze-onGlaciers}. We highlight these possibilities to motivate further research into these important controls.  

%https://agupubs.onlinelibrary.wiley.com/doi/10.1029/2022JF007028

%while in three-dimensions over realistic small-scale (<25 m) topography the maximum traction is sustained as \(u_b\) increases, with \(\gtrapprox\) 50\% of the bed flooded by cavities in a mature system \citep{Helanow2021ATopography}. However, 

\subsection{ Threshold velocities for regularised-Coulomb sliding} \label{s:threshold}

In a regularised-Coulomb relationship the transition velocity, \(u_t\), describes the sliding velocity at which resistance to motion transitions from power-law towards plastic behaviour and exerts significant control over the form of the sliding relationship. Lower (higher) \(u_t\) values result in higher (lower) traction values at low velocities and an earlier (later) approach to the maximum sliding resistance. In large-scale numerical models using regularised-Coulomb sliding, \(u_t\) is usually set as constant across the domain with values from 100-300 m a\textsuperscript{-1} common (e.g. \citealp{Joughin2019RegularizedAntarctica, Robinson2020Description1.0, Moreno-Parada2023SimulatingMaximum, Leger2025AAI, Kazmierczak2024AGlaciers}), and even reaching 10,000 m a\textsuperscript{-1} as part of sensitivity studies \citep{Trevers2024ApplicationGreenland}. Most of these studies justify the use of a regularised-Coulomb relationship through \citet{Zoet2020ABeds} and \citet{Helanow2021ATopography} (or earlier, \citealp{Schoof2005TheSliding, Gagliardini2007Finite-elementLaw}), however these two physically-based studies indicate somewhat lower values of \(u_t\) than appear in larger-scale models, with substantial variability in response to parameter variation and a strong dependence on effective pressure, \(N\), in particular.% (Fig. \ref{fig:transition}).

\begin{figure}
\centering{\includegraphics[width=0.35\textwidth]{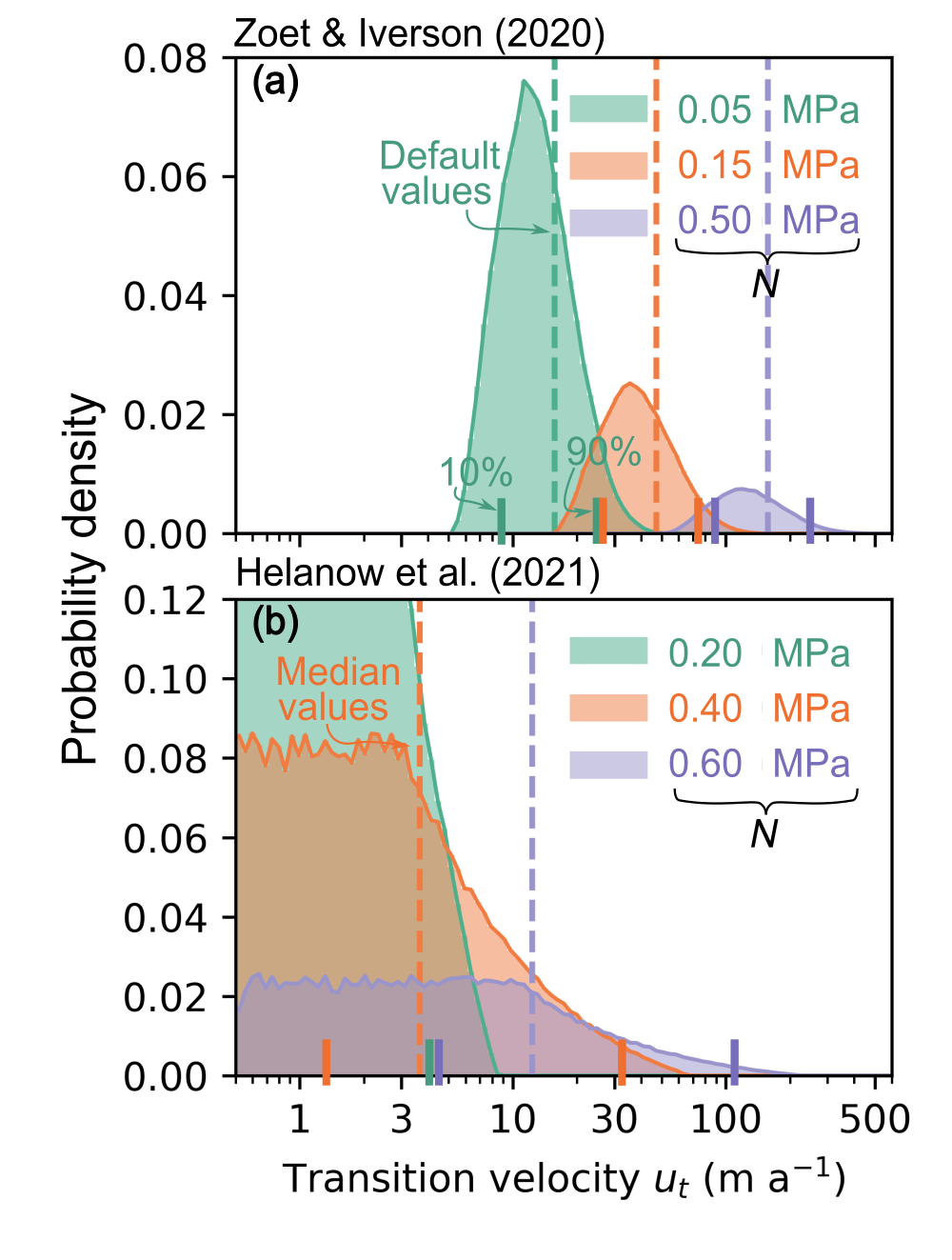}}
\caption{\textbf{Probability distributions for transition velocities, \(u_t\), from (a) \citet{Zoet2020ABeds} and (b) \citet{Helanow2021ATopography} from a Monte Carlo simulation.} We use a realistic range of values in Zoet and Iverson (2020, Eq. 2) centred upon `default' values in their Table S1, with the vertical dashed lines showing the \(u_t\) value produced with their default value for each effective pressure, \(N\). We use upper and lower values from Helanow and others (2021, Table S1) setting \(u_t=A_sC^nN^n\) in their Eq. 3 with the vertical dashed line showing \(u_t\) at their median values. Monte Carlo simulation run with 10\textsuperscript{5} samples. Vertical solid tick marks represent 10-90\% distribution bounds for each \(N\). Parameter values used are conservative, wider ranges result in wider distributions. See Appendix \ref{A:transition} for further details. Greater apparent distribution volumes at lower transition velocities are a result of the logarithmic x-scale.}  %with written permission from Peter Knight.}
\label{fig:transition}
\end{figure} 

With default parameter values from Zoet and Iverson (2020, Eq. 2), including an \(N\) of 0.15 MPa (the average of their values), the calculated \(u_t\) is 47.7 m a\textsuperscript{-1}. This decreases to 3.7 m a\textsuperscript{-1} for the median values from \citet{Helanow2021ATopography} using their default 0.4 MPa value, % and setting \(u_t = A_s C^nN^n\) in their Eq. 3, 
with original experiments in \citet{Helanow2021ATopography} not exceeding 100 m a\textsuperscript{-1}. (Note that an \(N\) of 0.45 MPa corresponds to a thickness of 1,000 m and a \(p_w=0.95p_i\).) A Monte Carlo sensitivity analysis (Fig. \ref{fig:transition}, Appendix \ref{A:transition}) furthermore shows large spread under reasonably anticipated parameter and effective pressure variation, though caution may be reasonable if applying this approach for \(N\) values increasingly distant from those used in the original experiments. %(We also note that we did not vary \(n\) in \citet{Helanow2021ATopography} for Fig. \ref{fig:transition} as their \(A_s\) value would need recalibrating and as is it produces an incredibly large spread).

Considering only \citet{Zoet2020ABeds} and \citet{Helanow2019SlidingBeds} it is clear that \(u_t\) can vary substantially as a result of reasonable parameter variation making it challenging to pin down the physical meaning of \(u_t\) to one, or a group of physical variables. %, or to suggest that \(u_t\) should be fixed as a spatially-consistent parameter to which clear physical meaning can be attributed (further discussed in Section \ref{s:physicalmeaning}). 
For instance, for the \citet{Zoet2020ABeds} case, a low \(N\) but small clast radius may produce the same \(u_t\) as larger \(N\) with larger clast radius, and it is not possible to say which parameter controls \(u_t\)  without information on both the clast size distribution \textit{and} the effective pressure -- a situation complicated further if overlapping hard-soft bed settings are considered. We are further unaware of studies that explicitly represent the (very strong) dependence of \(u_t\) on \(N\) within their models -- hydrology is most often implicitly incorporated within \(C\) in studies investigating variation in \(u_t\) (e.g. \citealp{Jouvet2019FutureGlacier, Trevers2024ApplicationGreenland}) or in \citet{Kazmierczak2024AGlaciers} included as in Eq. \ref{eq:constitutive_slip}, but not within \(u_t\) itself which remains fixed. This is a significant problem that does not have a simple resolution -- \(u_t\) may reasonably be anticipated to have a similar degree of spatial variation as \(C\), with a similar influence upon the basal traction relationship, yet there is no straightforward way to invert for both of these quantities simultaneously if the problem is under-determined. Studies that use high values of \(u_t\) (with no link between \(N\) and \(u_t\)) may furthermore not be physically based -- a high \(u_t\) may occur in practice if \(u_t\) performs as a proxy for a different physical processes, such as the decrease in effective pressure towards the grounding line (where velocities are higher). Clearly, further work is required to improve this situation. In the meantime, it is challenging to advocate for values of \(u_t\) much greater than 100 m a\textsuperscript{-1} or for spatially constant \(u_t\) values at all if the intention is that these \(u_t\) values represent physically meaningful quantities. 

%\subsection{ Implications for glacier and ice-sheet modellers} \label{s:implications}

\subsection{ Sliding and scale}
\label{s:scale}

Returning to the representative area element from the introduction. If the dimensions of a model grid cell prone to cavitation well exceed the scale of cavitation then the tangential component of traction, \(\boldsymbol{t}_f^\parallel\), acting over the smoothed real bed, \(\bar{b}_f\), can reasonably be viewed as being scale-independent as the grid cell becomes larger. %(though note we do not introduce an equivalent to \(l_c\) for sediments as this condition will be met already even at resolution <10 m). 
However, while \(\boldsymbol{t}_f^\perp\) can be ignored at sufficiently high model resolutions if the \emph{} model bed follows the geometry of the non-cavitating smoothed \emph{real} bed, \(\bar{b}_f\), the normal resistance will become increasingly important as grid-cell dimensions increases past \(l_{c}\) in settings where non-cavitating roughness is present. If \(\boldsymbol{t}_f^\parallel\) and \(\boldsymbol{t}_f^\perp\) are well represented by regularised-Coulomb and power-law relationships respectively, then this increase in \emph{model} grid-cell area %move from \(l_{c_1}\) to \(l_{c_2}\) 
will be concomitant with an increasing dominance of the power-law term. % in areas featuring non-cavitating roughness. 
Further, it is at present questionable if an upper limit exists %, say \(l_{c_3}\),
beyond which \(\boldsymbol{t}_f^\perp\) is no longer dependent on scale. This has been implied in previous studies \citep{Schoof2003TheDynamics} and in the case of relatively regular large-scale roughness (e.g. Fig. \ref{fig:3D}a) it seems like a reasonable prospect, but very large topographic features that span several km and are not picked up in bed topography products (e.g. Fig. \ref{fig:3D}b; \citealp{Barndon2025IceLandscapes}) may complicate the matter. Therefore, while this should be an area for future research, it may be reasonable to treat glacier sliding as a a phenomenon that is strongly scale-dependent and which cannot be universally captured through a set representative area element. %than bulk ice rheology.  

%https://doi.org/10.1007/s00161-003-0119-3

Typical resolutions for glacier and ice sheet models share only a small degree of overlap (Fig. \ref{fig:schematic}), so scale may also be a bifurcating factor between the two applications. However, heuristic studies of valley glaciers indicate power-law relationships with \(m\) values around 3-4 (Table \ref{tab:heuristic}), which at first glance implies a large contribution from a power-law informed \(\boldsymbol{t}_f^\perp\). This is certainly possible, valley glaciers typically have % a steeper nature with 
rough beds and side-wall buttressing as a result of their, by-definition, mountainous settings. %Bed topography products for valley glaciers can feature just as much relative uncertainty as ice-sheet bed topography products (TO DO, cite). 
There is however a second possible explanation: many valley glaciers simply do not reach the high threshold velocities commonly invoked in regularised-Coulomb studies (Section \ref{s:threshold}; \citealp{Laumann1992ReactionsLevel, Jouvet2019FutureGlacier}) such that one could interpret a power-law relationship from a data series that would exhibit a limiting traction value if velocity became high enough (although we emphasise that many valley glaciers should have some regions exceeding the threshold velocity discussed in Section \ref{s:threshold}). Caution could therefore be required if applying the findings of sliding studies conducted at valley glaciers %\citep{Gimbert2021DoSpeed, Gilbert2023InferringGlacier}
to ice sheets, and vice versa %\citep{Tsai2021ASliding}
, %in the same manner that care is needed in applying laboratory and numerical modelling studies to glacier and ice sheet settings, 
even though the individual sub-processes will remain valid in both settings. Last, while observations of point locations of the glacier bed are incredibly useful (i.e. Svartisen Glacier, Norway) some caution is required in coupling them to discretised models, where the point observation may not necessarily be representative of the grid cell it is contained within (cf. \citealp{Doyle2018PhysicalGreenland}; \citealp{Law2021ThermodynamicsSensing}).  

%By convention, the sliding of ice sheets is treated in the same manner as the sliding of valley glaciers from which foundational sliding theories are typically derived (e.g., \citealp{Hodge1974VariationsGlacier, Engelhardt1978BasalPhotography, Fischer1997StickslipGlacier, Willis1995Intra-annualReview, Bouchayer2024Multi-scaleSvalbard}). However, differences intrinsic to the two settings, and choices made in model development, may significantly influence the combination of sliding sub-processes occurring and those represented within a given grid cell. 

%Valley glaciers (i.e. those occupying a limited geographical area, not draining a large accumulation area ice field, and comprising only a handful of tributaries) have several features distinguishing them from both ice-sheet interiors and ice-sheet outlet glaciers. First, valley glaciers generally occupy a sufficiently small area such that, as a rule, major geological boundaries are not crossed beneath glacier, making a distinction between soft- and hard-bed sliding more reasonable (and indeed consequential, with surging glaciers generally occurring over inferred soft beds; \citealp{Cuffey2010TheGlaciers}, Section 12.2). 

%In addition, if the glacier is sufficiently thin, for example over a rise before an ice fall, the inner flow may effectively occupy the entire ice column -- i.e. no overlying ice will be only minimally influenced by bed conditions. 

\subsection{ Does a `single-term universal sliding law' exist?} \label{s:unified}

In most studies focussing on production model applications there is an inference that glacier sliding can be described through a single-term equation (typically selected from those in Table \ref{tab:eqs}) -- the spatial variability of which is driven through a single parameter (typically \(C\) within Table \ref{tab:eqs}) that can be reasonably obtained through an inversion procedure -- which provides an accurate description of sliding processes across the settings, scales, and time periods applicable to production models. (Eq. 6 of \citealp{Barnes2022TheAntarctica} is the only use of a compound sliding relationship in any model that we are aware of.) Furthermore, it is often assumed that such a `single-term universal sliding law' is neatly connected directly to a specific and limited set of processes or material parameters occurring at or defining the ice-bed interface (for instance, \citealp{Hank2025TheObservations}). This is not descriptive of every study, particularly those interested in relating diurnal or seasonal hydrology to velocity changes, but it is pervasive (see for example the introductions of the papers listed in Table \ref{tab:heuristic}, or commentaries such as \citealp{Minchew2020TowardLaw}). For the purposes of our discussion (and to avoid introducing more new terminology) we refer to such a universally applicable relationship for production modes with one spatially variable coefficient as a `single-term universal sliding law'. We emphatically note that it will of course be possible to describe glacier sliding in a unified manner provided sufficient complexity is permitted %We emphatically note that given sufficient complexity a unified manner of describing glacier sliding will exist 
-- our focus here is whether glacier sliding can be well-represented through a single-term single-spatially-tunable-coefficient relationship as outlined above. %I can maybe half-remember one study where they have a dual term relationship just can't remember what it's called !! Think in tc somewhere. Good to introduce as e.g. 'In fact, we are only aware of ... where this is not the case'

Unfortunately, the broad utility of a single, simple `single-term universal sliding law' as outlined above is not well-supported by the breadth of single-term sliding relationships required to reproduce the setting-specific behaviour and observations outlined through the heuristic studies in Section \ref{s:heuristic}, or through the range of behaviours requiring representation outlined in Sections \ref{s:slidingprocesses}, \ref{s:inout}. This makes it challenging to advocate for a `single-term universal sliding law' that can be applied indiscriminately across settings appropriate for production model use. This paper has explored the various possible reasons for the existence of this misfit and we emphasise three major potential causes: (i) the contribution of non-cavitating roughness in transitioning a given grid cell from a bounded to unbounded traction regime; (ii) variability in appropriate \(u_t\) and \(m\) values; and (iii) compensating errors introduced through misrepresentations in other aspects of the model. 

\subsubsection{ \track{Tuning coefficients or parameters with actual physical meaning?}} \label{s:physicalmeaning}

In addition, the diversity of bed conditions, -- both across ice sheet and glacier settings and within individual grid cells -- means that, even if form drag and compensating errors can be neglected, it is difficult to link a single sliding coefficient to physically interpretable parameters governing tangential slip. A process-based connection to parameters is sensible in controlled studies with a limited set of sub-processes: for example, \citet{Zoet2020ABeds} relate the threshold velocity, \(u_t\), to the protruding fraction of clast radii and to the thermal properties of ice and rock, subsequently approximated by a bulk till-friction angle, while \citet{Helanow2021ATopography} determine \(A\)
from very slow simulations in which cavitation does not occur. However, even between these two controlled frameworks, intercomparison of physically defined parameters is non-trivial. At the scale of a grid cell containing both soft- and hard-bed regions, the regionally averaged sliding coefficient \(C\) cannot be uniquely attributed to underlying bed conditions: equifinality allows the same value of \(C\) to arise from many distinct combinations of soft-bed and hard-bed parameters (see also Appendix \ref{a:slip}), even before considering ice–bed conditions outside the range of these studies (for example the diversity in sediment reviewed by \citealp{Evans2006SubglacialClassification}), hydrology, form drag, or compensating errors.

Studies attempting to relate observational geophysics to the sliding coefficient mirror this challenge. \citet{Kyrke-Smith2017CanModels} report a correlation between acoustic impedance and a power-law sliding coefficient at large (\(\sim\)7 km) but not small (\(\sim\)1 km) scales, consistent with earlier findings by \citet{Vaughan2003AcousticStreams} and \citet{Smith2015MappingMicroseismicity}, yet highlight the absence of a physical theory linking impedance to the parameters controlling slip. Similar work using radar specularity and reflectivity finds a weak and slightly more reasonable correlation in \citet{Das2023InAntarctica} and \citet{Haris2024WhatAntarctica}, respectively, but these proxies also lack direct mechanistic ties to the processes represented by \(C\). This means that even where correlations exist, interpreting them as causal or mapping \(C\) to a reduced set of physical parameters rather than to coarse geophysical attributes remains unresolved. Geostatistical approaches \citep{MacKie2021StochasticGlacier, Mackie2023GStatSimSimulation} offer an alternative by linking observations to \(C\) while retaining the substantial observational and process uncertainty inherent to basal conditions. % (Section \ref{s:threshold}).

In some respects, the situation is analogous to our understanding and implementation of ice rheology, which is more tractable to isolate in laboratory settings. Although ice deformation is better represented by multi-term flow relations \citep{Goldsby2001SuperplasticObservations} and is further complicated by anisotropy, impurities, and evolving microstructure, the Nye–Glen isotropic flow law remains the pragmatic choice for production models. We routinely describe ice behaviour through the bulk parameter \(A\) in Eq. \ref{eq:glenn} (optionally modified by an enhancement factor), rather than through explicit variables such as grain size or fabric. Adopting an analogous outlook for glacier sliding (consistent with other geophysical problems such as reservoir hydrology or slope stability) we suggest that an agnostic basal traction coefficient is a more appropriate descriptor of basal conditions than deterministic quantities such as `roughness' or `till friction angle', which rarely map in a straightforward manner onto the heterogeneous processes contributing to basal drag.

\subsubsection{ What is the optimum basal sliding relationship?}

A single-term sliding relationship with effectively one spatially tunable coefficient remains the most practicable form for implementation in large-scale, production models. Advocating for dual- or even tri-term formulations (e.g. Eq.~\ref{eq:slidingcompound1D}) is perhaps more physically motivated, but introduces additional parameters that are not yet able to be uniquely constrained through available observations and may simply expand the space of equifinal solutions. At the same time, changes in the mathematical form of the sliding relationship do lead to substantial differences in model output (Section~\ref{s:background}), shifting the problem from identifying an `optimum' law to choosing a simple one whose inevitable misfits are as low as possible for the application at hand.

The criteria for a `successful' production-model sliding relationship is then that it minimises systematic biases arising from this necessary simplification of basal processes. For production models, %there is understanding %broad implicit agreement 
%that 
the basal traction relationship has two further constraints on its form. The relationship should pass through the  \((u_b,\tau_b) = (0,0)\) origin, and not exhibit rate-weakening over the velocity range of interest (i.e. be single-valued). Variation between commonly used single-term parameterisations then arises primarily from (i) whether \(\tau_b\) will reach an effective upper limit over the modelled velocity range, (ii) the exponent \(m\) as a control on how rapidly \(\tau_b\) increases with \(u_b\), and (iii) how the effective pressure \(N\) is included. If we accept that the sliding relationship used in an ice-sheet model is not directly tied to a small, well-resolved set of basal processes (Section \ref{s:physicalmeaning}, and not a new viewpoint: \citealp{Schweizer1992TheBed}), then there is justification for reproducing this qualitative behaviour with either a single-term power law or a regularised-Coulomb-type law. In the former case, large values of \(m\) can make the curve resemble a bounded law over the finite range of velocities realised in a model, while in the latter case small threshold velocities \(u_t\) ensure that the Coulomb-like plateau is attained within that range.

This paper outlines many outstanding issues relating to sliding, and might therefore be expected to offer a definitive solution, yet no single solution within the confines of a `single-term universal sliding law' is immediately forthcoming. Future methodology that astutely handles a multi-term relationship, or multiple spatially tunable coefficient presents a promising route forwards, but also a major challenge. In the interim, we tentatively suggest that power-law sliding with a setting-dependent exponent \(m\) is a reasonable option in many (though not all) settings, sidestepping the additional \(u_t\) parameter needed in a regularised-Coulomb relationship. (Note that in this case although \(u_t\) is removed from the sliding relationship and greater \(m\) values will steepen the earlier portion of the curve, the transition velocity remains an important and uncertain control.) Guided by the limited number of heuristic and process-based studies available (Table~\ref{tab:heuristic}) and consistent with our earlier discussion (Sections~\ref{s:appeq8}--\ref{s:unified}), lower values (\(m \lesssim 4\)) are better suited to coarser resolutions where non-cavitating roughness contributes significantly to overall traction, while higher values of \(m\) are more appropriate in regions where the bed is expected, or known, to be extensively smooth. In this view, \(C\) remains the primary object of inversion, while \(m\) may be assigned based on additional constraints such as topographic roughness or setting. In areas known to be extensively planar below scale of cavitating roughness a bounded regularised-Coulomb relationship may be as, or more,  appropriate, with care required for the \(u_t\) value. We hope, however, that these suggestions primarily motivate the substantial further work required to refine the functional form(s) of the sliding relationship, to quantify their setting-specific variation and suitability, and to characterise the errors -- particularly compensating errors -- introduced by any given choice.

%Based primarily on heuristic studies (Section \ref{s:heuristic}), whose empirically-validated nature makes them more suitable for application to glacier and ice-sheet models. we suggest that power-law sliding (or equivalently, pseudo-plastic and Weertman-\textit{type} sliding), with a setting-dependent \(m\), presents a reasonable first choice. 

%While we do not doubt the efficacy of regularised-Coulomb in some settings (Table \ref{tab:heuristic}), it is simpler to manipulate a power-law relationship to approximate regularised-Coulomb using a high value of \(m\) than vice versa. %As models for glaciers or ice sheets frequently invert for traction coefficients over regions of inferred frozen beds \citep{Goelzer2020TheISMIP6, Seroussi2020ISMIP6Century}, it is further likely that a regularised-Coulomb relationship would misrepresent these areas. 

Last, it is advantageous to include an effective pressure term in cases where the error associated with including an approximation for \(N\) is lower than the error associated with excluding it. This is straightforward to include in a regularised-Coulomb or power-law relationship. A height above buoyancy approach modified to keep effective pressure below \(\sim0.05\rho_igh\) (rather than allowing it up to \(\rho_igh\); Section \ref{s:hydrology}) is the simplest way to do this though more involved methods may also be suitable. Setting \(N\lesssim0.05\rho_igh\) presents issues for cold-based regions where there is no subglacial hydrology system and hence no meaningful \(N\), but this is acceptable if cold-based ice regions are of lower dynamic importance than faster-flowing warm-bedded regions. We do not advocate for a non-unity exponent for \(N\) given the lack of process- or observation-based evidence (Section \ref{s:hydrology}). %Our analysis (Section \ref{s:inout}) does not find a dependence of the power-law component on effective pressure, but its inclusion in a production model using a power-law relationship is useful nonetheless if we accept (as above) the sliding relationship is a representation of the bulk behaviour, rather than directly tied to the physical processes.  

\section{ Conclusions}
\label{s:conclusion}

We have outlined glacier sliding as arising from a potentially complex set of processes operating in the vicinity of the glacier bed, the relative importance of which can vary substantially within and between glacier settings, and which -- in production model settings -- can be defined as the negative of `standard' bulk ice deformation. The complexity inherent to sliding, together with the limited resolution at which basal conditions can be observed or represented, has led to a variety of partly disconnected theoretical and modelling treatments. Here -- sitting alongside ongoing uncertainties -- we have attempted to outline a clear framework that links these perspectives together and allows for the relative importance of subprocesses and errors to receive further future quantification. As this complexity is effectively irreducible under realistic observational capabilities, and muddied further by compensating errors to a presently unclear degree, the sliding law used in a production model is best regarded as a bulk descriptor of this combination of processes rather than a direct statement about any specific set of mechanisms. (Nonetheless, improving understanding of processes and compensating errors offers a good route towards reducing overall errors.)

This approach implies that smooth-to-rough may be a more consequential distinction than a traditional soft-to-hard-bed separation, with the as-yet poorly quantified upper length boundary for pervasive cavitation proving an important divider between bounded (regularised-Coulomb) and unbounded (power-law) basal traction. A dual- (or even tri-) term relationship (Eq. \ref{eq:slidingcompound1D}) may offer a more faithful representation of this behaviour and is an important future objective, but in the interim a single-term relationship remains the more practical choice in many production-model settings, where it is often only feasible to invert for one spatially variable parameter. 

%A dual (or even tri) term relationship  may offer improved realism, a single-term sliding relationship remains the most practical choice for production model settings where it is often only feasible to invert for one effective basal parameter. Both power-law and regularised-Coulomb relationships are capable of emulating this range of behaviour (Section \ref{s:unified}) but, in view of the difficulty of determining threshold velocities such as \(u_t\) (Section~\ref{s:threshold}), and in line with several heuristic studies (Table~\ref{tab:heuristic}), a power-law form with a setting-dependent exponent \(m\) may offer a workable first approximation. In this scheme \(C\) remains the principal quantity to be inferred, while \(m\) may be assigned based on information such as local topographic roughness (Sections~\ref{s:appeq8}--\ref{s:unified}).

Our discussion highlights several more open objectives:

\begin{enumerate}
    \item Improved understanding of ice rheology and anisotropy (including for temperate ice), and their importance for the stress state of the sliding layer (Sections \ref{s:form}, \ref{s:BIL})
    \item A clearer methodology for relating stick--slip behaviour to continuous sliding relationships across different settings (Section~\ref{s:stickslip})
    \item Improved understanding and consensus-building for how effective pressure \(N\) should be stated and parametrised within production models (Section~\ref{s:hydrology})
    \item Much improved constraints on the transition scale between seldom and frequent cavitation, as well as any setting-specific dependence (Section~\ref{s:scalecavities})
    \item Quantitative measures of how larger-scale roughness influences grid-scale basal traction, and how these relationships vary with model resolution (Section~\ref{s:deformation})
    \item Further assessment of the importance of compensatory error terms and whether this will fundamental alter the anticipated form of a sliding relationship (Section~\ref{s:errors})
    \item The sensitivity of model outcome on spatially variable and \(N\)-informed \(u_t\) (Section \ref{s:threshold})
    \item Better isolating the importance of grid-cell resolution on the appropriate sliding relationship, and whether this implies a bifurcation between glacier and ice-sheet applications (Section \ref{s:scale})
    \item Determining whether there is a single-term sliding law can minimise error across diverse settings, or whether setting-dependent relationships are preferable (and if so, how to appropriately implement this as methodology; Section~\ref{s:unified})
\end{enumerate}

%Overall, our aim has been to provide a clear structure within which existing ideas about glacier sliding can be interpreted and compared, and to offer a basis for evaluating how simple sliding relationships are used in numerical models.

%...is [of all friction laws??] regularized-Coulomb sliding relationships, though it is notably still absent from large-scale ice sheet models used in ISMIP predictions. In regularized-Coulomb relationship increasing subglacial water pressure is almost equatable with decreasing the traction coefficient \citep{Zoet2020ABeds, Helanow2021ATopography}. However, measuring a representative subglacial water pressure over a given area is a fiendish task given almost all field and modelling studies point to varying subglacial water pressure from scales of 10s of centimeters onward (cites). This is made more challenging by borehole-based pressure sensors only sampling a single point in space. Pressure sensors in tandem (e.g., \citealp{Doyle2021WaterGreenland}), or in groups (e.g., \citealp{Rada2018ChannelizedYukon}, add other Rada cite) can give indications of behaviour in the regions in between, but we are still a long way from a full understanding of this. 

\bibliography{references}   % reads igsrefs.bib
\bibliographystyle{igs} 

\section{ Acknowledgements} 

RL thanks Norges Forskningsr\aa d for funding as part of the SINERGIS project (Norwegian Research Council Grant 314614, Simulating Ice Cores and Englacial Tracers in the Greenland Ice Sheet) and the ETH Zurich Postdoctoral Fellowships program. 

\appendix
\renewcommand{\thesection}{A\arabic{section}}
\renewcommand{\thefigure}{A\arabic{figure}}
\setcounter{figure}{0}  % Reset figure counter for appendix

\clearpage        % flush floats, optional but usually sensible
\onecolumn        % switch to one-column mode

\section{ Notation} \label{A:notation}

Notation provided in Table A\ref{tab:notation}

\begin{longtable}{lll}
\caption{Notation used throughout the paper. Where vector and scalar forms are shown in the paper (e.g. \(\boldsymbol{\tau}_b\) and \(\tau_b\)), only the vector form is given here. [] in Units implies quantity is dimensionless.}
\label{tab:notation} \\
\toprule
\textbf{Symbol} & \textbf{Description} & \textbf{Units} \\
\midrule
\endfirsthead

\toprule
\textbf{Symbol} & \textbf{Description} & \textbf{Units} \\
\midrule
\endhead

\midrule
\multicolumn{3}{r}{\emph{Continued on next page}} \\
\bottomrule
\endfoot

\bottomrule
\endlastfoot

\multicolumn{3}{l}{\textit{Stress, strain, and velocity}} \\
$\boldsymbol{\tau}$ & Deviatoric stress tensor & Pa \\
$\boldsymbol{\tau}_b$ & Basal traction vector at \emph{real} sliding-bulk later interface or \emph{model} bed & - \\
$\boldsymbol{\tau}_f$ & Basal traction vector at \emph{real} floor & - \\
$\boldsymbol{t}_f$ & traction projected at the \emph{real} floor & - \\
$\dot{\boldsymbol{\epsilon}}$ & Strain rate tensor & a\textsuperscript{-1} \\
$\boldsymbol{u}$ & Velocity vector & m a\textsuperscript{-1} \\
$\boldsymbol{u}_b$ & Basal velocity vector at \emph{real} sliding-bulk later interface or \emph{model} bed & - \\
$\boldsymbol{u}_f$ & Basal velocity vector at \emph{real} floor & - \\[3pt]

\multicolumn{3}{l}{\textit{Pressures and forces}} \\
$p$ & Pressure & Pa \\
$p_i$ & Ice overburden pressure ($\rho_i g H$) & - \\
$p_w$ & Water pressure & - \\
$N$ & Effective pressure ($p_i - p_w$) & - \\
$p_h$ & Hydrostatic pressure component & - \\
$p'$ & Non-hydrostatic pressure component & - \\
$F_{\boldsymbol{s}_i}^{\Gamma_b}$, $F_{\boldsymbol{s}_i}^{\tilde{\Gamma}_b}$ & Shear forces acting tangentially across $\Gamma_b$ and $\tilde{\Gamma}_b$ & N \\[3pt]

\multicolumn{3}{l}{\textit{Regions, boundaries, surfaces, and surface-defined vectors}} \\
$R_r$ & \emph{Real} region &  \\
$R_m$ & Mirroring \emph{model} region &  \\
$b_f$ & \emph{Real} ice-bed interface position &  \\
$\bar{b}_f$ & \emph{Real} ice-bed interface position smoothed to remove cavitating roughness &  \\
$b_b, \tilde{b}_b$ & \emph{Real} sliding-bulk interface and \emph{model} bed positions &  \\
$\Gamma_f, \Gamma_b, \Gamma_s$ & \emph{Real} ice-bed interface, bulk-sliding interface, and sides defining \(R_r\) &  \\
$\tilde{\Gamma}_b$ & \emph{Model} bed defining \(R_m\) &  \\
$V_r$ & Volume of \(R_r\) & m\textsuperscript{3} \\
$a_b, \tilde{a}_b$ & Areas of $\Gamma_b$, $\tilde{\Gamma}_b$ & m\textsuperscript{2} \\
$\boldsymbol{s}_i \in \{ \boldsymbol{s}_x, \boldsymbol{s}_y \}$ & Unit vectors aligned with the local tangnet plane of \(b_b\) and $x$ and $y$ axis &  \\
$\boldsymbol{n}_f, \boldsymbol{n}_b$ & Upwards pointing unit vectors of normals of $b_f$ and $b_b$ &  \\[3pt]
$\theta$ & Maximum up-flow bed-slope & \textsuperscript{o} \\[3pt]

\multicolumn{3}{l}{\textit{Sliding and ice deformation relationship parameters}} \\

$n$ & Nye-Glen isotropic flow law exponent & [] \\
$A$ & Ny-Glen isotropic flow law rate factor & Pa$^{-n}$ a$^{-1}$ \\
$m$ & Sliding relationship exponent if separate from $n$ & [] \\
$C$ & Sliding coefficient & Pa m\textsuperscript{$-m$} a\textsuperscript{$m$} \\
$C_p$ & Sliding coefficient for plastic component & Pa \\
$C_{rC}, C_{pl}$ & Sliding coefficients for regularised-Coulomb and power-law components & Pa m\textsuperscript{$-m_{rC,pl}$} a\textsuperscript{$m_{rC,pl}$} \\
$m_{rC}, m_{pl}$ & Sliding relationship exponents for regularised-Coulomb and power law components & [] \\
$\mu_s, \mu_t$ & Static and transient coefficients of friction & [] \\
$\psi$ & State variable in rate-and-state relationship & [] \\
$\tau_c$ & Yield stress in pseudo-plastic relationship & Pa \\
$\tau_p$ & Area-averaged plastic resistance in \citet{RoldanBlasco2025ImpactSliding} & - \\
$\alpha$ & Scaling factor for basal-temperature dependent sliding & [] \\
$u_t$ & Threshold velocity in pseudo-plastic and regularised-Coulomb sliding relationships & m a\textsuperscript{-1} \\[3pt]

\multicolumn{3}{l}{\textit{Temperature}} \\
$T*=T_{\textrm{pmp}}-T_b$ & Difference between the pressure melting point, $T_{\textrm{pmp}}$, and basal temperature $T_b$ & \textsuperscript{o}C \\ [3pt]

\multicolumn{3}{l}{\textit{Other}} \\
$l_{c}$ & Transition length-scale for cavitation & m \\
$H_i$ & Total ice thickness & - \\
$\iota$ & The proportion of total ice thickness included in the sliding layer & [] \\
$\boldsymbol{\epsilon}_{\boldsymbol{\tau}_b}$, $\boldsymbol{\epsilon}_c$, $\boldsymbol{\epsilon}_t$ & Error terms for \emph{real}-\emph{model} mismatch, compensating errors, and total, respectively & \\
$\theta_{p_s}$, $\theta_{p_d}$, $\theta_{p_c}$ & Parameter sets influencing \(\boldsymbol{t}_f^\parallel\), \(\boldsymbol{t}_f^\perp\), and \(\boldsymbol{\epsilon}_c\), respectively &  \\

\end{longtable}

\twocolumn      % switch back to two-column mode

\section{ Differential sliding relationships} \label{A:differential}

The main focus of this paper is algebraic relationships between \(\tau_b\) and \(u_b\) -- or in other words where there is a direct and immediate link between the two quantities -- and therefore which sliding sub-processes are represented in the coefficients used. However, there are other more involved methods of relating \(\tau_b\) and \(u_b\). These allow further variables to be incorporated into sub-process parameterisations, but may not be appropriate for production models. For example, a differential term could be included for time-varying subglacial water pressure making Eq. \ref{Eq:1}
\begin{equation}
    \tau_b = f\left(u_b, \frac{\partial p_w}{\partial t}\right)
    \label{eq:differential}
\end{equation}

where \(t\) is time. A further discussion of relationships incorporating \(\frac{\partial p_w}{\partial t}\) as a function of spatial variations in \(p_w\) and hydraulic diffusivity is given in \citet{Tsai2021ASliding}. Technically, one may also wish to define relationships that include a dependently or independently calculated \(N\) (e.g., \citealp{Cook2020CoupledGreenland} and even if not dependent on \(\frac{\partial p_w}{\partial t}\)) as `implicit', though we refrain from doing so here. 

\begin{figure*}
    \centering{\includegraphics[width=0.6\textwidth]{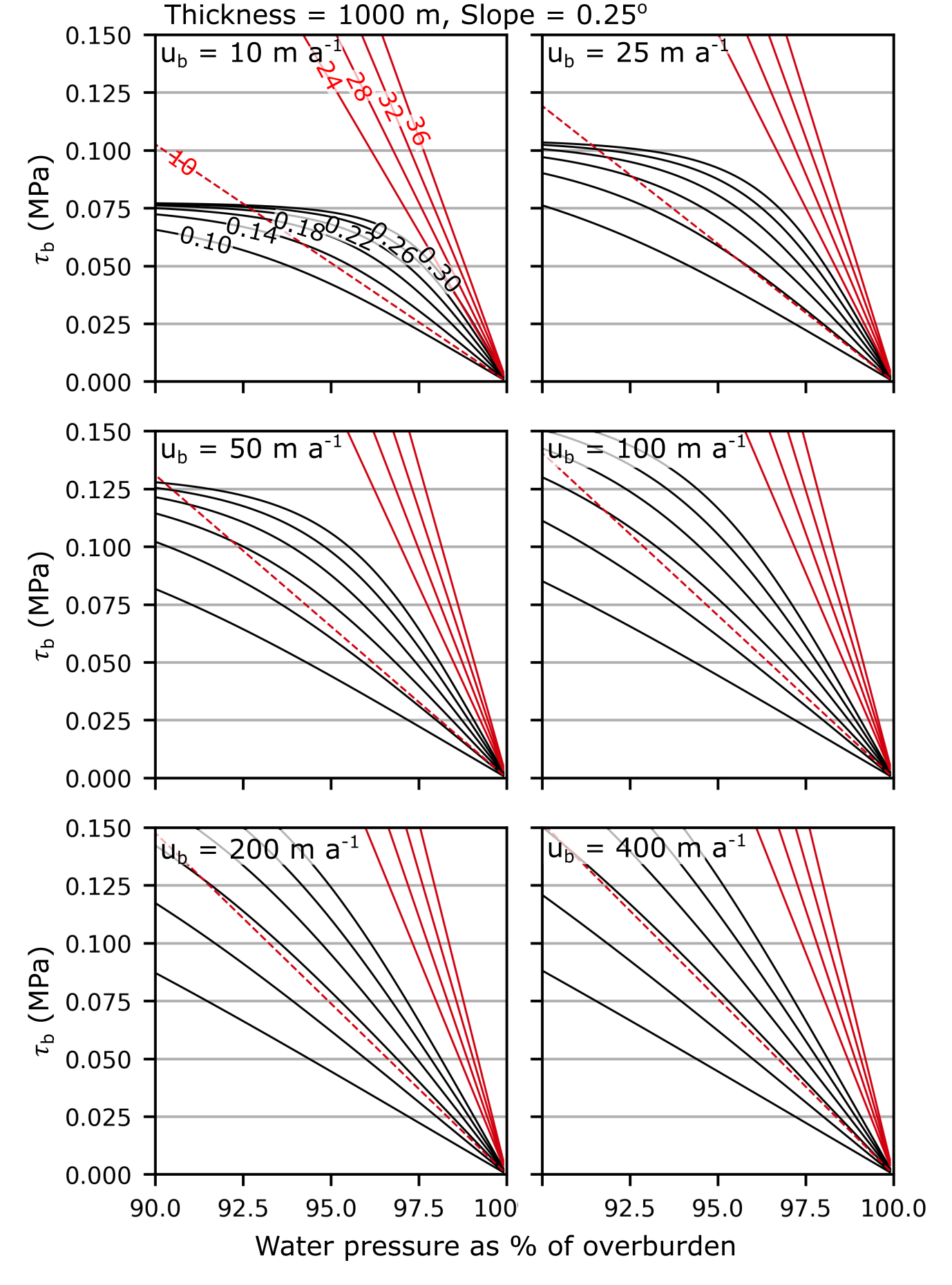}}
    \caption{\textbf{The difference between \citet{Zoet2020ABeds} and \cite{Helanow2021ATopography} at high water pressures.}. \citet{Zoet2020ABeds} in red and \citet{Helanow2021ATopography} in black using default parameters reported in those studies with thickness = 1000 m and slope = 0.25\textsuperscript{o} for \(u_b\) between 10 m a\textsuperscript{-1} and 400 m a\textsuperscript{-1}. Values in black refer to the \(C\) parameter of \citet{Helanow2021ATopography}. Values in red refer to friction angle in \citet{Zoet2021TransientSlip}.}
    \label{fig:helanow-zoet}
\end{figure*}

\section{ Reasoning behind line placements in Fig. \ref{fig:scale}} \label{A:scale}

\textbf{Form drag, sediment} from \citet{Zoet2020ABeds} and \citet{Minchew2020TowardLaw} based on the smallest clast sizes used in \citet{Zoet2020ABeds} but assumed to continue to sub-milimetre scales (the ice crystal size used is milimetre-scale) and extended to a reasonable but conservative upper limit for subglacial debris size. 
\textbf{Skin friction, sediment} from \citet{Zoet2020ABeds} and \citet{Minchew2020TowardLaw} based on the size of the shear-ring apparatus used in \citet{Zoet2020ABeds} and extending to sub milimetre-scale of subglacial till sample used. Assumed to continue to sub-milimetre scales and to be valid at higher spatial scales given low topograhic variation. 
\textbf{Weertman sliding} from  \citet{Weertman1957OnGlaciers} with 1 cm to 10 m scale clearly stated in the text. 
\textbf{Budd} smaller-scale segment from laboratory tests in \citet{Budd1979EmpiricalSliding} and larger-scale segment from application to west Antarctica in \citet{Budd1984ASheet}. 
\textbf{Nye-Kamb sliding} Based on landscape used in Fig. 2 of \citet{Nye1970GlacierApproximation}. 
\textbf{Rate-and-state, Argentière} using assumed seismic rupture length in \citet{Helmstetter2015BasalMotion}.
\textbf{Cavitation theory} from \citet{Lliboutry1968GeneralGlaciers}, \citet{Gagliardini2007Finite-elementLaw}, and \citet{Helanow2021ATopography} amongst many others. Cavitation theory is often non-dimensionalised but has not exceeded the 25 m scale used in \citet{Helanow2021ATopography}. The dashed arrow is extended to 100 m to denote a plausible but untested upper limit to cavitation.
\textbf{Intermediate scale processes} taken from the lower resolution and upper domain size used in \citet{Law2023ComplexIceb} and extended upwards with a dashed arrow to indicate an untested upper limit for the processes described. 
\textbf{Argentière Wheel} based on the Argentière wheel experiments of \citet{Vivian1973SubglacialFrance}, \citet{Gimbert2021DoSpeed} and \citet{Gilbert2023InferringGlacier} amongst others where the glacier width is \(\sim\)300 m and 20 m is roughly double the cavity length.
\textbf{Skin drag} from \citet{Kyrke-Smith2018RelevanceAntarctica} based on the lowermost resolution of radar data (40 m) to the 5 km upper limit of Bedmap2 \citep{Fretwell2013Bedmap2:Antarctica}. \textbf{Form drag, topographic} from \citet{Bingham2017DiverseFlow} and \citet{Kyrke-Smith2018RelevanceAntarctica} using a lower radar data resolution of 40-100 m and an upper domain area of 20 km. 
\textbf{'Unified'}, from 1 km lower limit suggested in \citet{Tsai2021ASliding} to a plausible upper limit of 10 km.
\textbf{Rate-and-state, Siple Coast} Taken as broad area over which seismicity is recorded at Siple Coast in Fig. 4 of \citet{Podolskiy2016Cryoseismology} in contrast to single events of \citet{Helmstetter2015BasalMotion}.
\textbf{Form drag, resolution/scale dependent} based on arguments within this paper (Section \ref{s:form}). 
\textbf{Typical glacier and ice sheet model resolution} based on standard resolutions from detailed glacier studies to coarse resolution paleo simulations. \textbf{Typical basal ice crystal size} from \citet{Thorsteinsson1995CrystalGreenland} and \citet{Cook2007TheIceland}.

\section{ Sliding relationships in geodynamics} \label{A:geodynamics}

In geodynamics problems where rate-and-state friction theory is the more common way of viewing slip, models tend to use either a free-slip between mechanical layers (e.g., \citealp{Sizova2010SubductionExperiments}; \citealp{Nakakuki2013DynamicsFormation}), a pseudo-plastic yielding based on the coefficient of frictional sliding where boundaries between layers are not made explicit but a slip horizon is created within a continuum (e.g., \citealp{Tackley2000Self-consistentYielding}; \citealp{Schmalholz2015ShearAlps}), or focus on shear-localization through a non-Newtonian rheology with grain-size evolution in an initially rheologically homogeneous media (\citealp{Bercovici2003TheConvection} and references therein). These settings tend to have much more planar interfaces and do not feature the topographic variability that characterises some glacier beds, but provide interesting comparisons for glacier sliding theory.

\begin{figure*}
\centering{\includegraphics[width=0.65\textwidth]{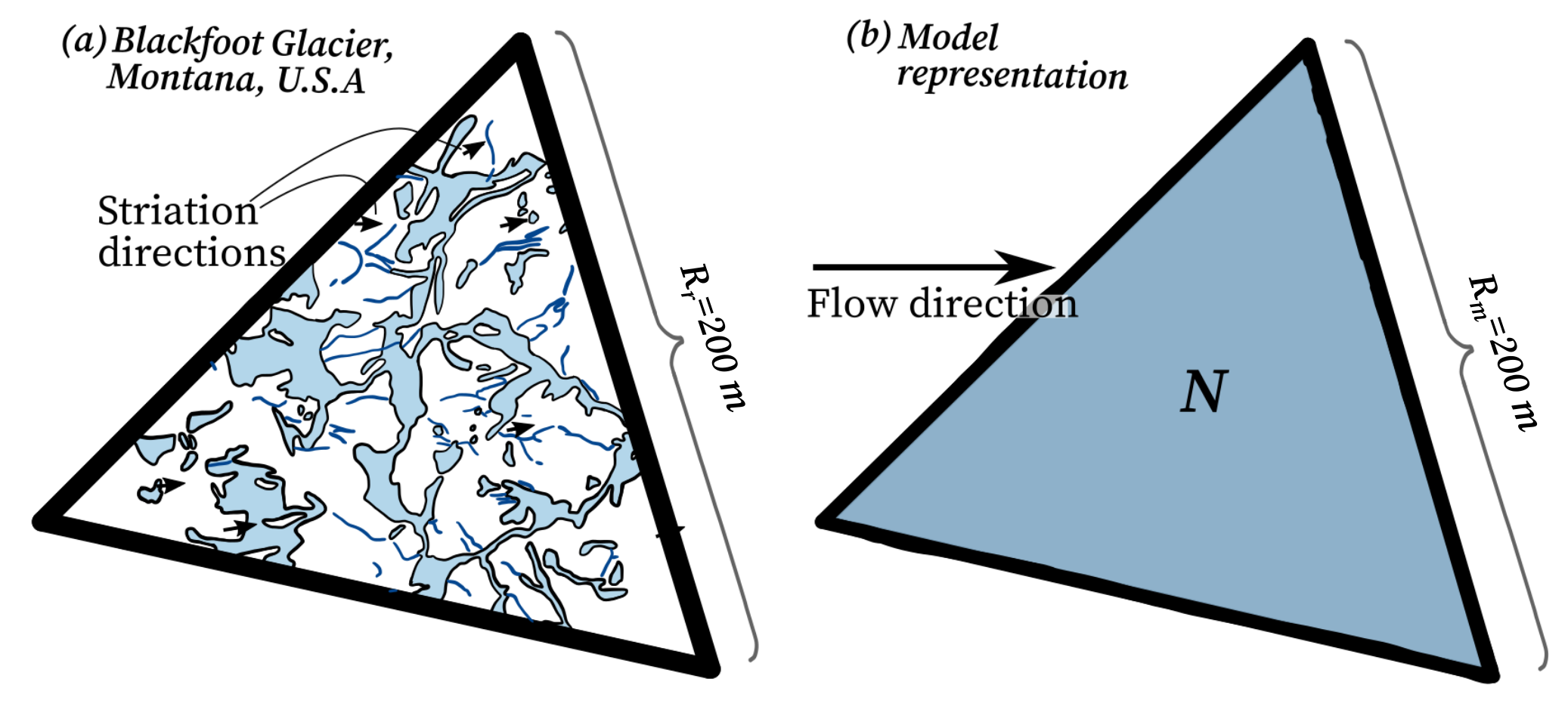}}
\caption{\textbf{Subglacial hydrology as observed in the field or represented in an ice flow model element. } \textbf{a} Cavities (pale blue), channels (dark blue lines) and striation flow markers from fieldwork at Blackfoot Glacier \citep{Walder1979GeometryCavities}. \textbf{b} Simplification to a single mesh element.}
\label{fig:hydro}
\end{figure*}

\section{ Inner-flow framework in early Fowler papers} \label{A:fowler}

\citet{Fowler1977GlacierDynamics}, \citet{Fowler1978OnAnalysis}, and \citet{Fowler1981AJSTOR} (hereafter collectively Fowler1977) partially address the \emph{real}-\emph{model} dichotomy through an `inner'-`outer' flow framework for approximating the rough component of a \textit{real} glacier bed through a smooth \textit{model} bed in two-dimensional, and we build from this work to explicitly incorporate variability in basal slip and rheological properties over three dimensions. In Fowler1977 ice motion is separated into an `inner' flow and an `outer' flow (Fig. 2.3 of Fowler, 1977). Their inner flow, which includes components of basal slip over a frictionless surface and ice deformation, is theorised to closely follow the \emph{real} bed topography (though that the \emph{real} bed in their situation is still limited to a frictionless two-dimensional surface with low bed slopes). Their outer flow accounts for internal deformation at the typical depth and length scale of the entire glacier and is considered to be `sliding' over a smoothed representation of the \emph{real} bed (i.e. the inner-outer boundary defines model bed). The outer and inner flows are then joined by a sliding relationship in an asymptotic matching region under the condition that the inner flow `feels' the outer flow as a uniform shearing flow, and the outer flow `feels' the inner flow as a tangential stress at the smoothed bed boundary.

\section{ Sliding layer force balance extension} \label{a:forcebalance}

%As \(\boldsymbol{s}_x\) and \(\boldsymbol{s}_y\) are tangential to \(\Gamma_{b}\) and \(\tilde{\Gamma}_b\), \(F_{\boldsymbol{s}_i}^{\Gamma_{b}}\) and \(\tilde{F}_{\boldsymbol{s}_i}^{\tilde{\Gamma}_{b}}\) will define shear forces. 
Considering only \(R_r\) for the time being, we integrate momentum conservation over the fixed control volume \(R_r\). Assuming that the flow is steady and non-inertial, we neglect time derivatives. Projecting the integral form of the momentum conservation on directions \(\boldsymbol{s}_i\), \(i \in \{x, y\}\) yields
\begin{equation}
    \int_{R_r} (\nabla \cdot \boldsymbol{\tau} + \rho_i \boldsymbol{g} - \nabla p ) \cdot \boldsymbol{s}_i \, dV = 0
\end{equation}
where \(\boldsymbol{\tau}=\boldsymbol{\sigma}+p\boldsymbol{I}\) is the full deviatoric stress tensor (\(\boldsymbol{I}\) is the identity matrix), \(p=-\frac{1}{3}\textrm{tr}(\boldsymbol{\sigma})\) is the total pressure, \(\rho_i\) is the density of ice, and \(\boldsymbol{g}\) is the gravity vector. We can further set \(p = p_h + p'\) where \(p_h\) and \(p'\) are hydrostatic and non-hydrostatic pressure components, respectively, and \(\nabla p_h = \rho_i \boldsymbol{g}\), simplifying \(\rho_i \boldsymbol{g} - \nabla p\) to \(-\nabla p'\). Applying the divergence theorem to the deviatoric stress component (expressing that the net deviatoric force within the volume is balanced by surface tractions acting on its boundary) and to the non-hydrostatic stress gradient, \(\nabla p'\) gives
%SEE 4.18.7 of https://wiki.epfl.ch/emem-2022/documents/CHAPTER-4---Stress-and-Integral-Formulations-o_2010_Introduction-to-Continuu.pdf
\begin{equation}
    \oint_{\partial R_r} (\boldsymbol{\tau} \cdot \boldsymbol{n}-p'\boldsymbol{n}) \cdot \boldsymbol{s}_i \, dA = 0\, .
    \label{eq:forcebalancesmol}
\end{equation}

If we omit the side faces following our established assumption in Section \ref{s:innervolume} we can write

%[\comment{Outline that this assumption is responsible for the tau max tau0 issue}]

\begin{equation}
\begin{aligned}
    \oint_{\partial R_r} (\boldsymbol{\tau} \cdot \boldsymbol{n} - p'\boldsymbol{n}) \cdot & \boldsymbol{s}_i \, dA 
    \approx \int_{\Gamma_{b}} (\boldsymbol{\tau} \cdot \boldsymbol{n}_{b}- p'\boldsymbol{n}_b) \cdot \boldsymbol{s}_i \, dA  \\
    &\quad + \int_{\Gamma_f} \left(\boldsymbol{\tau} \cdot (-\boldsymbol{n}_f)- p'(-\boldsymbol{n}_f)\right) \cdot \boldsymbol{s}_i \, dA
    \label{eq:onlynotsides}
\end{aligned}
\end{equation}
where \(\boldsymbol{n}_f\) refers to the upwards-pointing normal at \(b_f\) and which can be defined as in Eq. \ref{eq:normalnormal}. While the ice surface will be stress free, omitting forces on \(\Gamma_s\) means stress will not decrease as the thickness of the sliding layer increases. This is acceptable for low sliding layer thickness (and in keeping with Folwer1977) but would become problematic at greater thicknesses. We additionally drop the \(\approx\) from here, making the approximation implicit. We decompose the traction projected at the ice base, \(\boldsymbol{t}_f = \boldsymbol{\tau} \cdot (-\boldsymbol{n}_f)- p'(-\boldsymbol{n}_f)\), into its normal and tangential components as Eq. \ref{eq:normaltangential} (repeated below)
\begin{equation}
    \boldsymbol{t}_f\cdot\boldsymbol{s}_i = \underbrace{\boldsymbol{t}_f^{\perp} \cdot\boldsymbol{s}_i}_{\text{normal component}} + \underbrace{\boldsymbol{t}_f^{\parallel} \cdot \boldsymbol{s}_i}_{\text{tangential component}} \, 
\end{equation} 
where \(\boldsymbol{t}_f^{\perp} = (\boldsymbol{t}_f \cdot \boldsymbol{n}_f)\boldsymbol{n}_f\) and \(\boldsymbol{t}_f^{\parallel} = \boldsymbol{t}_f - (\boldsymbol{t}_f \cdot \boldsymbol{n}_f)\boldsymbol{n}_f
\).
The total force acting in the direction \(\boldsymbol{s}_i\) at the \emph{real} floor is then
\begin{multline}
    F_{\boldsymbol{s}_i}^{\Gamma_{f}} = \int_{\Gamma_f} (\boldsymbol{\tau} \cdot (-\boldsymbol{n}_f)-p'(-\boldsymbol{n}_f) \cdot \boldsymbol{s}_i \, dA = \int_{\Gamma_f} \boldsymbol{t}_f^{\parallel} \cdot \boldsymbol{s}_i \, dA + \int_{\Gamma_f} \boldsymbol{t}_f^{\perp} \cdot \boldsymbol{s}_i \, dA \, .
    \label{eq:forcefloor}
\end{multline}
Placing the expansions from Eqs. \ref{eq:onlynotsides} and \ref{eq:forcefloor} into Eq. \ref{eq:forcebalancesmol} gives

%[\comment{TODO Need to add in -p' here (but check carefully)}]

\begin{multline}
    \int_{\Gamma_{b}} (\boldsymbol{\tau} \cdot \boldsymbol{n}_{b} - p'\boldsymbol{n}_b) \cdot \boldsymbol{s}_i \, dA = 
    - \int_{\Gamma_f} \boldsymbol{t}_f^{\perp} \cdot \boldsymbol{s}_i \, dA 
    - \int_{\Gamma_f} \boldsymbol{t}_f^{\parallel} \cdot \boldsymbol{s}_i \, dA \, .
\end{multline}
%Because the free ice surface satisfies \(\boldsymbol{\sigma}\cdot\boldsymbol{n}_s \approx 0\), the integrated tangential traction on planes parallel to \(\Gamma_b\) necessarily decreases with height, with the rate set by the depth-integral of \((\nabla p') \cdot \boldsymbol{s}_i\) (and the small lateral terms we neglect), and vanishes at the surface. [Ivan says of this: This I don't get, needs reformulating (or removal)]

Taking total shear forces at \(\Gamma_{b}\) in the \(\boldsymbol{s}_i\) directions
\begin{equation}
    F_{\boldsymbol{s}_i}^{\Gamma_{b}}=\int_{\Gamma_{b}}(\boldsymbol{\tau} \cdot \boldsymbol{n}_{b} - p'\boldsymbol{n}_b) \cdot \boldsymbol{s}_i \, dA = \int_{\Gamma_{b}}\boldsymbol{\tau}_b  \, dA
\end{equation}
means \(F_{\boldsymbol{s}_i}^{\Gamma_{b}}\) can be expressed as the sum of forces operating at the ice-bed interface and \(F_{\boldsymbol{s}_i}^{\perp}\), the non-hydrostatic pressure gradient integral within the volume:
\begin{equation}
    \label{eq:Force}
    F_{\boldsymbol{s}_i}^{\Gamma_{b}} = F_{\boldsymbol{s}_i}^{\parallel} + F_{\boldsymbol{s}_i}^{\perp} \,
\end{equation}
where \(F_{\boldsymbol{s}_i}^{\parallel}\) and \(F_{\boldsymbol{s}_i}^{\perp}\) refer to resistance from slip and from normal forces over \(\Gamma_f\) in the direction of \(\boldsymbol{s}_i\) respectively, as
\begin{equation}
\label{eq:FparFperp}
    F_{\boldsymbol{s}_i}^{\parallel} = \int_{\Gamma_f} \boldsymbol{t}_f^{\parallel} \cdot \boldsymbol{s}_i \, dA \,, \ \ F_{\boldsymbol{s}_i}^{\perp} = \int_{\Gamma_f} \boldsymbol{t}_f^{\perp} \cdot \boldsymbol{s}_i \, dA \, .
\end{equation}
We refer to \(F_{\boldsymbol{s}_i}^{\perp}\) as the `drag' force or form drag from here while \(F_{\boldsymbol{s}_i}^{\parallel}\) is referred to as the `slip' force. %\(F_{\boldsymbol{s}_i}^{\textrm{body}}=\rho_i \boldsymbol{g} \cdot \boldsymbol{s}_i V_{r}\) accounts for the gravitational driving force of the volume itself which will become increasingly large relative to the overall driving force in the column projected vertically above \(R_r\) as \(\eta\) in Eq. \ref{eq:seed} increases.
This gives \(\boldsymbol{\tau}_b\) averaged over \(\Gamma_{b}\) as in Eq. \ref{eq:taubAio}

\section{ Slip extension} \label{a:slip}

To define multiple spatially independent slip subprocesses within \(R_r\) we use an indicator function, \(I_{R_i}(x, y)\), defining 0-thickness sub-regions, \(R_i\), at the ice-bed interface of the \emph{real} region, \(R_r\), where
\begin{equation}
    I_{R_i}(x,y) = \begin{cases}
    1 & \text{if } (x,y) \in R_i \, , \\
    0 & \text{otherwise} \, .
    \end{cases} 
    \label{eq:indicator}
\end{equation}
with each \(R_i \subseteq R_r\) (and not \(R_m\) which is necessarily treated as homogenous. % [\comment{possible issue with 3D region -> 2D region here?}]). 
It may be possible to extend this approach to potentially spatially overlapping slip processes, but we do not include this here. This gives a composite traction field, \(\boldsymbol{t}_f^\parallel(x, y, \boldsymbol{u}_f, \theta_{p_{s}})\), arising from an independent set of vectorised slip processes, \(\boldsymbol{\mathcal{S}}_{R_i} ( \boldsymbol{u}_f, \theta_{p_{s}})\), as
\begin{equation}
    \boldsymbol{t}_f^\parallel(x, y, \boldsymbol{u}_f, \theta_{p_{s}}) = \sum_{i \in \{\alpha, \beta, \gamma, ...\}} I_{R_i}(x, y)\boldsymbol{\mathcal{S}}_{R_i} ( \boldsymbol{u}_f, \theta_{p_{s}})\,
    \label{eq:totalslip}
\end{equation}
where the subscript \(R_i\) refers to a slip process occurring in \(R_\alpha\), \(R_\beta\) and so on (Fig. \ref{fig:inout}), \(\boldsymbol{u}_f = (u_{f, x}, u_{f, y})\) refers to the slip velocity tangential to the \emph{real} ice-bed interface within each \(R_i\) which can be defined using local basis vectors, and \(\theta_{p_s}\) refers to the set of additional parameters that may also influence the relationship (e.g. normal pressure, grain size, ice debris content) but which remain undefined here. If one wished to make the processes contained within \(\boldsymbol{\mathcal{S}}_{R_i}\) anisotropic \citep{Hindmarsh2000SlidingBeds, Barndon2025IceLandscapes} then \(\theta_{p_{s}}\) could also be vectorised. 

If we wish to be more definite over the nature of the slip relationship(s) giving rise to \(F_{\boldsymbol{s}_i}^\parallel\) we can set a generic constitutive relationship. For example, choosing regularised-Coulomb (Table \ref{tab:eqs}) as
\begin{equation}
    \boldsymbol{\mathcal{S}}_{R_i} = -I_{R_i}(x, y)NP_{R_i} \left( \frac{||\boldsymbol{u}_f||}{||\boldsymbol{u}_f||+{u_t}_{R_i}} \right)^{\frac{1}{m_{R_i}}} \dfrac{\boldsymbol{u}_f}{||\boldsymbol{u}_f||} 
    \label{eq:constitutive_slip}
\end{equation}
where \(P_{R_i}\), and \(m_{R_i}\), and \({u_t}_{R_i}\) are constants for a given region \(R_i\) and which can represent the regularised-Coulomb relationship or in a similar manner for a power-law relationship. See Section \ref{s:threshold} for further discussion regarding \(u_t\). 

As stated in Section \ref{s:inclslip}, evidence in Section \ref{s:slidingprocesses} suggests that a regularised-Coulomb relationship (Eq. \ref{eq:constitutive_slip}) may be the most appropriate relationship at length scales below 25 m. So, if we only use Eq. \ref{eq:constitutive_slip} then we can reduce the RHS of Eq. \ref{eq:totalslip} to a single term with no indicator function to give Eq. \ref{eq:constitutive_slipsumaggregate} (reprinted below) as
\begin{equation}
    \boldsymbol{t}_f^\parallel \approx \boldsymbol{\mathcal{S}}_{R_a} \approx -P_{R_a}N_{R_a} \left( \frac{||\boldsymbol{u}_f||}{||\boldsymbol{u}_f||+{u_t}_{R_a}} \right)^{\frac{1}{m_{R_a}}} \frac{\boldsymbol{u}_f}{||\boldsymbol{u}_f||}
    \label{eq:constitutive_slipsumaggregate2}
\end{equation}
where the subscript \(R_a\) refers to the aggregate or representative value or function over \(\Gamma_f\) and \(P_{R_a}\), \(N_{R_a}\), \(u_{t_{R_a}}\), and \(m_{R_a}\) are the representative parameters which can be defined as e.g., 
\begin{equation}
    P_{R_a}(x, y)=\sum_{i \in \{\alpha, \beta, \gamma, ...\}} \frac{1}{A_{b_i}}I_{R_i}(x,y)P_{R_i} \, 
    \label{eq:repr}
\end{equation}
where \(A_{b_i}\) is the area of each sub region of the ice-bed interface. It may be possible to define \(N_{R_a}\) in a similar manner; see Section \ref{s:hydrology}. %, though complications could arise if e.g. high \(N\) areas of \(R\) exert an outsized influence (Sections \ref{s:hydrology}, \ref{s:threshold}). 
While \(m_{R_a}\) cannot be calculated as a weighted average, reducing the set to a single power is a reasonable approximation in keeping with the use of a constant value over all glacier and ice sheet settings in production models, hence the \(\approx\) in Eq. \ref{eq:constitutive_slipsumaggregate2}.%, but still warrants further detailed testing. 

\begin{figure*}
    \centering{\includegraphics[width=0.7\textwidth]{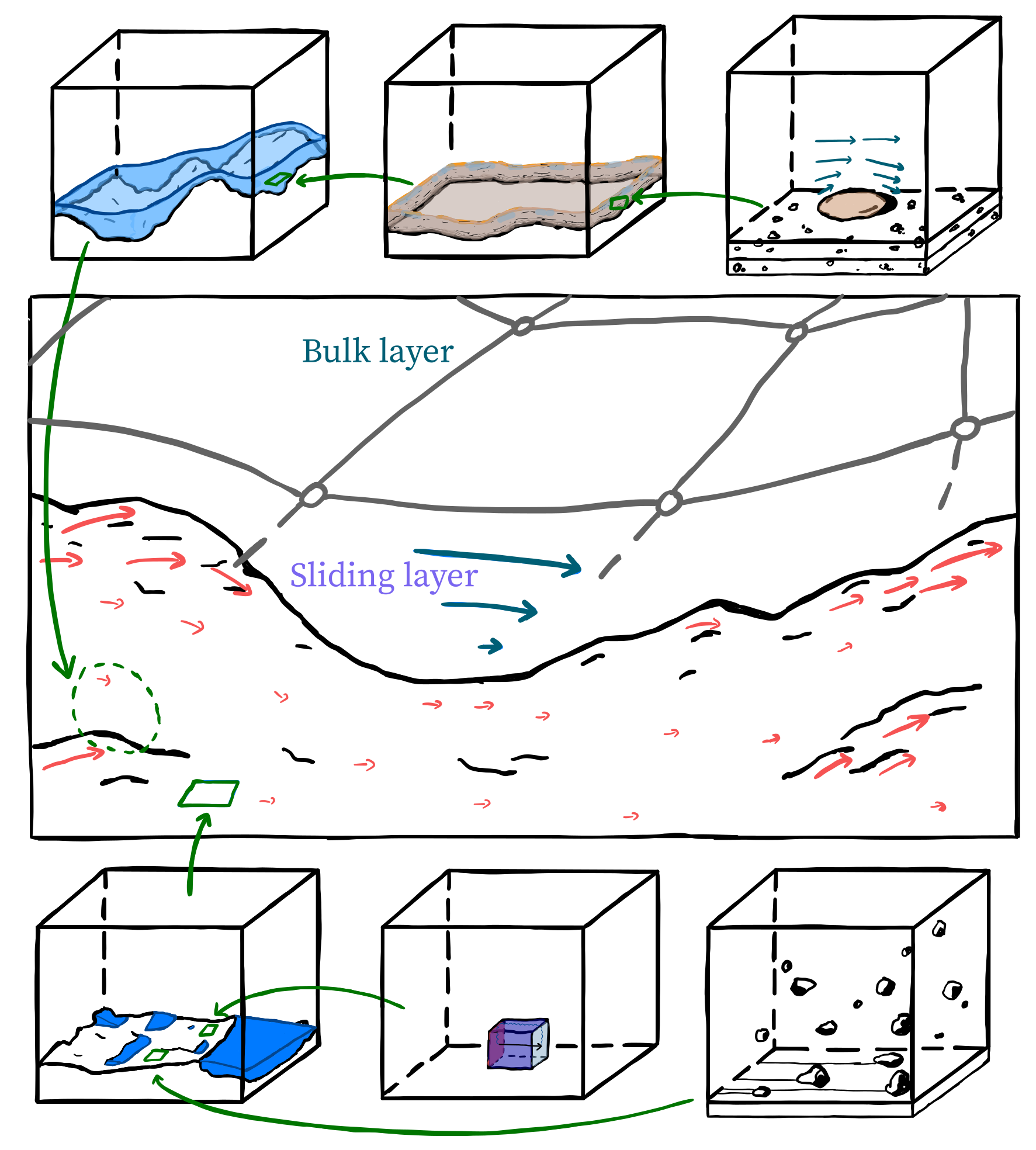}}
    \caption{\textbf{Schematic of a set of sliding processes operating within an inner flow.} Sub-processes in cubes should be recognisable from Fig. \ref{fig:modes}. Grey mesh with nodes represents model grid cells at the basal boundary.}
    \label{fig:innerouter3d}
\end{figure*}

\section{ Dimensional analysis of drag force component} \label{A:dimensional}

In order to arrive at Eq \ref{eq:dimensional} we can set \(F_D=f(U_\textrm{far}, e, A, h, n)\). Then using Buckingham \(\pi\) theorem we can write the dimensionless parameters
\begin{equation}
    \pi_1 = \frac{F_D}{A^{-\frac{1}{n}} U_\textrm{far}^{\frac{1}{n}} e^{\frac{2n-1}{n}}} \, , \, \pi_2=\frac{h}{e} \, , \, \pi_3=n
\end{equation}
and putting \(\pi_1=f(\pi_2, \pi_3)\) gives 
\begin{equation}
    \label{eq:dimensioning}
F_D=A^{\frac{-1}{n}}U_{\textrm{far}}^{\frac{1}{n}}e^{\frac{2n-1}{n}}\Phi\left(\frac{h}{e}, n\right) \, .
\end{equation}
which is the Eq. \ref{eq:dimensional} in the main text. 

\section{ Simple-shear 1D column for frozen beds} \label{A:frozencolumn}

Considering a simple shear column only (and separately to the system defined in Section \ref{s:innervolume}) we treat a frozen-bed zone as a thin shear layer of thickness \(h\) above a rigid bed, within which horizontal velocity increases from zero at the bed to \(u(h)\) at the top of the layer. We take vertical variation in shear stress as negligible compared with its overall magnitude over this limited depth range the, so we approximate \(\tau_{xz}(z)\approx \tau_b\) for \(0\le z\le h\). Glen’s law then reduces to
\begin{equation}
\frac{\partial u}{\partial z} = A\,\tau_b^{\,n},
\end{equation}
and integrating from \(z=0\) (no slip) to \(z=h\) gives
\begin{equation}
u(h)= A\,\tau_b^{\,n}\,h.
\end{equation}
Rearranging gives a local relationship for a frozen-bed under these conditions of
\begin{equation}
\tau_b = \left(\frac{u(h)}{A h}\right)^{1/n} \propto u(h)^{\,1/n},
\end{equation}
i.e. a power-law with an exponent inherited from the creep rheology alone. In the full inner–outer description this motivates the use of
\begin{equation}
\boldsymbol{\tau}_b \propto \boldsymbol{u}_b^{1/n}
\end{equation}
for frozen basal contact zones if no slip is permitted.

\begin{figure}
\centering{\includegraphics[width=0.4\textwidth]{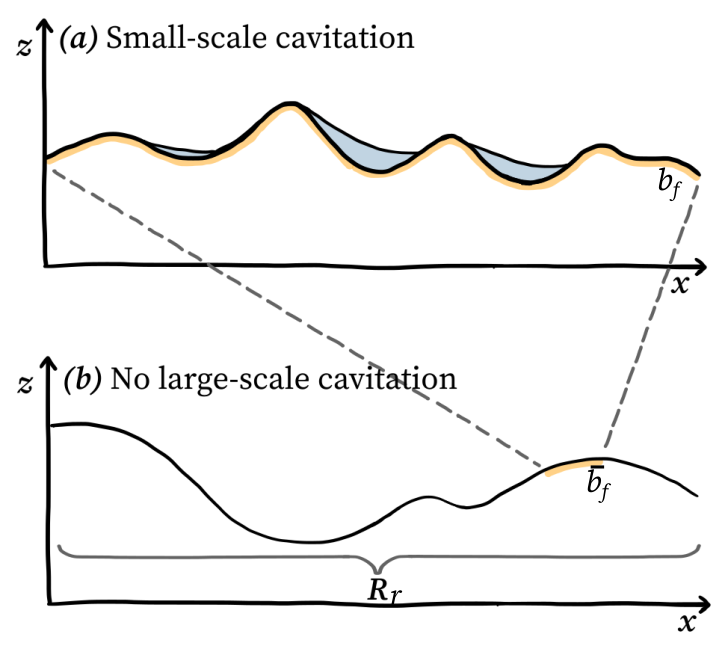}}
\caption{\textbf{A spatial bound to Iken's bound?} \textbf{a} Cavitation (blue) occurring where topography is sufficiently small adapted from \citet{Schoof2005TheSliding}. \textbf{b} Cavitation can occur at a sufficiently small scale (yellow highlight) but will not occur in larger wavelength depressions. z is height from baseline and x is distance along flow. The thick black line is the glacier bed. \(R\) represents a possible region in consideration for the sliding relationship. This figure is drawn in two dimensions for comparison with previous studies, but the same principles apply in three dimensions with realistic topography.}
\label{fig:iken}
\end{figure} 

\section{ Calculating threshold velocities} \label{A:transition}

The transition velocity $u_t$ represents the diagnostic sliding speed at which basal resistance transitions between dominance by viscous form drag and frictional slip.

We consider two models for $u_t$. For \citet{Zoet2020ABeds}, the transition velocity is 
\begin{equation}
    u_t = \left[\frac{1}{\eta (Ra)^2 k_0^3} + \frac{4 C_1}{(Ra)^2 k_0} \right] \left( \frac{n_f N}{2 + n_f k} \right)
    \label{eq:utzi}
\end{equation}

where $k_0 = \pi / (2R)$ with further parameter values in Table \ref{tab:transition} (their Eq. 2). And for \citet{Helanow2021ATopography} \(u_t\) is
\begin{equation}
    u_t = A_s C^n N^n \quad 
    \label{eq:uth}
\end{equation}

where \(C\) depends on bed morphology, \(A_s\) depends on ice rheology and bed morphology, adapted from their Eq. 3 to match \(u_t\) in Table \ref{tab:eqs} and Zoet and Iverson (2020, Eq. 3). Note values in \citet{Helanow2021ATopography} are in MPa whereas values in \citet{Zoet2020ABeds} are in Pa. 

The Monte Carlo simulations were carried out by sampling each parameter uniformly within physically and observationally motivated ranges (Table \ref{tab:transition}). For each fixed value of effective pressure $N$, $10^5$ independent random parameter combinations were generated. The transition velocity $u_t$ was computed from the above expressions using these parameters. Effective viscosity $\eta$ was varied in log space, while other parameters were sampled linearly centred upon the default of median parameters. Default parameter values (noted in parentheses) were used to compute a reference value for $u_t$ for comparison.

Larger clasts and increased \(k\) can result in much lower \(u_t\) for Zoet and Iverson (2020). \(C\) and \(A_s\) may exhibit variation beyond the values provided in Helanow and others (2020, Table S1), particularly if ice rheology parameters are varied. Varying \(n\) for Helanow and others (2020) results in substantially wider variability, but is not included, as technically \(C\) and \(A_s\) will also vary with \(n\). \(N\) values are based on a limited range either side of that used in the original studies as it is unclear the extent to which Eqs. \ref{eq:utzi}, \ref{eq:uth} are valid as \(N\) departs from default values. The ice height in \citet{Helanow2021ATopography} is \(\sim\)55 m, with a \(p_w\) only 20\% of \(p_i\), so confirmation may be required if applying this approach to a thick ice sheet setting with high \(p_w\) as a proportion of overburden. The distributions in Fig. \ref{fig:transition} are therefore likely conservative. 

\begin{table*}
    \centering
    \begin{tabular}{|p{1.4cm}|p{3.75cm}|p{3.75cm}|p{4.5cm}|}
    \hline
    \textbf{Symbol} & \textbf{Meaning} & \textbf{Range (Default)} & \textbf{Source} \\
    \hline
    $\eta$ & Effective ice viscosity & $10^{13}$--$10^{14}$ Pa s ($3.2\times10^{13}$) & Z \& I 2020, \citet{Cuffey2010TheGlaciers} \\
    $R$ & Clast radius & 0.01--0.02 m (0.0153) & Z \& I 2020, wider variability likely \citep{Evans2006SubglacialClassification}  \\
    $a$ & Fraction of clast radius protruding from bed surface. & 0.2--0.3 (0.25) & Z \& I 2020 \\
    $N_f$ & Bearing-capacity factor for till & 26--40 (33) & Z \& I 2020 \\
    $k$ & Till-strength reduction resulting from ice-pressure shadow in the lee of clasts & 0.05--0.2 (0.1) & Z \& I 2020 \\
    $k_0$ & Derived geometric constant & $\pi/(2R)$ & Z \& I 2020 \\
    $C_1$ & Regelation parameter & $6\times10^{-16}$ m$^2$ Pa$^{-1}$ s$^{-1}$ & Z \& I 2020 \\
    $N$ & Effective pressure & 0.05--0.5 MPa (fixed per run) & Both models \\
    $A_s$ & Sliding law prefactor & 274--47759 m a$^{-1}$ MPa$^{-n}$ (median) & H et al. 2021 \\
    $C$ & Sliding friction coefficient & 0.10--0.28 (median) & H et al. 2021 \\
    $n$ & Sliding exponent & 1--4 (3) & H et al. 2021 \\
    \hline
    \end{tabular}
    \caption{Parameters used in Monte Carlo simulations of the transition velocity $u_t$ (and distinct from Table \ref{tab:notation}). }
    \label{tab:transition}
\end{table*}
%\appendix\section{ Appendix E}

%\begin{figure}
%    \centering{\includegraphics[width=0.5\textwidth]{figs/A2.png}}
%    \caption{\textbf{Possibility of a different sliding relationship required for each %region \(R\) considered.} Very faint coloured lines in background from Fig. %\ref{fig:schematic} with each grey-scale line representing a sliding relation%ship for a given region \(R\) under uniform hydrology.}
%    \label{fig:slidingfun}
%\end{figure}

\end{document}